\documentclass[11pt,english,twoside]{report}
\usepackage[T1]{fontenc}
\usepackage[latin1]{inputenc}
\usepackage{graphicx}
\usepackage{setspace}
\usepackage{cite}

\makeatletter

 \usepackage{verbatim}

\usepackage{epsfig}

\usepackage{fancyhdr}
\fancypagestyle{plain}{
\fancyhead{}

}

\usepackage{babel}
\makeatother

\begin{document}

\thispagestyle{empty}

\begin{titlepage}
\title{
\hfill{\normalsize IFT-16-2005}\\
 \hfill{\normalsize July-December 2005}\\
\vspace{.5cm} 
{\LARGE Improving reliability of perturbative QCD predictions
at moderate energies} 
\vspace{.5cm}}
\author{Piotr A. R\c{a}czka\\
 Institute of Theoretical Physics, Department of Physics\\
Warsaw University, ul.~Ho\.{z}a 69, 00-681 Warsaw, Poland.}
\date{\quad{}}
\maketitle
\begin{abstract}
The problem of precise evaluation of the perturbative QCD
predictions at moderate energies is considered. Substantial
renormalization scheme dependence of the perturbative predictions
obtained with the conventional renormalization group improved
perturbative approximants is discussed, using as an example the
QCD effective charge appearing in the static interquark potential
and the QCD corrections to the Gross-Llewellyn-Smith sum rule in
the deep inelastic neutrino-nucleon scattering. A new method for
evaluating the QCD predictions at moderate energies is proposed,
based on a modified couplant, which is constructed to be free
from Landau singularity and less renormalization scheme dependent
than the conventional couplant.  The modified couplant contains
an adjustable parameter, which may be used to improve the
modified perturbative predictions in low orders by utilizing some
information outside of perturbation theory. This parameter is
fixed by matching the modified predictions to the
phenomenological expression for the effective charge in the
interquark potential. The renormalization scheme dependence of
the perturbative predictions in the modified expansion is then
discussed in detail, including the predictions selected by the
Principle of Minimal Sensitivity applied to the modified
expansion. It is shown that the predictions obtained in the
modified expansion are much more stable with respect to change of
the renormalization scheme parameters than the predictions
obtained in the conventional approach. It is also found that the
modified predictions display somewhat weaker energy dependence
then the predictions obtained with the conventional expansion,
which may help accommodate in a consistent way some low values of
the strong coupling parameter measured at low energies with the
values measured at high energies. 
\end{abstract}



\end{titlepage}


\newpage

\pagenumbering{roman}

\tableofcontents{}

\newpage

\chapter{Introduction}

\pagenumbering{arabic}

One of the most important predictions of modern theory of
strong interactions, the quantum chromodynamics (QCD), is that in
the processes occuring at high energies these interactions become
perturbative \cite{0213gross73,0215politz73}.  This phenomenon is
usually described by introducing an effective (i.e.\ running) coupling
parameter that is decreasing with increasing energy, and the
theory is said to be asymptotically free. The energy dependent
coupling parameter plays central role in the formulation of the
so called renormalization group improved perturbation expansion,
which is now a standard tool used to evaluate perturbative
predictions for strong interaction effects. Due to the asymptotic
freedom property of QCD this renormalization group improved
expansion works very well at high energies.

However, already the early perturbative calculations of strong
interaction effects in higher orders of perturbation expansion
have shown that application of perturbative methods in QCD is
more subtle than in other field theories. This is partially
related to the fact that in QCD there is no unique and natural
definition of the coupling parameter with a clear physical
interpretation. There are many viable definitions of the coupling
parameter in QCD, corresponding to different choices of the
renormalization scheme; each choice gives (numerically) a slightly
different perturbative prediction, as was discussed in
\cite{0201bard78,0202cel79,0203cel79,0204cel81,0205dhar81,0231grun80,0208grun84,0206stev81a,0207stev81b,0207stev82,0209dhar83a,0210dhar83b,0211dhar84,0229blm83}. 
In a given order of perturbation expansion this renormalization
scheme (RS) dependence of perturbative predictions is formally a
higher order effect, and its significance diminishes at very
large energies, where the running coupling parameter becomes
small. It turns out, however, that in the region of moderate
energies --- say, of the order of few GeV --- the renormalization
scheme dependence becomes numerically quite large, even with a
conservative choice of the scheme parameters, as has been pointed
out by the author of this report in
\cite{0281par92a,0281par92b,0283par95a,0285par95b}. This is
a rather unfortunate circumstance, because there are many strong interaction
effects of considerable interest, for which the characteristic
energy scale is not very high. Another problem, which  plagues the
conventional renormalization group improved perturbation
expansion at low and moderate energies is the presence of the so called Landau 
 singularity --- the running coupling parameter may become infinite
at some nonzero energy and then below this energy the
perturbative predictions simply do not exist.

Various attempts have been made to improve the reliability of the
QCD predictions at moderate energies.  In the case of physical
quantities defined at timelike momenta it was observed in
\cite{0335ledib92} that one
may reduce the renormalization scheme dependence of the
predictions by resumming some higher order corrections with the
contour integral technique in the complex momentum space
\cite{0333arad82,0333piv92}; this was verified in detail in
\cite{0337par96a,0339par96b,0341par98}. 
Other propositions included a resummation of the series expansion
for physical quantities via the Pad\'{e} approximants
\cite{0145ellis96,0147ellis96b,0149gardi97,0151brod97};
modification of the expansion by enforcing certain analyticity
properties in the complex energy plane
\cite{0157milton97,0159solov99,0163shirkov96,0165shirkov97,0167shirkov98,0169milton98,0177milton00}; 
and application of more complicated resummation methods
\cite{0181cvet98,0183cvet00,0153elias98,0328ahma02a,0328ahma02b,0328ahma03,0185cvet00,0328fisch00,0328fisch02,0328cvet01,0328lee02,0328contre04}.  It appears, however,
that neither of the proposed methods provides a complete and
satisfactory solution of the problem.

The aim of this report is to present an alternative approach. Our
starting point is the observation that strong RS dependence of
perturbative predictions obtained in the conventional approach is
largely a consequence of the very strong RS dependence of the
conventional running coupling parameter. We propose therefore to
introduce a modified running coupling parameter, obtained by
integrating the renormalization group equation with a modified,
nonpolynomial $\beta$-function. The sequence of modified
$\beta$-function is a generalization of the sequence of
polynomial $\beta$-function approximants used in the conventional
expansion. This sequence is constructed in such a way so that the
nonpolynomial approximants for each order satisfy certain general
constraints and ensure the reduced renormalization scheme
dependence of the effective coupling parameter. In particular,
they are chose to ensure the absence of singularity at nonzero
positive energy. The modified perturbation expansion for physical
quantities is then constructed by replacing the conventional
coupling parameter by the modified running coupling
parameter. Using as an example the QCD effective charge appearing
in the static interquark potential and the corrections to the
Gross-Llewellyn-Smith (GLS) sum rule in deep inelastic
neutrino-nucleon 
scattering, we show that perturbative predictions obtained at
moderate energies in such a modified expansion are much more
stable with respect to change of the renormalization scheme. We
also find that the predictions obtained from the modified
expansion have a weaker dependence on the characteristic energy
scale than the predictions obtained form the conventional
expansion.  Making some fits to the experimental data for the GLS
sum rule we show that this effect might have interesting
consequences for the QCD phenomenology.

This report is organized as follows: to prepare the stage for
further discussion, we briefly summarize in Chapter 2 the main facts
about the renormalization scheme dependence of the conventional
perturbative predictions in QCD. As a concrete example, we
discuss in detail the RS dependence of the perturbative
expression for two quantities, the QCD effective charge related to  the
static interquark potential, and the QCD correction to the GLS sum
rule, in order to 
show explicitly that at moderate energies the RS dependence is
indeed quite substantial. We then briefly discuss previous
attempts to improve the reliability of perturbative QCD
predictions. In Chapter 3 we describe the construction of an
improved running coupling parameter that has much weaker RS
dependence than the conventional coupling parameters. We first
give some general constraints that such a modified coupling
parameter should satisfy in order to give more stable
predictions, and then we present a concrete model of the modified
$\beta$-function.  In Chapter 4 we discuss perturbative
predictions for physical quantities, obtained with the modified
coupling parameter in various renormalization schemes. In
particular, we consider the defining 
equations for the so called Principle of Minimal Sensitivity
(PMS) scheme in the modified expansion, which plays important
role in our approach.  An interesting aspect of our approach is
that the sequence of the modified $\beta$-functions contains a
free parameter, which 
gives us a natural way of utilizing the information of
nonperturbative or phenomenological character to improve the
accuracy of perturbative predictions in low orders of
perturbations expansion.  We propose certain
procedure for fixing this parameter, involving phenomenological
expression for the effective charge related to the interquark
potential.  In Chapter 5 we discuss 
the problem of uniqueness of the predictions obtained in the
modified expansion: we consider an alternative model of the
modified running coupling parameter and we evaluate perturbative
predictions with this coupling parameter. We argue that predictions
obtained in the modified expansion  should be
insensitive to the choice of the concrete form of the modified
coupling parameter even in low orders of the perturbation
expansion, provided that the PMS scheme is used. In 
Chapter 6 we examine, how the use of the 
modified expansion might affect the QCD phenomenology.  We give
an argument that for all  physical quantities (that belong to the
general class discussed in this report) the modified
predictions obtained in the PMS scheme lie below the conventional
PMS predictions and evolve less rapidly as a function of the
characteristic energy scale. Using the QCD correction to the GLS sum rule we
illustrate that this might be a welcome trend, improving the
consistency between low and high energy determinations of the
strong coupling parameter. In the Appendix A we give a solution
of certain inequality, which is used to select ''reasonable''
scheme parameters in the discussion of the renormalization scheme
dependence in Chapter 2. In the Appendix B we describe yet
another modified coupling parameter, which was not included in
our analysis of perturbative predictions for physical quantities;
it is distinguished by the fact, that it is 
described by very simple formulas and may be interesting in its
own right.

Basic ideas presented in this report  have been previously 
communicated briefly by the author in \cite{0530par00} and \cite{0187par05}.

All the symbolic and numerical calculations reported here have been performed
with \textsc{Mathematica}.

\chapter{Renormalization scheme dependence of perturbative QCD predictions}

\section{General form of perturbative approximants in various renormalization
schemes}

In order to prepare the stage for further discussion and introduce
appropriate notation let us recall some basic facts about the renormalization
scheme dependence of perturbative QCD predictions in finite order
of perturbation expansion. We shall concentrate on the class of simplest
QCD predictions that may be expressed in the form of a dimensionless
quantity $\delta$ depending on a single variable with dimension of
$\mbox{(energy)}^{2}$, which we shall further denote as $Q^{2}$.
We shall also assume that the effects of nonzero quark masses are
approximated by the step-function $Q^{2}$-dependence of the number
$n_{f}$ of the {}``active'' quark flavors, and we shall restrict
our attention to the class of mass and gauge parameter independent
renormalization schemes. Under these assumptions the $N$-th order
perturbative expression for $\delta$ may be written --- apart
from a multiplicative 
constant --- in the form:
\begin{eqnarray}
\delta^{(N)}(Q^{2},\mu^{2}) = 
\hat{a}_{(N)}(\mu^{2})\,\left[1+
\hat{r}_{1}(\mu^{2},Q^{2})\,\hat{a}_{(N)}(\mu^{2})\,+ \qquad\qquad 
\right.\nonumber\\
\qquad\qquad \left. 
+\, \hat{r}_{2}(\mu^{2},Q^{2})\,\hat{a}_{(N)}^{2}(\mu^{2})\,+
...+\,\hat{r}_{N}(\mu^{2},Q^{2})\,\hat{a}_{(N)}^{N}(\mu^{2})\right],
\label{eq:201deltamu}
\end{eqnarray}
 where $\mu$ is the undetermined scale appearing in the process of
renormalization and $\hat{a}_{(N)}(\mu^{2})$ 
is the coupling parameter (a ''couplant''), related to the gauge
coupling $\hat{g}_{(N)}(\mu^{2})$: 
$\hat{a}_{(N)}(\mu^{2})=\hat{g}_{(N)}^{2}(\mu^{2})/4\pi^{2}= 
\hat{\alpha}_{s}^{(N)}(\mu^{2})/\pi$. The couplant satisfies
satisfying the $N$-th 
order renormalization group (RG) equation:
\begin{eqnarray}
\mu^{2}\frac{d\hat{a}_{(N)}}{d\mu^{2}}&=&\beta^{(N)}(\hat{a}_{(N)})
\nonumber \\
&=&-\frac{b}{2}\hat{a}_{(N)}^{2}\,\left[1+c_{1}\,\hat{a}_{(N)}+
c_{2}\,\hat{a}_{(N)}^{2}+...+c_{N}\,\hat{a}_{(N)}^{N}\right].
\label{eq:203murge}
\end{eqnarray}
In Equation  (\ref{eq:201deltamu}) we assumed that  the expansion for
$\delta$ starts with the couplant $a$ in the first power, which
 is the most common 
case. The expansion for $\delta$ may of course begin with
$a^{p}$, where $p\neq1$; it is straightforward to generalize our
discussion to include this case, but we may also note that  our considerations
could be directly applied to $\delta^{1/p}$. In the following we
shall usually omit the index $N$ in $\hat{a}_{(N)}$, assuming that
$\delta^{(N)}$ is always evaluated with the couplant satisfying the
$N$-th order RG equation. We introduced here the ''hat'' notation
in order to distinguish the parameters $\hat{r}_{i}(\mu^{2},Q^{2})$
and $\hat{a}(\mu^{2})$ from the corresponding parameters $r_{i}$
and $a(Q^{2})$ in the renormalization group improved expression for
$\delta$.

Under the assumptions listed above the coefficients $b$ and
$c_{k}$ in the $\beta$-function 
are ordinary numbers and the differential equation (\ref{eq:203murge})
may be transformed into a transcendental equation of the form
\begin{equation}
\frac{b}{2}\ln\frac{\mu^{2}}{\Lambda^{2}}=c_{1}\ln\frac{b}{2}+
\frac{1}{\hat{a}_{(N)}}+
c_{1}\ln\hat{a}_{(N)}+F^{(N)}(\hat{a}_{(N)},c_{2},...,c_{N}),
\label{eq:205murgint}\end{equation}
 where
\begin{equation}
F^{(N)}(a,c_{2},...,c_{N})=\int_{0}^{a}da'\,
\left[\frac{b}{2\beta^{(N)}(a')}+\frac{1}{a'^{2}}-\frac{c_{1}}{a'}\right].
\label{eq:207betint}\end{equation}
 The parameter $\Lambda$ appearing in Equation ~(\ref{eq:205murgint})
is the usual dimensional parameter used to distinguish between different
integral curves of Equation  (\ref{eq:203murge}), and the arbitrary integration
constant has been chosen according to the conventions set by \cite{0201bard78}.
The role of subtractions introduced in the definition of the function
$F^{(N)}$ is to make the integrand finite in the limit $a\rightarrow0$,
which simplifies both analytic and numerical evaluation of the integral
in various cases.

The form of the coefficients $\hat{r}_{i}$ is constrained by the
fact that physical predictions of the theory should be (at least formally)
independent of $\mu$, i.e. 
\begin{equation}
\mu^{2}\frac{d\delta^{(N)}}{d\mu^{2}}=O(a^{N+2}).
\label{eq:208muind}\end{equation} 
 Under our assumptions the coefficients $\hat{r}_{i}$ may be written
in the form \begin{eqnarray}
\hat{r}_{1}(\mu^{2},Q^{2}) & = &
r_{1}^{(0)}+\frac{b}{2}\ln\frac{\mu^{2}}{Q^{2}}\,,\label{eq:209genr1}\\ 
\hat{r}_{2}(\mu^{2},Q^{2}) & = &
r_{2}^{(0)}+(c_{1}+2r_{1}^{(0)})\,\frac{b}{2}\ln\frac{\mu^{2}}{Q^{2}}+
\left(\frac{b}{2}\ln\frac{\mu^{2}}{Q^{2}}\right)^{2},\quad\mbox{etc.}
\label{eq:209genr2}\end{eqnarray}
 where the parameters $r_{i}^{(0)}$ are ordinary numbers.

The results of perturbative QCD calculations are usually expressed
in the modified minimal subtraction ($\overline{\mbox{MS}}$) renormalization
scheme \cite{0201bard78}, but there are many other possible renormalization
schemes, which correspond to different choices of the finite parts
of the renormalization constants. One may for example choose the momentum
subtraction (MOM) prescription \cite{0202cel79,0203cel79,0204cel81,0205dhar81},
where the renormalized coupling constant is defined by absorbing radiative
corrections to some vertex functions at certain configurations of the
momenta. However, any choice of the finite parts of the renormalization
constants leads to a well defined renormalized theory, so we may take
a broader view that  these finite parts may may be in principle
arbitrary; in other words, there is in fact a continuum of
renormalization 
schemes \cite{0206stev81a,0207stev81b}. The coupling parameter in
some general renormalization scheme X is related to the $\overline{\mbox{MS}}$
couplant by a finite renormalization: 
\begin{equation}
\hat{a}_{\overline{{{\rm MS}}}}(\mu^{2})=
\hat{a}_{{{\rm X}}}(\mu^{2})\left[1+A_{1}\hat{a}_{{{\rm
	X}}}(\mu^{2})+
A_{2}\hat{a}_{{{\rm X}}}^{2}(\mu^{2})+...\right],
\label{eq:211finrena}\end{equation}
 where the constants $A_{i}$ are related to the finite parts of the
renormalization constants and in principle they may take arbitrary
values. If the physical quantity $\delta$ is expanded in terms of
$a_{{{\rm X}}}$, the expansion coefficients take the form: \begin{eqnarray}
\hat{r}_{1}^{{{\rm X}}} & = & \hat{r}_{1}^{\overline{{{\rm
	MS}}}}+A_{1},\nonumber \\ 
\hat{r}_{2}^{{{\rm X}}} & = & \hat{r}_{2}^{\overline{{{\rm
	MS}}}}+2A_{1}\hat{r}_{1}^{\overline{{{\rm
	MS}}}}+A_{2},\quad\mbox{etc.}\label{eq:213newrk}\end{eqnarray} 
 In general in the new scheme also the coefficients in the $\beta$-function
would be different. Under our assumptions the coefficients $b$
and $c_{1}$ are independent 
of the renormalization scheme, but the coefficients $c_{i}$ for $i\geq2$
are in general scheme dependent; in the NNL order we have for example
\begin{equation}
c_{2}=c_{2}^{\overline{{{\rm MS}}}}+A_{1}c_{1}+A_{1}^{2}-A_{2}\,.
\label{eq:215newc2}\end{equation}
 The change of the renormalization scheme affects also the parameter
$\Lambda$ \cite{0203cel79}: 
\begin{equation}
\Lambda_{{{\rm X}}}=\Lambda_{\overline{{{\rm
	MS}}}}\exp\left(-\frac{A_{1}}{b}\right).
\label{eq:217newlam}\end{equation}
 This relation is \emph{exact} to all orders of the perturbation expansion.

In the $N$-th order of the perturbation expansion we have $N$ parameters
($A_{i}$) characterizing the freedom of choice of the renormalization
scheme and $2N-1$ expansion coefficients ($r_{i}$ and $c_{k}$),
which depend on these parameters. It is clear therefore that  the
expansion coefficients are not totally independent --- for $N\geq2$
there must be $N-1$ combinations of these coefficients, which are
independent of the renormalization scheme
\cite{0207stev81b,0208grun84,0209dhar83a,0210dhar83b,0211dhar84}. 
For $N=2$ the relevant scheme invariant combination may be written
in the form:
\begin{equation}
\rho_{2}=c_{2}+\hat{r}_{2}-c_{1}\hat{r}_{1}-\hat{r}_{1}^{2}.
\label{eq:219rho2}\end{equation}
 It should be noted that  the invariant shown above corresponds to
the definition adopted in \cite{0208grun84,0209dhar83a,0210dhar83b,0211dhar84},
which is slightly different from the definition introduced in
\cite{0207stev81b}. 
It is not surprising that  we have some freedom in defining the scheme
invariant combinations of the expansion parameters, because any combination
of the scheme invariants is of course itself a scheme invariant. It
seems, however, that the invariants defined by
\cite{0208grun84,0209dhar83a,0210dhar83b,0211dhar84} 
give a more natural measure of the magnitude of the higher order radiative
corrections. This may be seen in the following way: Let us take for
example the NNL order approximant $\delta^{(2)}$, calculate the derivative
$Q^{2}d\delta^{(2)}(Q^{2})/dQ^{2}$ and expand the result in terms
of $\delta^{(2)}$ itself. We obtain
\begin{equation}
Q^{2}\frac{d\delta^{(2)}(Q^{2})}{dQ^{2}}=
-\frac{b}{2}(\delta^{(2)})^{2}\left[1+c_{1}\delta^{(2)}+
\rho_{2}(\delta^{(2)})^{2}+\sum_{k=3}\hat{\rho}_{k}(\delta^{(2)})^{k}\right]. 
\label{eq:220ec}\end{equation}
 This looks very much like the renormalization group equation
 (\ref{eq:203murge}), 
except that instead of the coefficient $c_{2}$ we have the scheme
invariant combination $\rho_{2}$ (exactly in the form proposed in
\cite{0208grun84,0209dhar83a,0210dhar83b,0211dhar84}), while the
higher order expansion coefficients $\hat{\rho}_{k}$ in this expression
are --- of course --- scheme dependent. If we would do the same calculation
with $\delta^{(3)}$, we would obtain a similar equation, where the
coefficients \emph{up to and including} the order $N=3$ would be
the scheme invariants $\rho_{i}$ defined by
 \cite{0208grun84,0209dhar83a,0210dhar83b,0211dhar84}, 
while the coefficients of higher order would be scheme dependent.
This shows that  the invariants proposed in
 \cite{0208grun84,0209dhar83a,0210dhar83b,0211dhar84} 
have indeed some universal meaning.

In phenomenological applications it is usually assumed that  renormalization
scale $\mu$ is proportional to the characteristic energy scale of
the process: $\mu^{2}=\lambda^{2}Q^{2}$, where $\lambda$ is some
constant. In this way we obtain the so called renormalization group
improved expression for the physical quantity $\delta$
\begin{eqnarray}
\delta^{(N)}(Q^{2})&=&a_{(N)}(Q^{2})\left[1+r_{1}a_{(N)}(Q^{2})\,+\right. 
  \nonumber \\
  &&\left. + \,r_{2}a_{(N)}^{2}(Q^{2})+ ...+ r_{(N)}a_{(N)}^{N}(Q^{2})\right],
\label{eq:221dnrgim}
\end{eqnarray}
 where the coefficients $r_{i}$ are now independent of $Q^{2}$ and
in the arbitrary scheme X they take the form 
\begin{eqnarray}
r_{1} & = & r_{1}^{(0)\overline{{{\rm MS}}}}+b\ln\lambda+A_{1},
\label{eq:223newrgir1}\\
r_{2} & = & r_{2}^{(0)\overline{{{\rm
 MS}}}}+(c_{1}+2r_{1}^{(0)\overline{{{\rm MS}}}})\,
 b\ln\lambda+\left(b\ln\lambda\right)^{2}+\nonumber \\ 
 &  & +\,2A_{1}(r_{1}^{(0)\overline{{{\rm
	MS}}}}+b\ln\lambda)+A_{2}\,,\quad\mbox{etc.}
\label{eq:224newrgir2}
\end{eqnarray}
The whole $Q^{2}$-dependence of $\delta$ comes then from the $Q^{2}$-dependence
of the couplant $a_{(N)}(Q^{2})=\hat{a}_{(N)}(\lambda^{2}Q^{2})$,
which is determined by the implicit equation 
\begin{eqnarray}
\frac{b}{2}\ln\frac{Q^{2}}{\Lambda_{\overline{{{\rm
	  MS}}}}^{2}}&=&-A_{1}-b\ln\lambda+c_{1}\ln\frac{b}{2}+
	  \nonumber \\
&& +\, \frac{1}{a_{(N)}}+c_{1}\ln a_{(N)}+F^{(N)}(a_{(N)},c_{2},...,c_{N}),
\label{eq:225intrgim}
\end{eqnarray}
 where we used the exact relation \cite{0203cel79} to introduce
 $\Lambda_{\overline{{{\rm MS}}}}$ 
 as a reference phenomenological parameter. The use of
 $a(Q^{2})$ 
instead of $\hat{a}(\mu^{2})$ in (\ref{eq:221dnrgim}) is equivalent
to resummation of some of the $(\ln\frac{\mu^{2}}{Q^{2}})^{k}$ terms
appearing in $\hat{r}_k(\mu^2, Q^2)$ to all orders. Of course
 $a_{(N)}(Q^{2})$ satisfies the equation 
\begin{equation}
Q^{2}\frac{d a_{(N)}}{dQ^{2}}=\beta^{(N)}(a_{(N)}(Q^2))
\label{eq:225q2rge}
\end{equation}

If we change the renormalization scheme, the finite renormalization
of the couplant is compensated by the change in the coefficients $r_{i}$,
but in finite order of the perturbation expansion such compensation
may be of course only approximate, so the actual numerical value obtained
for $\delta^{(N)}$ does depend on the choice of renormalization scheme.
The differences between values of $\delta^{(N)}$ calculated in various
renormalization schemes are formally of the order $a^{N+2}(Q^{2})$,
but numerically for $Q^{2}$ of the order of few $\mbox{GeV}^{2}$
they may become quite significant, which creates a practical problem
when we want to confront theoretical predictions with the experimental
data. This is the problem that we want to address in this report.

In order to study the renormalization scheme dependence we need a
convenient parameterization of the available degrees of freedom in
choosing the perturbative approximants. The next-to-leading (NL) order
approximants contain in principle two arbitrary parameters,
$A_{1}$ and
$\lambda$, which however appear in the expression for $\delta^{(1)}$
and in the Equation  (\ref{eq:225intrgim}) in the combination
$A_{1}+b\ln\lambda$. This means that  in the NL order the freedom
of choice of the approximants may be characterized by only one parameter;
we found it convenient to choose as such a parameter the coefficient
$r_{1}$ in the renormalization group improved expression for $\delta$.
Thus in the NL order we have: 
\begin{equation}
\delta^{(1)}(Q^{2},r_{1})=a(Q^{2},r_{1})\left[1+
r_{1}a(Q^{2},r_{1})\right],
\label{eq:227d1con}
\end{equation}
 with $a(Q^{2},r_{1})$ determined by the equation 
\begin{equation}
\frac{b}{2}\ln\frac{Q^{2}}{
\Lambda_{\overline{{{\rm MS}}}}^{2}}=
r_{1}^{(0)\overline{{{\rm MS}}}}-r_{1}+c_{1}\ln\frac{b}{2}+
\frac{1}{a}+c_{1}\ln a + F^{(1)}(a), 
\label{eq:229a1con}
\end{equation}
where 
\begin{equation}
F^{(1)}(a)=-c_{1}\ln(1+c_{1}a).
\end{equation}

It should be emphasized, however, that although under our assumptions
the parameters $A_{1}$ and $\lambda$ have the same \emph{effect}
on $\delta^{(1)}$, they have in fact a completely different character.
For example, in a more general class of renormalization schemes the coefficient
$A_{1}$ may depend 
on quark masses and the gauge parameter. The procedure of fixing
the choice of the renormalization
scheme (i.e.\ $A_{1}$) differs from the procedure of fixing the renormalization
scale ($\lambda$), 
even in the NL order. This observation has important consequences for
the discussion of the ''reasonable'' schemes for NL order approximants.
For example, it shows that  there is no such a thing as a ''reasonable''
value of the scale parameter; as has been emphasized in
\cite{0281par92a}, a given value of $r_{1}$ --- which 
provides a unique specification of the NL order approximant --- may
correspond to a ''reasonable'' value of $\lambda$ with one subtraction
procedure (say, $\overline{\mbox{MS}}$), and a completely ''unreasonable''
value of $\lambda$ in a scheme defined by some other subtraction
procedure (for example momentum subtraction).

In the next-to-next-to-leading (NNL) order there appears an additional
degree of freedom in choosing the renormalization scheme, corresponding
the freedom of choice of the parameter $A_{2}$. Following \cite{0207stev81b}
we shall parameterize this degree of freedom by the coefficient $c_{2}$
in the NNL order $\beta$-function. If the parameters $r_{1}$ and
$c_{2}$ are fixed, the value of the coefficient $r_{2}$ may be determined
from the scheme invariant combination $\rho_{2}$:
\begin{equation}
r_{2}(r_{1},c_{2})=\rho_{2}-c_{2}+c_{1}r_{1}+r_{1}^{2}.
\label{eq:230r2r1c2}
\end{equation}
 Thus in the NNL order we have:
\begin{eqnarray}
\delta^{(2)}(Q^{2},r_{1},c_{2})= a(Q^{2},r_{1},c_{2})
\left[1+r_{1}a(Q^{2},r_{1},c_{2})\,+\right. \qquad\qquad \nonumber \\
\qquad \left. +\,r_{2}(r_{1},c_{2})\,
  a^{2}(Q^{2},r_{1},c_{2})\right],
\label{eq:231d2con}
\end{eqnarray}
 where $a(Q^{2},r_{1},c_{2})$ is determined by the equation 
\begin{equation}
\frac{b}{2}\ln\frac{Q^{2}}{\Lambda_{\overline{{{\rm
	  MS}}}}^{2}}=r_{1}^{(0)\overline{{{\rm
	MS}}}}-r_{1}+c_{1}\ln\frac{b}{2}+\frac{1}{a}+c_{1}\ln
a+F^{(2)}(a,c_{2}),
\label{eq:233a2con}
\end{equation}
 where again we have shown explicitly the dependence of various terms
on the scheme parameters $r_{1}$ and $c_{2}$. For $c_{2}>c_{1}^{2}/4$
the function $F^{(2)}(a,c_{2})$ has the form: 
\begin{eqnarray}
F^{(2)}(a,c_{2})&=&-\frac{c_{1}}{2}\ln(1+c_{1}a+c_{2}a^{2})+
\nonumber \\
&&+\,
\frac{2c_{2}-c_{1}^{2}}{\sqrt{4c_{2}-c_{1}^{2}}}
\arctan\left(\frac{a\sqrt{4c_{2}-c_{1}^{2}}}{2+c_{1}a}\right).  
\label{eq:235f2con}
\end{eqnarray}
 The expression for $F^{(2)}(a,c_{2})$ for other values of $c_{2}$
may obtained via analytic continuation in $c_{2}$. The coefficients
of the $\beta$-function have the following values: $b=(33-2n_{f})/6$
\cite{0213gross73,0215politz73}, $c_{1}=(153-19n_{f})/(66-4n_{f})$
\cite{0217casw74,0219jones74,0221egor79} and
\cite{0223tara80,0225larin93} 
\[
c_{2}^{\overline{{{\rm MS}}}}=\frac{77139-15099\, n_{f}+
325\, n_{f}^{2}}{288(33-2\, n_{f})}
\]

\section{Propositions for the optimized choice of the renormalization scheme}

Besides the $\overline{\mbox{MS}}$ \cite{0201bard78} and the
momentum subtraction schemes
\cite{0202cel79,0203cel79,0204cel81,0205dhar81} there were other
propositions concerning the optimal way to choose the
renormalization scheme. We now briefly review major approaches
that have been considered. 

An interesting proposition, directly referring to diagrammatic
calculations and motivated by analogy with QED, was formulated in
\cite{0229blm83}: it was proposed to choose the renormalization
scale (represented in our notation by the coefficient $\lambda$)
in the NL order expression in such a way that the contribution to
the physical quantity from the vacuum polarization effects due to
quarks (represented in the coefficient $r_1^{(0)\overline{\rm
MS}}$ in the formula (\ref{eq:209genr1}) by the term depending on
the number $n_f$ of ''active'' quarks) is absorbed into the
definition of the renormalized coupling constant (via the
$n_f$-dependent term in the coefficient $b$). Extensions of this
so called BLM procedure to the case of scale fixing in the
presence of some higher order corrections were discussed in
\cite{0229neub95,0229bene95,0229ball95,0229brods97,0229brod01,0229xhorn03}.
Unfortunately, it proved difficult to extend the BLM method to
higher orders in such a way that it would lead to a unique choice
of all the renormalization scheme parameters relevant in the
given order of the perturbation expansion, as was discussed in
\cite{0229grun92,0229grun92b,0229chyl95,0229xrath96,0229mikha04}.

A more radical proposition, originating from the work of
\cite{0231grun80}, was to choose the constants $A_{k}$ in the
Equation (\ref{eq:211finrena}) in such a way that all the
expansion coefficients $r_{i}$ in the $N$-th order approximant
$\delta^{(N)}$ for the physical quantity $\delta$ are identically
zero, $r_{i}\equiv0$, without any reference to any explicit
subtraction procedure. In other words, in this scheme the
renormalized couplant coincides with the physical quantity
$\delta$.  In our parameterization this corresponds to the choice
of the scheme parameters $r_{1}=0$, $c_{k}=\rho_{k}$. This choice
was dubbed in \cite{0207stev81b} the Fastest Apparent Convergence
(FAC) scheme.

Another interesting proposition was to choose the scheme
parameters according to the so called Principle of Minimal
Sensitivity \cite{0206stev81a,0207stev81b,0207stev82} (for a more
recent discussion see \cite{0207xmattin94}): since the physical
predictions of the theory should in principle be independent of
the choice of the renormalization scheme, we should give
preference to those values of the scheme parameters, for which
the finite order perturbative predictions are least sensitive to
the local changes in these parameters. Since this approach would
play an important role in our further discussion, and our
parameterization of the scheme dependence is slightly different
from that assumed in the original paper \cite{0207stev81b}, we
shall describe it here in some detail.

In the NL order the scheme parameter $\bar{r}_{1}$
corresponding to the PMS scheme is a solution of the equation:
\begin{equation}
\frac{\partial}{\partial
  r_{1}}\delta^{(1)}(a(Q^{2},r_{1}),r_{1})\mid_{r_{1}=\bar{r}_{1}}=0,  
\label{eq:237dd1dr1}
\end{equation}
 where we emphasized the fact that $\delta^{(1)}$ depends on
 $r_{1}$ both explicitly and implicitly via the
 $r_{1}$-dependence of $a(Q^{2})$.  Performing the
 differentiation and taking into account that in the NL order we
 have
\begin{equation}
\frac{\partial a}{\partial r_{1}}=\frac{2}{b}\beta^{(1)}(a),
\label{eq:239da1dr1}\end{equation}
 we obtain 
\begin{equation}
\bar{a}^{2}+(1+2\bar{r}_{1}\bar{a})\frac{2}{b}\beta^{(1)}(\bar{a})=0
\label{eq:240d1pmseq}\end{equation}
 Solving this equation for $\bar{r}_{1}$ we find 
\begin{equation}
\bar{r}_{1}=-\frac{c_{1}}{2(1+c_{1}\bar{a})}.
\label{eq:241r1pmscond1}\end{equation}
 Inserting this expression into the Equation (\ref{eq:229a1con})
for the NL order couplant we obtain a transcendental equation for
$\bar{a}$. Solving this equation we obtain the numerical value of
$\bar{a}$ for the chosen value of  $Q^2$, which then allows us to
determine the 
numerical value of $\bar{r}_1$, and hence the numerical value of
$\delta^{(1)}$ in the PMS scheme for this $Q^2$.  For small
$\bar{a}$ 
(i.e.\ for large $Q^2$) we may use an approximate expression for
$\bar{r}_1$:
\begin{equation}
\bar{r}_{1}=-\frac{c_1}{2}+ O(\bar{a}).
\label{eq:241nlpmsapp}\end{equation}

In the NNL order the parameters $\bar{r}_{1}$ and $\bar{c}_{2}$
corresponding 
to the PMS scheme are solutions of the system of two equations: 
\begin{equation}
\frac{\partial}{\partial
  r_{1}}\delta^{(2)}(a(Q^{2},r_{1},c_{2}),r_{1},c_{2})\mid_{r_{1}= 
\bar{r}_{1},\,c_{2}=\bar{c}_{2}}=0, 
\label{eq:243dd2dr1}\end{equation}
 and
\begin{equation}
\frac{\partial}{\partial
  c_{2}}\delta^{(2)}(a(Q^{2},r_{1},c_{2}),r_{1},c_{2})\mid_{r_{1}=
\bar{r}_{1},\,   
  c_{2}=\bar{c}_{2}}=0, 
\label{eq:245dd2dc2}\end{equation}
 where it is understood that  in performing partial differentiation
of $\delta^{(2)}$ over $r_{1}$ and $c_{2}$ also of the implicit
dependence of $a(Q^{2},r_1,c_2)$ on these parameters is taken into
account. In the NNL
order we have 
\begin{equation}
\frac{\partial a}{\partial r_{1}}=\frac{2}{b}\beta^{(2)}(a),
\label{eq:243da2dr1}\end{equation}
so the partial derivative over $r_1$ has the form: 
\begin{eqnarray}
\frac{\partial}{\partial r_{1}}\delta^{(2)}&=&a^2 + (c_1+2r_1) a^3 + (1+
2r_1 a + 3r_2 a^2)\frac{2}{b}\beta^{(2)}(a) 
\label{eq:243dd2dr1-b}\\
&=&-(2c_1 r_1 + c_2 + 3r_2)a^4 - 
(2r_{1}c_{2}+3c_1 r_{2})a^5- \nonumber \\
&& - 3r_{2}c_{2}a^{6}.
\label{eq:243dd2dr1-c}
\end{eqnarray}
The equation (\ref{eq:243dd2dr1}) is therefore equivalent to: 
\begin{equation}
3\rho_2 -2\bar{c}_2 +5c_1\bar{r}_1 + 3 \bar{r}_1^2 +
(2\bar{r}_{1}\bar{c}_{2}+3\bar{r}_{2}\bar{c}_{1})\bar{a}+
3\bar{r}_{2}\bar{c}_{2}\bar{a}^{2}=0,  
\label{eq:247d2pmseqr1}\end{equation}
where in the $O(\bar{a}^0)$ term we expressed $\bar{r}_2$ in
terms of $\bar{r}_1$ and $\bar{c}_2$. The partial derivative of
$\delta^{(2)}$ over $c_2$ has the form: 
\begin{equation}
\frac{\partial}{\partial c_{2}}\delta^{(2)}= -a^{3}+
(1+2r_{1}a+3r_{2}a^{2})\frac{\partial a}{\partial c_{2}}\,,
\label{eq:248dd2dc2calc}\end{equation}
where 
\begin{eqnarray}
\frac{\partial a}{\partial c_{2}}&=&-\frac{2}{b}
\beta^{(2)}(a)\frac{\partial F^{(2)}}{\partial c_{2}} \nonumber \\
&=&\beta^{(2)}\int_0^a\,\frac{1}{(\beta^{(2)})^2}\frac{\partial
  \beta^{(2)}}{\partial c_2}.
\label{eq:248da2condc2-a}
\end{eqnarray}
More explicitly 
\begin{equation}
\frac{\partial a}{\partial c_{2}}=a^{3}+H(a,c_{2}),
\label{eq:249da2condc2-b}\end{equation}
 where for $c_{2}>\frac{c_{1}^{2}}{4}$ we have
\begin{eqnarray}
H(a,c_{2}) & = &
 \frac{4c_{2}}{(4c_{2}-c_{1}^{2})^{3/2}}
\left[a^{2}(1+c_{1}a+c_{2}a^{2})
\arctan\frac{a\sqrt{4c_{2}-c_{1}^{2}}}{2+c_{1}a}\,-\right.\nonumber \\ 
 &  & \left.-\,\frac{\sqrt{4c_{2}-c_{1}^{2}}}{2}
a^{3}(1+\frac{c_{1}}{2}a)\right].
\label{eq:251da2condc2h}\end{eqnarray}
For other values of $c_{2}$ the relevant expression is obtained
via analytic continuation in $c_{2}$. It is easy to verify that 
$H(a,c_{2})=O(a^{5})$. 
The equation (\ref{eq:245dd2dc2}) for the
PMS parameters may be therefore written in the form
\begin{equation}
2\bar{r}_{1}+3\bar{r}_{2}\bar{a}+\frac{H(\bar{a},\bar{c}_{2})}{\bar{a}^4} 
\left(1+2\bar{r}_{1}\bar{a}+3\bar{r}_{2}\bar{a}^{2}\right)=0. 
\label{eq:253d2pmseqc2}\end{equation}

Solving the PMS equations (\ref{eq:247d2pmseqr1}) and
(\ref{eq:253d2pmseqc2}) we obtain the parameters $\bar{r}_{1}$
and $\bar{c}_{2}$ singled out by the PMS method, expressed in
terms of $\bar{a}$. Inserting this solution into the implicit
equation for the NNL order couplant (\ref{eq:233a2con}) we obtain
the parameter $\bar{a}$ for the chosen value of $Q^2$. Inserting
these values into the expression (\ref{eq:231d2con}) we then
obtain the PMS prediction for $\delta^{(2)}$ at this $Q^2$.

The equations (\ref{eq:247d2pmseqr1}) and (\ref{eq:253d2pmseqc2})
are quite complicated, but for small $\bar{a}$ (i.e.\ for large
$Q^2$) it is easy to solve them in an approximate way
\cite{0235penn82,0237wrig83,0239stev83,0241migna83}. If we look
for $\bar{r}_{1}$ and $\bar{c}_{2}$ in the form of a series
expansion in $\bar{a}$, then from  (\ref{eq:253d2pmseqc2}) we
see, that
\begin{equation}
\bar{r}_{1}=O(\bar{a}).
\label{eq:255d2pmsapr1}
\end{equation}
From (\ref{eq:247d2pmseqr1}) we then immediately find that  
\begin{equation}
\bar{c}_{2}=\frac{3}{2}\rho_{2}+O(\bar{a}).
\label{eq:257d2pmsapc2}\end{equation}

The choice of the PMS scheme has nice conceptual motivation, but
it may be also advantageous from the point of view of resummation
of the perturbation series. As is well known, the perturbation
series in QCD is only asymptotic (i.e.\ its radius of convergence
is equal to zero) and a straightforward summation of successive
corrections in any fixed renormalization scheme must inevitably
give infinite result. One may nevertheless give meaning to the
sum of such divergent series by applying various summation
methods, such as the method based on the Borel
transform.\footnote{The problem of large order behaviour in
various field theory models and appropriate summation methods has
been reviewed for example in
\protect{\cite{0229fisch94,0229fisch97}}.}  In \cite{0245stev84}
it was conjectured that the PMS method --- which involves a
''floating'' choice of the scheme parameters, depending on the
energy and the order of the expansion --- could provide an
''automatic'' resummation method, which would convert a divergent
series expansion into a convergent sequence of nonpolynomial
approximants. This idea was further discussed in
\cite{0245max83,0245max84,0245chyl90,0245chyl92,0245acoley04}.
Although in the case of realistic QCD series this conjecture is
far from being proved or disproved, it has been known for a long
time that re-expansion procedures based on the introduction of
some auxiliary parameters, which are ''floating'', i.e.\ they
are fixed in an order dependent way, could indeed provide an
effective resummation method for divergent series. For example in
\cite{0543sez79} a rigorous proof has been given that a method
based on an order dependent mapping gives convergent results in
the case of the factorially divergent expansion of a one
dimensional non-Gaussian integral, and a compelling numerical
evidence has been obtained that the same is true in the case of
the divergent perturbation expansion for the ground state of the
anharmonic oscillator; in a numerical study of the anharmonic
oscillator presented in \cite{0245caswell79} it has been shown,
that a sequence of approximants constructed according to the minimal
sensitivity criteria seems to be convergent; finally, in
\cite{0245halli79,0245halli81} a method for improving the
perturbation series for the anharmonic oscillator has been
proposed, which involves order dependent choice of certain
parameters, and it has been proved, that it results in a convergent
sequence.\footnote{A more general approach of the  problem of resumming the
divergent series with a nonpolynomial sequence of approximants
with auxiliary parameters has been presented in 
\protect{\cite{0245yuk90,0245yuk91,0245yuk92,0245yuk99}}. I am 
grateful to Prof.\ Yukalov for bringing these references to my 
attention.} 
For all these model divergent series rigorous proofs have been
obtained, that the resummation methods based on the PMS criteria
do indeed give convergent results 
\cite{0245buck93,0245duncan93,0247guida95,0249guida96}.

Regardless of the ingenuity of various propositions for the
optimal choice of the renormalization scheme one cannot escape
the fact that they are (at least so far) of heuristic
character. In the scheme parameter space in the vicinity of the
''optimal'' schemes we have many other schemes, which \emph{a
priori} look equally reasonable; predictions in such schemes also
should be somehow taken into account. Unfortunately, even if we
restrict ourselves to ''reasonably'' looking schemes, the
variation of the predictions obtained from the conventional
perturbation expansion at moderate energies appears to be quite
large.  We illustrate this problem in the following sections.

Before we move on, let us comment on a completely
different strategy of evaluating perturbative QCD predictions,
i.e.\ the so called method of effective charges, originating from
the works of
\cite{0231grun80,0208grun84,0209dhar83a,0210dhar83b,0211dhar84}
and recently further developed in \cite{0249x4max97,0249x5max00}.
In its simplest  form this approach is based on the manifestly
RS independent evolution equation for $\delta(Q^{2})$,
\begin{equation}
Q^{2}\frac{d\delta (Q^{2})}{dQ^{2}}=
-\frac{b}{2} \delta^{2}\left[1+c_{1}\delta +
\sum_{k=2} \rho_{k} \delta^{k} \right].  
\label{eq:257ec}\end{equation}
which is obtained by taking the limit $N\rightarrow\infty$ in the equation
(\ref{eq:220ec}). The great advantage of this approach is that the
generator of the evolution for $\delta(Q^{2})$ is a scheme independent
object, so all the scheme dependence of perturbative predictions is
apparently absent in this formulation from the very beginning. Let
us note, however, that if we use as an
expression for the generator a simple truncated
series in the given order, we end up with the expression coinciding
with the formula obtained in the FAC scheme; this expression
cannot be considered 
completely satisfactory, for example because it suffers from
the Landau singularity problem (at least for physical quantities
with positive $\rho_{k}$).
On the other hand, if we try to improve the series expansion for the
generator via some sequence of nonpolynomial approximants, then
in the low orders 
we encounter the problem of arbitrariness in the choice of these approximants,
which has a similar effect on the predictions as the effect of arbitrariness
in the choice of the renormalization scheme in the usual
expansion in terms of the 
effective couplant. Another drawback of this approach is that there
are physical quantities, which cannot be easily expressed in
terms of simple effective
charges. It seems, therefore, that this approach cannot compete at
present with the commonly used expansion in terms of an effective
couplant. 

Another interesting manifestly scheme independent approach has
been developed in \cite{0250xbrod94,0250xbrod96,0250xbrod99b},
where authors propose to make a direct comparison of physical
quantities. The attractive feature of this approach is that by
expanding one physical quantity in terms of another physical
quantity one may avoid to some extent the complications arising
from the scheme dependence. It was also observed that relating
physical quantities at appropriately chosen energy scales ---
called by authors ''commensurate scales'' --- one may eliminate
large contributions to the expansion coefficients that arise from
the terms that explicitly depend on the $\beta$-function
coefficients. Unfortunately, results obtained with 
commensurate scale relations cannot be easily combined with the
results obtained in the more conventional approaches to the QCD
phenomenology. In particular, in order to achieve their goals,
authors of \cite{0250xbrod94} adopt a somewhat unusual multi-scale
method, evaluating {\em each term} of the series expansion at a
different energy scale.

\section{An example: renormalization scheme dependence of $\delta_{{{\rm V}}}$}

The fact that finite order perturbative predictions obtained with
the conventional perturbation expansion exhibit at moderate
energies a strong dependence on the choice of the renormalization
scheme, despite a conservative choice of the scheme parameters,
has been demonstrated in several articles
\cite{0281par92a,0281par92b,0283par95a,0285par95b,0339par96b,0145ellis96,0147ellis96b}.
As a preparation for further discussion we illustrate this effect
once again, using as an example the perturbative QCD expression
for the effective charge $\delta_{{{\rm V}}}$, appearing in the
static interquark interaction potential \cite{0253sus77}. To our
knowledge, a complete analysis of the RS dependence of
$\delta_{{{\rm V}}}$ 
has not been performed so far. This potential is very important for the study
of heavy quarkonia\footnote{The present status of the theory and
phenomenology of heavy quarkonia has been reviewed in
\cite{0254bram05}.}, and it would play an important role in our
construction of an improved perturbation expansion. 

The static interquark potential  may be
defined in a nonperturbative and gauge invariant way via the
vacuum expectation of rectangular Wilson loop of size $r$ in
spatial dimensions and size $\tau$ along temporal axis
\cite{0253sus77}:
\begin{equation}
V(r)=-\lim_{\tau \,\rightarrow \,\infty} \frac{1}{i \tau} 
\ln <0\,|\,\mbox{Tr}\,\mathcal{P} \exp 
\left(ig\oint \,dx^{\mu}\,A_{\mu}^a T^a \right)|\,0>, 
\end{equation}
where $\mathcal{P}$ denotes a path ordering prescription. The
effective charge $\delta_{{\rm V}}$ enters the Fourier transform
$V(Q^2)$ of $V(r)$ in the following way:
\begin{equation}
V(Q^{2})=-4\pi^2 C_{{{\rm F}}}\frac{\delta_{{{\rm
	V}}}(Q^{2})}{Q^{2}},
\label{eq:259vdef}\end{equation}
where in the SU(N) gauge theory with fermions in the fundamental
 representation $C_{{{\rm F}}}=T_{{\rm F}} (N^2-1)/N$, with
 $T_{{\rm F}}$ defined by the normalization of the gauge group
 generators, $\mbox{Tr}(T^aT^b)=T_{{\rm F}}
 \,\delta^{ab}$. Perturbative expression\footnote{Recent
 theoretical studies of various contributions to $\delta_{\rm V}$
 have been summarized in \cite{0277sum05}.}  for $\delta_{{{\rm
 V}}}$ has the form (\ref{eq:201deltamu}) and is presently
 completely known up to and including the NNL order
 \cite{0255fisch77,0255xappel77,0261bill80,0263pet97a,0265pet97b,0269sch99,0267melles98,0271melles00}, and some terms in the
 expansion are known even beyond this order
 \cite{0273bram99,0275kniehl99}.  For $n_{f}=3$, which is the
 most interesting case from the phenomenological point of view,
 we have in the $\overline{\mbox{MS}}$ scheme
 $r_{1}^{(0)\overline{{{\rm MS}}}}=1.75$ and
 $r_{2}^{(0)\overline{{{\rm MS}}}}=16.7998$, as well as $b=9/2$,
 $c_{1}=16/9$ and $c_{2}^{\overline{{{\rm MS}}}}=4.471$, which
 implies $\rho_{2}^{{{\rm V}}}=15.0973$. 

\begin{figure}[p]
\begin{center}
\includegraphics[width=0.55\textwidth]{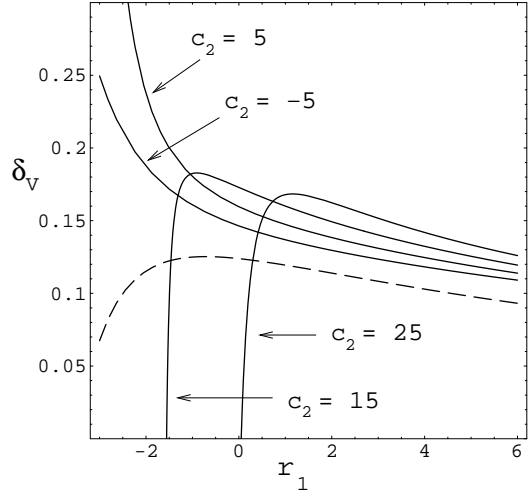}
\end{center}
  \vspace{-.7cm}\caption{\small $\delta_{{{\rm V}}}$ at
    $Q^{2}=3\,\mbox{GeV}^{2}$ ($n_{f}=3$), 
as a function of $r_{1}$, for several values of $c_{2}$, as given
by the conventional perturbation expansion with
    $\Lambda_{\overline{{{\rm MS}}}}^{(3)}=0.35\,\mbox{GeV}$. 
Dashed line indicates the NL order contribution. 
\label{fig:01condvr1}}
\end{figure}

\begin{figure}[p]
\begin{center}
\includegraphics[width=0.55\textwidth]{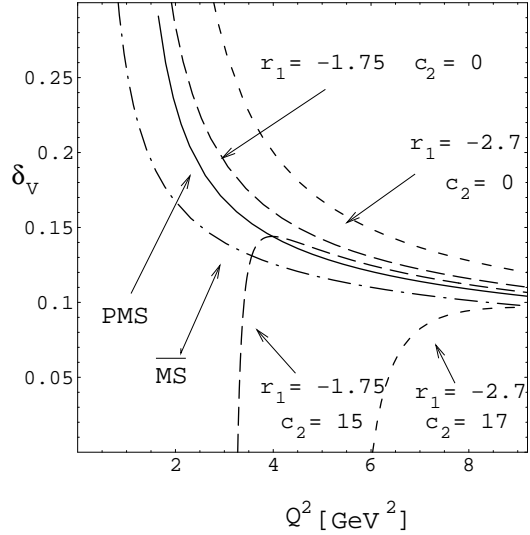}\end{center} 
  \vspace{-.7cm}\caption{\small $\delta_{{{\rm V}}}$ as a function of
    $Q^{2}$ (for $n_{f}=3$), 
as given by the conventional NNL order expansion in several renormalization
schemes, including the $\overline{\mbox{MS}}$ scheme (dash-dotted
line) and the PMS scheme (solid line). 
\label{fig:02condvqd}}
\end{figure}

In  Figure
 \ref{fig:01condvr1} we show $\delta_{{{\rm V}}}$ as a function
 of $r_{1}$, for several values of $c_{2}$, at
 $Q^{2}=3\,\mbox{GeV}^{2}$; the NL order prediction is also shown
 for comparison. (Here and in all other numerical calculations
 described in this report we assume\footnote{If we accept the
 world average $\alpha_{s}(M_{Z}^{2})=0.1182\pm0.0027$ obtained
 in \cite{0565beth04} and convert this parameter into
 $\Lambda^{(3)}_{\overline{{{\rm MS}}}}$, using the procedure
 described in detail in Chapter~6, we obtain
 $\Lambda^{(3)}_{\overline{{{\rm
 MS}}}}=0.35\,\mbox{GeV}\pm^{0.05}_{0.04}$.}  --- unless stated
 otherwise --- $\Lambda_{\overline{{{\rm
 MS}}}}^{(3)}=0.35\,\mbox{GeV}$.) As we see, the differences
 between predictions obtained for different values of the scheme
 parameters $r_1$ and $c_2$ are quite
 substantial. In Figure \ref{fig:02condvqd} we show the NNL order
 predictions for $\delta_{{{\rm V}}}$ as a function of $Q^{2}$,
 in several renormalization schemes, including the
 $\overline{\mbox{MS}}$ scheme and the PMS scheme. As we see, for
 $Q^{2}$ of the order of few $\mbox{GeV}^{2}$ the differences
 between predictions in various schemes are quite large, although
 they of course rapidly decrease with increasing $Q^{2}$.

\section{The problem of reasonable scheme parameters}

One could of course raise an objection that large differences
between predictions obtained in different schemes are a result of
a wrong choice of the scheme parameters. One could argue that if
we would choose ''reasonable'', ''natural'' scheme parameters,
then the differences between the schemes would not be very
big. In order to verify, whether this indeed might be the case we
need to give more precise meaning to the intuitive notion of a
``reasonable'' or ``natural'' renormalization scheme
parameters. Intuitively, a reasonable renormalization scheme is a
scheme in which the expansion coefficients for the physical
quantity $\delta$ {\em and} the $\beta$-function are not
unnaturally large. It was observed in
\cite{0281par92a,0283par95a,0285par95b}, that one could assess
the ''naturalness'' of the renormalization scheme by comparing
the expansion coefficients in this scheme with their scheme
invariant combinations $\rho_{k}$, relevant for the considered
physical quantity in the given order of perturbation theory: a
scheme could be considered ''natural'', if the expansion
coefficients are such that in the expression for the invariants
$\rho_{k}$ we do not have {\em extensive cancellations}. The degree of
cancellation in $\rho_{k}$ may be measured by introducing a
specific function of the scheme parameters, which in the simplest
variant may be chosen to be the sum of the absolute values of the terms
contributing to $\rho_{k}$. In the NNL order we have:
\begin{equation}
\sigma_{2}(r_{1},c_{2})=|c_{2}|+|r_{2}|+c_{1}|r_{1}|+r_{1}^{2}.
\label{eq:261sigma2}\end{equation}
 If we choose a scheme which corresponds to the scheme parameters
  that  give rise to extensive cancellations between various
terms contributing to $\rho_{2}$, then the sum of the absolute values
of the terms contributing to $\rho_{2}$ would be much larger than
$|\rho_{2}|$. The idea is then to estimate the magnitude of RS
  dependence by comparing the  predictions evaluated for
the schemes that have comparable degree of ''naturalness'' --- i.e.
that involve comparable degree of cancellation in $\rho_{2}$. The
problem of selecting natural schemes is in this way reduced to the
problem of deciding, what degree of cancellation is still acceptable.
As was pointed out in \cite{0283par95a}, one may answer this question
by referring to the PMS scheme. Using the approximate expressions
(\ref{eq:255d2pmsapr1}) 
and (\ref{eq:257d2pmsapc2}) for the parameters in the PMS scheme
we find that  in the weak coupling approximation
  $\sigma_{2}({{\rm PMS}})\approx2|\rho_{2}|$. Therefore, 
if we accept the PMS scheme as a ``reasonable'' scheme --- which
seems to be a sensible condition --- then we should also take
into account predictions in the whole set of schemes, for which the
scheme parameters satisfy the condition\footnote{Incidentally,
  this shows, why it is important to calculate the NNL order
  corrections to physical quantities --- this is the first order,
  in which we may introduce a constraint on the scheme parameters
  based on the scheme invariant combination!}
\begin{equation}
\sigma_{2}\leq2|\rho_{2}|.
\label{eq:263el2}\end{equation}
 An explicit description of the set of scheme parameters satisfying
the condition $\sigma_{2}\leq l|\rho_{2}|$, where $l$ is some constant
($l\geq1$) is given in the Appendix A.

In Figure  \ref{fig:03condvcpl} we show the contour plot of the NNL
order prediction for $\delta_{{{\rm V}}}$ at $Q^{2}=9\,\mbox{GeV}^{2}$
as a function of the scheme parameters $r_{1}$ and $c_{2}$. The
boundary of the set of the scheme parameters satisfying the condition
(\ref{eq:263el2}) is indicated with a dashed line. One may easily
verify that  scheme parameters for all the curves shown in Figure
\ref{fig:02condvqd} 
lie within this region. This suggests that  strong renormalization
scheme dependence of the NNL order perturbative predictions for
$\delta_{{{\rm V}}}$ 
cannot be simply attributed to the improper, ''unnatural'' choice
of the renormalization scheme.

\begin{figure}[tb]
\begin{center}  
\includegraphics[width=0.55\textwidth]{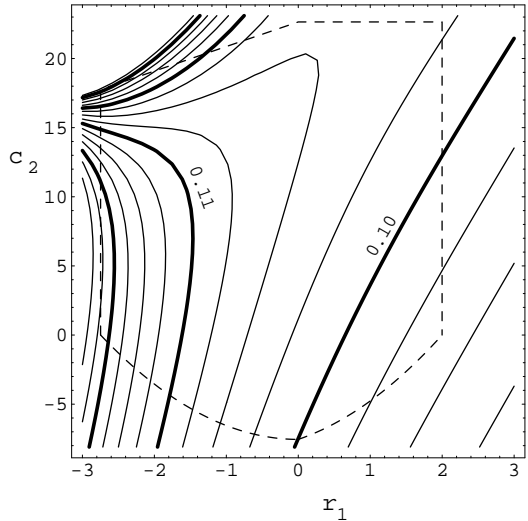}
\end{center} 
  \vspace{-.7cm}\caption{\small Contour plot of the NNL order
    predictions for $\delta_{{{\rm V}}}$ 
at $Q^{2}=9\,\mbox{GeV}^{2}$ (for $n_{f}=3$), as a function of the
scheme parameters $r_{1}$ and $c_{2}$. The dashed line indicates
the boundary of the set of parameters satisfying the condition
(\ref{eq:263el2}). 
\label{fig:03condvcpl}}
\end{figure}

Let us comment at this point that the frequently used procedure
of estimating the scheme dependence by varying the scale
parameter $\lambda$ in some range, say $0.5<\lambda<2$, may give
misleading results, as has been pointed out for example in
\cite{0281par92a}. The reason is that performing the variation of
$\lambda$ over the same range of values, but in \emph{different
schemes} (distinguished by different values of the coefficient
$A_{1}$ in Equation (\ref{eq:211finrena})), we cover different
ranges of the scheme parameter $r_{1}$.

\section{Renormalization scheme dependence of $\delta_{{{\rm GLS}}}$}

As an another example of a physical quantity of considerable
interest at moderate energies let us consider the QCD correction
to the Gross-Llewellyn-Smith (GLS) sum rule for the non-singlet
structure function $xF_3(x,Q^2)$ in the $\nu (\bar{\nu})$-nucleon deep
inelastic scattering \cite{0303gross69}:
\begin{eqnarray}
\lefteqn{\int_{0}^{1}dx\,\frac{1}{2}\left[F_{3}^{{\nu{{\rm
	  p}}}}(x,Q^{2})+F_{3}^{{\overline{\nu}{{\rm
	  p}}}}(x,Q^{2})\right]=\qquad\qquad\qquad } \nonumber \\
&&\qquad\qquad\qquad 3-3\left[\delta_{{{\rm
	GLS}}}(Q^{2})+\Delta_{{{\rm GLS}}}^{{{\rm
	HT}}}(Q^{2})\right],
\label{eq:265glsdef}
\end{eqnarray}
where $\Delta_{{{\rm GLS}}}^{{{\rm HT}}}$ is the nonperturbative
 (higher twist) contribution, and $\delta_{{{\rm GLS}}}$ denotes
 the perturbative QCD correction. The GLS integral is interesting,
 because it proved possible to measure it directly over large range
 of $Q^2$, without performing any extrapolation of $xF_3(x,Q^2)$
 over $Q^2$. 
The latest comprehensive analysis of the experimental data for
 the 
 GLS sum rule has been reported in \cite{0305kim98}. The
 perturbative correction $\delta_{{{\rm GLS}}}$ has the form
 (\ref{eq:201deltamu}) and 
is presently known up to the NNL order
 \cite{0201bard78,0309gori86,0311larin91,0313blum99}: for
 $n_{f}=3$ we have in the $\overline{\mbox{MS}}$ scheme
 $r_{1}^{(0)\overline{{{\rm MS}}}}=43/12$ and
 $r_{2}^{(0)\overline{{{\rm MS}}}}=18.9757$, which implies
 $\rho_{2}^{{{\rm GLS}}}=4.2361$.  The comparison of the
 conventional predictions for
 $\delta_{{{\rm GLS}}}$ in the $\overline{\mbox{MS}}$, PMS and EC
 schemes has been presented 
 in \cite{0317chyl92}, 
 and the problem of improving the perturbation expansion for this
 quantity has been 
 discussed in \cite{0319milt99,0321contre02}. 

\begin{figure}[tb]
\begin{center}  
\includegraphics[width=0.55\textwidth]{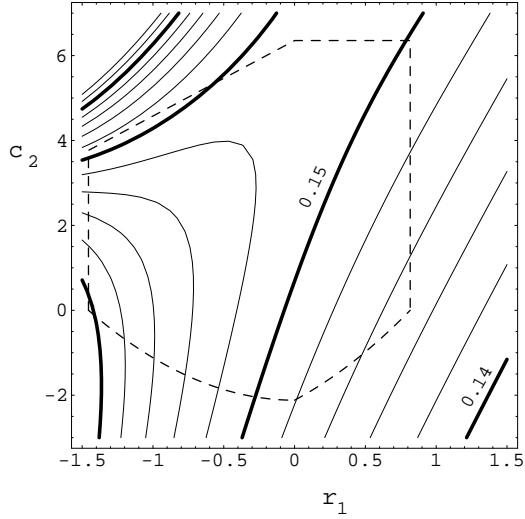}\end{center}
  \vspace{-.7cm}\caption{\small Contour plot of the NNL order
    predictions for $\delta_{{{\rm GLS}}}$ 
at $Q^{2}=5\,\mbox{GeV}^{2}$ (for $n_{f}=3$), as a function of the
scheme parameters $r_{1}$ and $c_{2}$. The dashed line indicates
the boundary of the set of parameters satisfying the condition
(\ref{eq:263el2}). 
\label{fig:04condgcpl}}
\end{figure}

We begin our discussion of the RS dependence of  $\delta_{{{\rm
 GLS}}}$ by considering the the contour plot of the NNL order
 prediction for $\delta_{{{\rm GLS}}}$ at
 $Q^{2}=5\,\mbox{GeV}^{2}$ as a function of the scheme
 parameters $r_{1}$ and $c_{2}$, as shown in the Figure
 \ref{fig:04condgcpl}. 
The boundary of the set of the
 scheme parameters satisfying the condition (\ref{eq:263el2}) is
 indicated with a dashed line. This set is of course smaller than
 the corresponding set for $\delta_{{{\rm V}}}$, because
 $\rho_{2}^{{{\rm GLS}}}$ is smaller than $\rho_{2}^{{{\rm
 V}}}$. Let us also observe that in the case of $\delta_{{{\rm
 GLS}}}$ the parameters of the $\overline{\mbox{MS}}$ scheme fall
 well outside the set singled out by the condition
 (\ref{eq:263el2}), since $\sigma_{2}(r_{1}^{\overline{{{\rm
 MS}}}},c_{2}^{\overline{{{\rm MS}}}})/|\rho_{2}^{{{\rm
 GLS}}}|=10.07$; nevertheless the predictions in this scheme
 do not show any pathological behaviour.

\begin{figure}[p]
\begin{center}  
\includegraphics[width=0.55\textwidth]{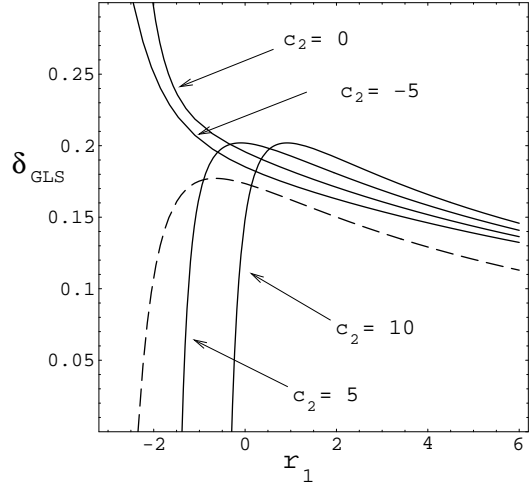}
\end{center}
  \vspace{-.7cm}\caption{\small $\delta_{{{\rm GLS}}}$ at
    $Q^{2}=3\,\mbox{GeV}^{2}$ (for $n_{f}=3$), 
as a function of $r_{1}$, for several values of $c_{2}$, as given
by the conventional perturbation expansion with
    $\Lambda_{\overline{{{\rm MS}}}}^{(3)}=0.35\,\mbox{GeV}$. 
Dashed line indicates the NL order prediction. 
\label{fig:05condgr1}}
\end{figure}

In Figure  \ref{fig:05condgr1} we show the dependence of $\delta_{{{\rm GLS}}}$
on the parameter $r_{1}$ at fixed value of $Q^{2}=3\,\mbox{GeV}^{2}$,
for several values of $c_{2}$. 

\begin{figure}[p]
\begin{center}  
\includegraphics[width=0.55\textwidth]{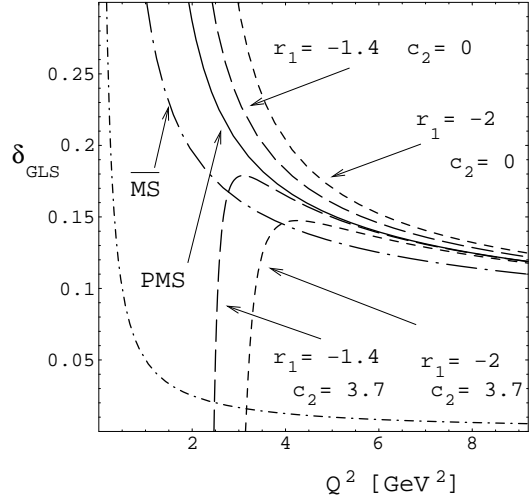}\end{center}
  \vspace{-.7cm}\caption{\small $\delta_{{{\rm GLS}}}$ as a function of
    $Q^{2}$ (for $n_{f}=3$), 
as given by the conventional NNL order expansion in several renormalization
schemes, including the $\overline{\mbox{MS}}$ scheme (long-dash-dotted
line) and the PMS scheme (solid line). The estimate of nonperturbative
contribution is shown for comparison (short-dash-dotted line). 
\label{fig:06condgqd}}
\end{figure}

In Figure \ref{fig:06condgqd} we show the NNL order predictions
for $\delta_{{{\rm GLS}}}$ as a function of $Q^{2}$ (for
$n_{f}=3$) in several renormalization schemes.  Since the set of
the scheme parameters satisfying the condition (\ref{eq:263el2})
is in the case of $\delta_{{{\rm GLS}}}$ relatively small, we
have included in Figure \ref{fig:06condgqd} also the predictions
for scheme parameters lying slightly outside this set, in order
to obtain a better picture of the scheme dependence. As we see,
also for this physical quantity the differences between the
predictions in various schemes satisfying the condition
(\ref{eq:263el2}) become quite substantial for $Q^{2}$ of the
order of few $\mbox{GeV}^{2}$.

It is of some interest to compare
the perturbative contribution $\delta_{{{\rm GLS}}}$ at various
$Q^{2}$ with the estimate of the nonperturbative contribution
$\Delta_{{{\rm GLS}}}^{{{\rm HT}}}(Q^{2})$, 
which has been discussed in several papers
\cite{0325fajf85,0327braun87,0329dasg96}. 
Following \cite{0305kim98} we shall assume: 
\begin{equation}
\Delta_{{{\rm GLS}}}^{{{\rm HT}}}(Q^{2})=
\frac{(0.05\pm0.05)}{Q^{2}}\,\mbox{GeV}^{2}.
\label{eq:267glsnpt}\end{equation}
As we see, the differences between predictions in various schemes
become large even at those values of $Q^2$, for which the
nonperturbative contribution is estimated to be small. 

\section{Possible significance of large scheme dependence at moderate $Q^{2}$}

The discussion of $\delta_{{{\rm V}}}$ and $\delta_{{{\rm GLS}}}$
shows that  for $Q^{2}$ of the order of few $\mbox{GeV}^{2}$ the
predictions obtained with the conventional renormalization group improved
perturbative approximants are surprisingly sensitive to the choice
of the renormalization scheme, even with a conservative choice of
the scheme parameters. The question is now of a proper interpretation
of this fact. A ''pragmatic'' point of view --- assumed in many
phenomenological analyses --- is to dismiss this fact as some technical
''oddity'' and continue to use the $\overline{\mbox{MS}}$ scheme.
This is presumably a healthy practical approach, justified to some
extent by the experience that  variation the scheme parameters in
the immediate vicinity of the $\overline{\mbox{MS}}$ parameters does
not lead to dramatic scheme dependence --- at least in those cases,
for which the radiative corrections have been evaluated, because there
is no proof that  this would be true for arbitrary physical quantity.
(However, the important question for the applications in phenomenology is
not only what is the value of the prediction, but 
also what is the estimated accuracy of the prediction --- the plots
we have shown suggest that  perturbative results obtained in the
$\overline{\mbox{MS}}$ 
scheme are less precise than commonly thought.) However, from a theoretical
point of view such an approach cannot be considered completely satisfactory.

A different and a rather extreme point of view would be that the observed
scheme dependence is a signal of a genuine breakdown of perturbation
expansion, which indicates the need for a fully nonperturbative approach
already at the moderate energies. Fortunately, this seems unlikely
to be the case. We see --- for example in the case of the GLS sum
rule --- that the energies, at which the differences between predictions
in various schemes start to become large are higher than the energies,
for which --- according to the available estimates --- the essentially
nonperturbative corrections may become very important. This suggests,
that strong renormalization scheme dependence has little to do with
the unavoidable breakdown of perturbation expansion at very low energies.

There is then a third possibility that  the strong renormalization
scheme dependence indicates the need to abandon the conventional perturbative
approach in favor of an improved, less scheme dependent formulation,
more suitable for moderate energies. One should keep in mind that 
the conventional renormalization group improved expansion is after
all just a specific resummation of the original, ''non-improved''
expansion. The substantial RS dependence of the conventional approximants
may simply show that  this resummation, despite being extremely useful
at high energies, is not very well suited for moderate energies. This
is the point of view that we want to pursue in this report.

\section{Previous attempts to modify the renormalization group
  improved expansion} 

The problem of improving the stability of perturbative predictions
was considered by many authors. It turned out that  in the case of
physical quantities evaluated at  timelike momenta --- such as
the QCD corrections to total cross section for the $e^{+}e^{-}$ 
annihilation into hadrons or the total hadronic decay width of the
$\tau$ lepton --- surprisingly good results may be 
obtained by resumming to all orders some of the so called
$\pi^{2}$ corrections, 
arising from the analytic continuation from spacelike to timelike
region \cite{0335ledib92,0337par96a,0339par96b,0341par98}. Such 
resummation is conveniently performed by expressing these physical
quantities as contour integrals in the complex-$Q^{2}$ space,
which are then evaluated  numerically \cite{0333arad82,0333piv92,0335ledib92}.
For example in the case of the $\mbox{e}^{+}\mbox{e}^{-}$ annihilation
into hadrons the ratio $R_{e^{+}e^{-}}$ of the hadronic and leptonic
cross sections 
\begin{equation}
R_{e^{+}e^{-}}(s)=\frac{\sigma_{{\rm tot}}(e^{+}e^{-} 
\rightarrow \mbox{hadrons})}{\sigma_{{\rm tot}}(e^{+}e^{-} 
\rightarrow \mu^{+}\mu^{-})}, 
\end{equation}
may be expressed as 
\begin{equation}
R_{e^{+}e^{-}}(s)=12 \pi \mbox{Im} \Pi(s+i\epsilon)
=-\frac{1}{2\pi i}\int_{C}d\sigma\frac{D(\sigma)}{\sigma}\,,
\label{eq:271reec}\end{equation}
where $\Pi(q^{2})$ is the transverse part of the correlator of
electromagnetic currents and $D(q^{2})$ is defined as
\begin{equation}
D(q^{2})=-12\pi^{2}q^{2}\frac{d}{dq^{2}}\Pi(q^{2})=
3\sum_{f}Q_{f}^{2}\left[1+\delta_{D}(-q^{2})\right]. 
\label{eq:272adler}\end{equation}
 ($Q_{f}$ are the charges of the ''active'' quarks and $\delta_{D}$
denotes the perturbative QCD correction, which has the form
(\ref{eq:201deltamu}).) The contour $C$ runs clockwise 
from $\sigma=s-i\epsilon$, around $\sigma=0$ to $\sigma=s+i\epsilon$.

However, the case of physical quantities evaluated at spacelike
momenta --- such as the QCD corrections to the deep inelastic
sum rules --- proved to be a more difficult problem. Some authors
\cite{0145ellis96,0147ellis96b,0149gardi97,0151brod97}
advocated the use of Pad\'{e} approximants to resum the series expansion
(\ref{eq:221dnrgim}) for the physical quantity $\delta$, observing
some reduction of the RS dependence. Another group of authors pursued
an approach stressing the importance of ensuring the correct analyticity
properties of perturbative approximants in the complex $Q^{2}$-plane.
The basic idea of this so called Analytic Perturbation Theory (APT)
 is to modify the perturbative predictions by expressing them
as dispersion integrals over the timelike momenta, with the spectral
density chosen in such a way so as to ensure the consistency with
the weak coupling perturbation expansion \cite{0157milton97}.  
In the leading order this is equivalent to the ''analytization''
of the running coupling constant
\cite{0163shirkov96,0165shirkov97,0167shirkov98}, 
i.e.\ replacing the lowest order expression $a(Q^{2})=2/(b\ln
Q^{2}/\Lambda^{2})$ 
by an ''analytic'' couplant 
\begin{equation}
a_{{{\rm an}}}(Q^{2})=\frac{2}{b}\left[\frac{1}{\ln
    Q^{2}/\Lambda^{2}}+\frac{\Lambda^{2}}{\Lambda^{2}-Q^{2}}\right],
\label{eq:273analrc}\end{equation}
 which is nonsingular for $Q^{2}=\Lambda^{2}$. It was shown that 
using this approach one obtains substantial reduction of the RS dependence
in the case of the Bjorken sum rule \cite{0169milton98} and the GLS
sum rule \cite{0319milt99}. Within APT one may also obtain some
improvement for quantities evaluated at timelike momenta
\cite{0173solov98,0175milton98,0177milton00}. (The APT approach
has been summarized in \cite{0159solov99}, and it has been
extended to a more general type of physical quantities in 
\cite{0178a-karanik01,0178b-bakul04,0178c-bakul05,0178d-bakul05}.)   
The problem of reducing the RS dependence was also discussed within a somewhat
involved approach formulated in \cite{0181cvet98,0183cvet00,0185cvet00}.

Unfortunately, none of these approaches is completely
 satisfactory.\footnote{Besides the efforts mentioned above there
 were also other works, in which authors studied how  the
 application of various resummation methods affects the 
 stability of perturbative 
 predictions 
 \cite{0153elias98,0328cvet01,0321contre02,0328lee02,0328contre04,0328ahma02a,0328ahma02b,0328ahma03}, 
 but these attempts concentrated only on the {\em renormalization
 scale} dependence. Unfortunately,  such an analysis does not
 give a proper  picture of 
 the full renormalization scheme ambiguity of perturbative
 predictions at higher orders, as we tried to explain in the
 previous sections.}  The Pad\'{e} approximants, being rational
 functions, tend to develop singularities for some particular
 values of the scheme parameters.  It also seems that the
 character of divergence of the QCD perturbation expansion at
 high orders is such that it cannot be resummed by approximants
 of this type \cite{0229fisch94,0229fisch97}. On the other hand,
 the ''analytization'' procedure introduces some essentially
 nonperturbative contributions (required to cancel the unphysical
 singularities of the conventional perturbative approximants ---
 as may be seen in the formula shown above), which generate
 $1/Q^{2}$ corrections at moderately large $Q^{2}$.  These
 corrections are troublesome, because for some physical
 quantities --- such as the $R_{e^{+}e^{-}}$ ratio --- they are
 incompatible with the operator product expansion, and for other
 quantities --- like $\delta_{{{\rm GLS}}}$ --- such corrections
 interfere with the available estimates of nonperturbative
 contributions, causing a double counting problem. The
 ''analytization'' procedure also introduces a very strong
 modification of the low-$Q^{2}$ behaviour, making all the
 expressions \emph{finite} at $Q^{2}=0$ --- and this modification
 is very ''rigid'', i.e.\ it is uniquely defined by the weak
 coupling properties of the theory, which is slightly
 suspicious. Finally, if we decide to apply the ,,analytization''
 procedure directly to physical quantities, then we effectively
 abandon the concept of expressing all the predictions of the
 theory in terms of one universal effective coupling parameter.

In this note we make an attempt to formulate an alternative method
for improving perturbative QCD approximants, which gives more reliable
results than the conventional expansion and yet at the same time does
not depart too far from the standard perturbative framework.

\chapter{The modified couplant }

\section{Strong scheme dependence of the conventional couplant}

A closer look at the conventional RG improved expansion immediately
reveals that  one of the main reasons for the significant RS dependence
of finite order predictions is the very strong RS dependence of the
running coupling parameter itself. An extreme manifestation of this
scheme dependence is the fact that  in a large class of schemes the
couplant becomes singular at finite nonzero $Q^{2}$, with the location
and the character of the singularity depending on the choice of the
scheme. In the NL order the running coupling parameter becomes singular
at $Q^{2}=(Q_{{{\rm NL}}}^{\star})^{2}$, where: 
\begin{equation}
Q_{{{\rm NL}}}^{\star}=\Lambda_{\overline{{{\rm
	MS}}}}\left(\frac{b}{2c_{1}}\right)^{
\frac{c_{1}}{b}}\mbox{exp}\left[\frac{r_{1}^{\overline{{{\rm 
	MS}}}}-r_{1}}{b}\right], 
\label{eq:301qstar1}\end{equation}
 and for $Q^{2}$ close to $(Q_{{{\rm NL}}}^{\star})^{2}$ it behaves
as 
\begin{equation}
a_{(1)}(Q^{2})\sim...\left(bc_{1}\ln\frac{Q^{2}}{(Q_{{{\rm
	NL}}}^{\star})^{2}}\right)^{-\frac{1}{2}}.
\label{eq:303a1sing}\end{equation}
 In the NNL order the running coupling parameter has an infrared stable
fixed point for $c_{2}<0$, while for $c_{2}\geq0$ it becomes singular
at $Q^{2}=(Q_{{{\rm NNL}}}^{\star})^{2}$; for example, for $4c_{2}-c_{1}^{2}>0$
the location of singularity is given by: 
\begin{eqnarray}
Q_{{{\rm NNL}}}^{\star}&=&\Lambda_{\overline{{{\rm
	MS}}}}\,\,\mbox{exp}\left[\frac{1}{b}\left(r_{1}^{\overline{{{\rm
	MS}}}}-r_{1}+c_{1}\ln\frac{b}{2\sqrt{c_{2}}}+\right.
	\right. \nonumber \\
&&\left. \left. +\,
	\frac{2c_{2}-c_{1}^{2}}{\sqrt{4c_{2}-c_{1}^{2}}}
\arctan\frac{\sqrt{4c_{2}-c_{1}^{2}}}{c_{1}}\right)\right], 
\label{eq:305qstar2}
\end{eqnarray}
 and for $Q^{2}$ close to $(Q_{{{\rm NNL}}}^{\star})^{2}$ the
couplant behaves as 
\begin{equation}
a_{(2)}(Q^{2})\sim....\left(bc_{2}\ln\frac{Q^{2}}{(Q_{{{\rm
	NNL}}}^{\star})^{2}}\right)^{-\frac{1}{3}}.
\label{eq:307a2sing}\end{equation}

In higher orders we find similar type of behavior, depending on
the signs of $c_{i}$, i.e.\ either the infrared stable fixed
point or a singularity at nonzero (positive) $Q^{2}$, with the
character of the singularity determined by the highest order term
retained in the $\beta$-function. Although the singularity itself
occurs in the range of $Q^{2}$ which is normally thought to
belong to the nonperturbative regime, its strong RS dependence
affects also the behaviour of the couplant at $Q^{2}$ above the
singularity, resulting in a strong RS dependence of the running
coupling parameter for those values of $Q^{2}$ that lie in the
perturbative domain, which cannot be properly compensated by the
corresponding changes in the expansion coefficients $r_{i}$. It
seems unlikely that one would be able to improve stability of the 
perturbative predictions in a significant way without somehow
solving this problem.

Fortunately, within the perturbative approach we have some freedom
in defining the actual expansion parameter. The idea is then to exploit
this freedom and try to construct an alternative couplant, which would
be much less scheme dependent than the conventional couplant, hopefully
giving also much less RS dependent predictions for physical quantities.
From the preceding discussion it is obvious that  first of all we
would like this modified couplant to be free from the Landau singularity.

\section{General properties of the modified couplant}

The idea of modifying the effective coupling parameter $a(Q^{2})$
in order to remove the Landau singularity has of course a long history,
dating back to the early days of QCD \cite{0441celma78,0443corn79,0445arbuz86,0447arbuz88,0449krasnik96,0451krasnik01,0455rich79,0457mox83,0461bgt80,0463bt81,0465abe83,0467kisel01,0471sanda79,0473deo83,0477sim93,0479sim95,0481bad97,0485sol94,0487siss94,0489sol94,0491sol95,0493jon95,0501dok96,0503grun96,0505web98,0509aleks98,0511magra00,0513aleks00,0515aleks01,0517aleks05,0521nest00,0523nest00b,0525nest01,0527nest01b,0529nest03,0163shirkov96,0165shirkov97,0167shirkov98}. It turns out, however,
that various proposed models --- although very inspiring --- are not
directly useful from our point of view. Let us formulate more explicitly
the conditions that  the modified couplant should satisfy in order
to be suitable for our approach.\footnote{The ideas presented here have been
first briefly formulated by the author of this report in \cite{0530par00}.}

Our first constraints are related to our assumption that  we want
to stay as close as possible to the usual perturbative framework.
This means in particular that  we want to retain the concept of an
effective coupling parameter satisfying the renormalization group
equation, although the conventional, polynomial generator $\beta^{(N)}$
in this equation would have to be replaced by an appropriately chosen
nonpolynomial function $\tilde{\beta}^{(N)}$. First of all, in order
to ensure the perturbative consistency of the modified perturbation
expansion with the conventional expansion in the N-th order we shall
require

\begin{description}
\item [(I)] $\tilde{\beta}^{(N)}(a)-\beta^{(N)}(a)=O(a^{N+3}).$
\end{description}
If this condition is satisfied, then the expression obtained by replacing
the conventional couplant $a_{(N)}(Q^{2})$ in the expansion for $\delta^{(N)}$
by the modified couplant $\tilde{a}_{(N)}(Q^{2})$ (we may temporarily
denote such an expression as $\tilde{\delta}^{(N)}$) differs from
the original expression only by terms that are formally of higher
order than the highest order term retained in the conventional approximant.

Secondly, in order to ensure that  the modified running coupling parameter
remains nonsingular for all real positive $Q^{2}$, we shall require
that:

\begin{description}
\item [(IIa)] $\tilde{\beta}^{(N)}(a)$ is negative and nonsingular for
all real positive $a$ and for $a\rightarrow+\infty$ it behaves like
$\tilde{\beta}^{(N)}(a)\sim-\xi a^{k}$, where $\xi$ is a positive
constant and $k\leq1$, 
\end{description}
or

\begin{description}
\item [(IIb)] $\tilde{\beta}^{(N)}(a)$ has zero for real positive $a_{0}$
and is negative and nonsingular for $0<a<a_{0}$. 
\end{description}
Solving explicitly the RG equation in the case (IIa) with the asymptotic
form of the generator, we find in the limit $Q^{2}\rightarrow0$ for
$k<1$: 
\begin{equation}
a(Q^{2})\sim\left(\ln\frac{\Lambda^{2}}{Q^{2}}\right)^{\frac{1}{1-k}},
\label{eq:309amodsingA}
\end{equation}
 while for $k=1$ we obtain 
\begin{equation}
a(Q^{2})\sim\left(\frac{\Lambda^{2}}{Q^{2}}\right)^{\xi/2}.
\label{eq:311amodsingB}\end{equation}

We could have formulated the condition (IIa) in a much stronger way,
strictly enforcing some type of low-$Q^{2}$ behavior of the couplant.
However, our aim is to improve the QCD predictions at \textit{moderate}
energies, where the perturbation expansion is still meaningful and gives
dominant contribution, although its application may be nontrivial;
we do not intend to make any claims about the predictions at \textit{very
low} $Q^{2}$, where noperturbative effects dominate and a proper
treatment requires much more than just an improved running expansion
parameter. Therefore the exact asymptotic behaviour of the couplant
$a(Q^{2})$ at \textit{very low} energies is not very important for
our considerations; what we basically need is that the modified coupling
parameter does not become singular at nonzero positive $Q^{2}$.

Our next condition is related to the observation that  in general
it is not difficult to obtain a running coupling parameter without
the Landau singularity by introducing essentially nonperturbative
(i.e.\ exponentially small in the limit $a\rightarrow0$) terms in
the $\beta$-function; indeed, several models of this sort have been
discussed in the literature
\cite{0455rich79,0461bgt80,0471sanda79,0163shirkov96,0165shirkov97,0167shirkov98,0521nest00}.    
Unfortunately, such models usually generate $1/Q^{2n}$ corrections
at large $Q^{2}$, which are unwelcome. As we have already mentioned,
if such corrections are present, the existing estimates of nonperturbative
effects based on the operator product expansion become useless because
of the risk of double counting. The $1/Q^{2}$ correction in the large-$Q^{2}$
expansion is particularly unwelcome, because for many physical quantities
there is no room for such a term in the operator product expansion.
It seems therefore desirable to consider only those nonpolynomial
functions $\tilde{\beta}^{(N)}$ that  satisfy the condition:

\begin{description}
\item [(III)]$\tilde{\beta}^{(N)}(a)$ is analytic in some neighborhood
of $a=0$. 
\end{description}
Analytic $\beta$-functions are preferable also from another point
of view: the sequence of perturbative approximants in the modified
perturbation expansion with such a generator may always be
interpreted as a 
pure resummation of the original expansion, without any terms being
added by hand.

Our next constraint is related to the following observation: the finiteness
of $a(Q^{2})$ for all real positive $Q^{2}$ does not automatically
guarantee that  at \textit{moderate} (but not very low) $Q^{2}$ ---
corresponding to \textit{moderately large} $a$ --- the couplant would
grow less rapidly compared to the conventional couplant and that it
would be less RS dependent in this range. (It is easy to
construct 
explicit counterexamples.) To ensure such a behaviour we shall impose
an additional constraint on $\tilde{\beta}^{(N)}(a)$, which may be
loosely formulated in the following form:

\begin{description}
\item
  [(IV)]$\tilde{\beta}^{(N)}(a)-\beta^{(N)}(a)=
K^{(N)}(c_{1},c_{2},...,c_{N})a^{N+2+p}+O(a^{N+3+p})$,  
where $p$ is some positive integer and the coefficient $K^{(N)}$
is positive for $N=1$ and for $N\geq2$ it is a slowly varying function
of its parameters, with the property  that $K^{(N)}>0$ when the coefficients
$c_{2}$,...$c_{N}$ 
are large and positive and $K^{(N)}<0$ for $c_{2}$,...$c_{N}$ large
negative. 
\end{description}

The conditions (I)-(IV) are rather general, so that at any given order
there are many functions satisfying these criteria. Concrete models
of $\tilde{\beta}^{(N)}(a)$ would typically contain some arbitrary
parameters (this is the price we have to pay for getting rid of Landau
singularity). One may restrict this freedom by fixing these parameters
with help of some information coming from outside of perturbation
theory. However, 
if in every order of perturbation expansion there would appear new,
uncorrelated free parameters, the predictive power of such an approach
would be very limited. For this reason we shall further restrict the
class of possible functions $\tilde{\beta}^{(N)}(a)$. We shall require,
that:

\begin{description}
\item [(V)]$\tilde{\beta}^{(N)}(a)$ should not contain free parameters
specific to some particular order of the expansion (i.e.\ $N$).  
\end{description}
In other words, the free parameters appearing in the modified $\beta$-function
should characterize the whole \textit{sequence} of functions
$\tilde{\beta}^{(N)}(a)$, 
not just a model $\beta$-function at some particular order.

\section{Concrete model of the modified couplant}

Nonpolynomial $\beta$-functions that do not involve exponentially
small contributions and which lead to the expressions for $a(Q^{2})$
that are nonsingular for real positive $Q^{2}$ have been considered
before \cite{0441celma78,0443corn79,0445arbuz86,0447arbuz88,0449krasnik96,0451krasnik01,0533gardi98}.
The simplest idea is to use the Pad\'{e} approximant. In the NL order
we then obtain \begin{equation}
\tilde{\beta}^{(1)}(a)=-\frac{b}{2}a^{2}\frac{1}{1-c_{1}a}\,.
\label{eq:313bet1pad}\end{equation}
 Unfortunately, this expression becomes \emph{singular} for real positive
$a$, so there would not be much of an improvement in the behaviour
of $a$. In higher orders the situation becomes even worse --- not
only the Pad\'{e} approximants develop singularities for real positive
$a$, but also the locations of these singularities are scheme dependent.
For example, the {[}1/1{]} approximant in the NNL order has the
form:
\begin{equation}
\tilde{\beta}^{(2)}(a)=-\frac{b}{2}a^{2}\frac{1+\left(c_{1}-
\frac{c_{2}}{c_{1}}\right)a}{1-\frac{c_{2}}{c_{1}}a}\,.
\label{eq:315bet2pad}\end{equation}
 This shows that  in order to obtain a satisfactory model $\beta$-function
one has to go beyond the straightforward Pad\'{e} approximants or
their simple modifications.

Let us therefore construct a completely new model $\beta$-function.
Looking for a systematic method to produce nonpolynomial functions
with the correct weak-coupling expansion coefficients and appropriate
asymptotic behavior we choose an approach inspired by the so called
method of conformal mapping
\cite{0537ciul61,0539ciul62,0540a-legui77,0540b-khu77,0540c-khu79}. 
This method is essentially a convenient procedure for performing an
analytic continuation of the series expansion beyond its radius of
convergence and it has been a popular tool in resumming the divergent
series via the Borel transform \cite{0540a-legui77,0540b-khu77,0540c-khu79}.
(However, our use of the conformal mapping would be closer to the
attempts to resum the divergent series by an order dependent mapping
\cite{0543sez79}.) The basic idea of the mapping method is very simple:
we take a function $u(a)$, which maps the complex-$a$ plane onto
some neighborhood of $u=0$ in the complex-$u$ plane and which has
the following properties: (a) it is analytic in some neighborhood
of $a=0$, with the expansion around $a=0$ of the form $u(a)=a+O(a^{2})$;
(b) it is positive and nonsingular for real positive $a$, with an
inverse $a(u)$; (c) in the limit $a\rightarrow\infty$ it goes to
a real positive constant. In the following we shall use a very simple
mapping 
\begin{equation}
u(a)=\frac{a}{1+\eta a},
\label{eq:316confmap}\end{equation}
 where $\eta$ is a real positive parameter, but of course other choices
are possible.

Let us now apply the mapping method to obtain $\tilde{\beta}(a)$
with the asymptotics corresponding to $k=1$ in the condition (III).
We take the expression
\[
a^{2}+c_{1}a^{3}+c_{2}a^{4}+...+c_{N}a^{N+2}\]
 and rewrite it as 
\[
\kappa a-\kappa a+a^{2}+c_{1}a^{3}+c_{2}a^{4}+...+c_{N}a^{N+2}.
\]
 Then we omit the first term
\[
-\kappa a+a^{2}+c_{1}a^{3}+c_{2}a^{4}+...+c_{N}a^{N+2},
\]
and we substitute everywhere in this expression $a=a(u)$; expanding the
resulting function in powers of $u$, we obtain 
\[
-\kappa
 u+\tilde{c}_{0}u^{2}+\tilde{c}_{1}u^{3}+\tilde{c}_{2}u^{4}+...
+\tilde{c}_{N}u^{N+2}+...
\]  
 It is easy to see that  the function 
\begin{eqnarray}
\lefteqn{\tilde{\beta}^{(N)}(a)=-\frac{b}{2}\left[\kappa a-\kappa
  u(a)+\tilde{c}_{0}u(a)^{2}+\tilde{c}_{1}u(a)^{3}+\right.
\qquad\qquad\qquad\qquad\qquad\qquad}
  \nonumber \\ 
&&\qquad\qquad\qquad\qquad\qquad
 \left. +\,\,\tilde{c}_{2}u(a)^{4}+...+\tilde{c}_{N}u(a)^{N+2}\right].
\label{eq:317betE1gen}
\end{eqnarray}
 does indeed satisfy the conditions (I)--(III) and (V) described in the
previous section. For the concrete mapping (\ref{eq:316confmap})
we have
\begin{equation}
\tilde{c}_{0}=1-\eta\kappa,\qquad\tilde{c}_{1}=c_{1}+2\eta-\eta^{2}\kappa,
\label{eq:318ctilda1}\end{equation}
 and
\begin{equation}
\tilde{c}_{2}=c_{2}+3c_{1}\eta+3\eta^{2}-\eta^{3}\kappa.
\label{eq:319ctilda2}\end{equation}
More explicitly, the modified $\beta$-function that we propose to
use has in the NNL order the form
\begin{eqnarray}
\lefteqn{\tilde{\beta}^{(2)}(a)=-\frac{b}{2}\left[\kappa a-\frac{\kappa
  a}{1+\eta a}+(1-\eta\kappa)\frac{a^2}{(1+\eta a)^2}+\right.} 
\label{eq:317modbetE1}\\
&&\left. +\,(c_{1}+2\eta-\eta^{2}\kappa)\frac{a^3}{(1+\eta a)^3}+
(c_{2}+3c_{1}\eta+3\eta^{2}-\eta^{3}\kappa)\frac{a^4}{(1+\eta
  a)^4}\right] \nonumber
\end{eqnarray}

It remains to verify that  the function (\ref{eq:317modbetE1})
satisfies also the condition (IV). Expanding
$\tilde{\beta}^{(N)}(a)$ in terms of $a$ we find 
\begin{equation}
\tilde{\beta}^{(1)}(a)-\beta^{(1)}(a)=
\frac{b}{2}\, \eta \left[3c_{1}+3\eta-\eta^{2}\kappa\right]a^{4}+O(a^{5}), 
\label{eq:321deltaE1b1}\end{equation}
 and
\begin{equation}
\tilde{\beta}^{(2)}(a)-\beta^{(2)}(a)=
\frac{b}{2}\,\eta\left[4c_{2}+6c_{1}\eta+4\eta^{2}-
\eta^{3}\kappa\right]a^{5}+O(a^{6}),
\label{eq:323deltaE1b2}\end{equation}
 which shows that  our simple model $\beta$-function does indeed
satisfy also the condition (IV), at least in the NL and NNL
order.\footnote{This 
is to some extent a lucky coincidence --- there does not seem to be
any simple constructive method to satisfy the condition (IV).}

Our candidate for the modified $\beta$-function contains two free
parameters: the parameter $\kappa$, which determines the exponent
in the low-$Q^{2}$ 
behaviour of the couplant $a(Q^{2})$, and the parameter $\eta$, the inverse of
which characterizes the range of values of $a$, for which the nonpolynomial
character of the $\tilde{\beta}^{(N)}(a)$ becomes essential. For
any values of these parameters the function $\tilde{\beta}^{(N)}(a)$
is a viable replacement for $\beta^{(N)}(a)$. However, by a proper
choice of these parameters we may improve the quality of the modified
perturbation expansion in low orders. As a first try, we shall assume
in further calculations reported in the following sections the
value $\kappa=2/b$, 
which ensures $\xi=2$ in the Equation  (\ref{eq:311amodsingB}), implying
$1/Q^{2}$ behaviour of $a(Q^{2})$ at low $Q^{2}$, as is suggested
by some theoretical approaches \cite{0545kogu75}. In a more precise
approach one could consider adjusting this parameter in some range,
together with the parameter $\eta$. The procedure for choosing a
preferred value for $\eta$ is described further in the text.

\begin{figure}[tb]
\begin{center}  
\includegraphics[width=0.55\textwidth]{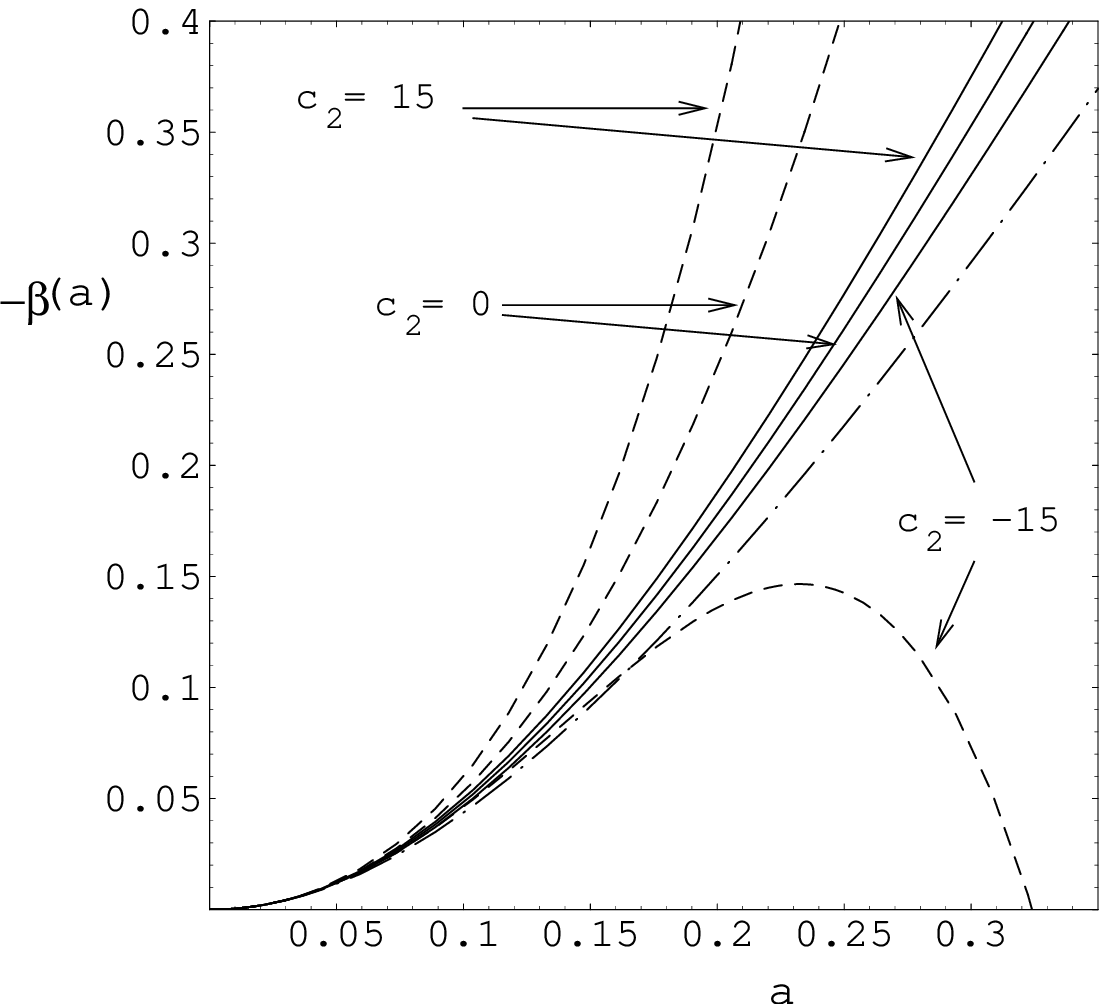}
\end{center}
  \vspace{-.7cm}
\caption{\small The modified $\beta$-function, as given by Equation
    (\ref{eq:317modbetE1}) 
with $\kappa=2/b$ and $\eta=4.1$, in the NL order (dash-dotted line)
and the NNL order for three values of $c_{2}$ (solid lines), compared
with the conventional $\beta$-function (dashed lines). 
\label{fig:07imbetE1}}
\end{figure}

The plot of this model $\tilde{\beta}$ in NL and NNL order for a
particular value of $\eta$ is shown in Figure  \ref{fig:07imbetE1}.
It is easy to guess from this figure, what the plots for smaller values
of $\eta$ would look like, since in the limit $\eta\rightarrow0$
the function $\tilde{\beta}^{(N)}(a)$ coincides with the conventional
$\beta$-function.

\begin{figure}[p]
\begin{center}  
\includegraphics[width=0.55\textwidth]{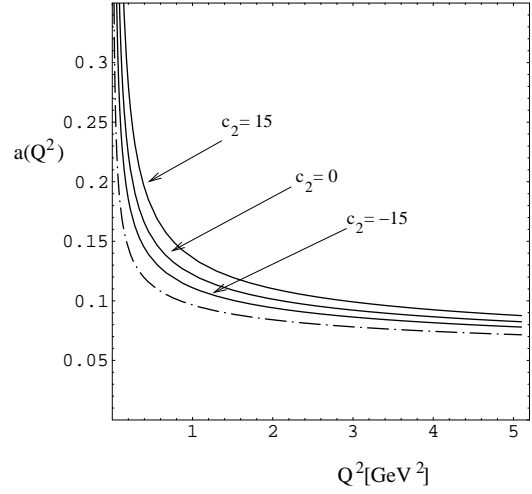}
\end{center}
  \vspace{-.7cm}
\caption{\small The modified couplant $a(Q^{2})$
    in the NL order (dash-dotted line) 
and NNL order (solid lines) for three values of $c_{2}$, as given
by Equation  (\ref{eq:205murgint}) with appropriate functions $F^{(N)}$
for $r_{1}=r_{1}^{\overline{{{\rm MS}}}}$,
    $\Lambda_{\overline{{{\rm MS}}}}^{(3)}=0.35\,\mbox{GeV}$, 
$\kappa=2/b$ and $\eta=4.1$. 
\label{fig:08imE1rcqd}}
\end{figure}

\begin{figure}[p]
\begin{center}  
\includegraphics[width=0.55\textwidth]{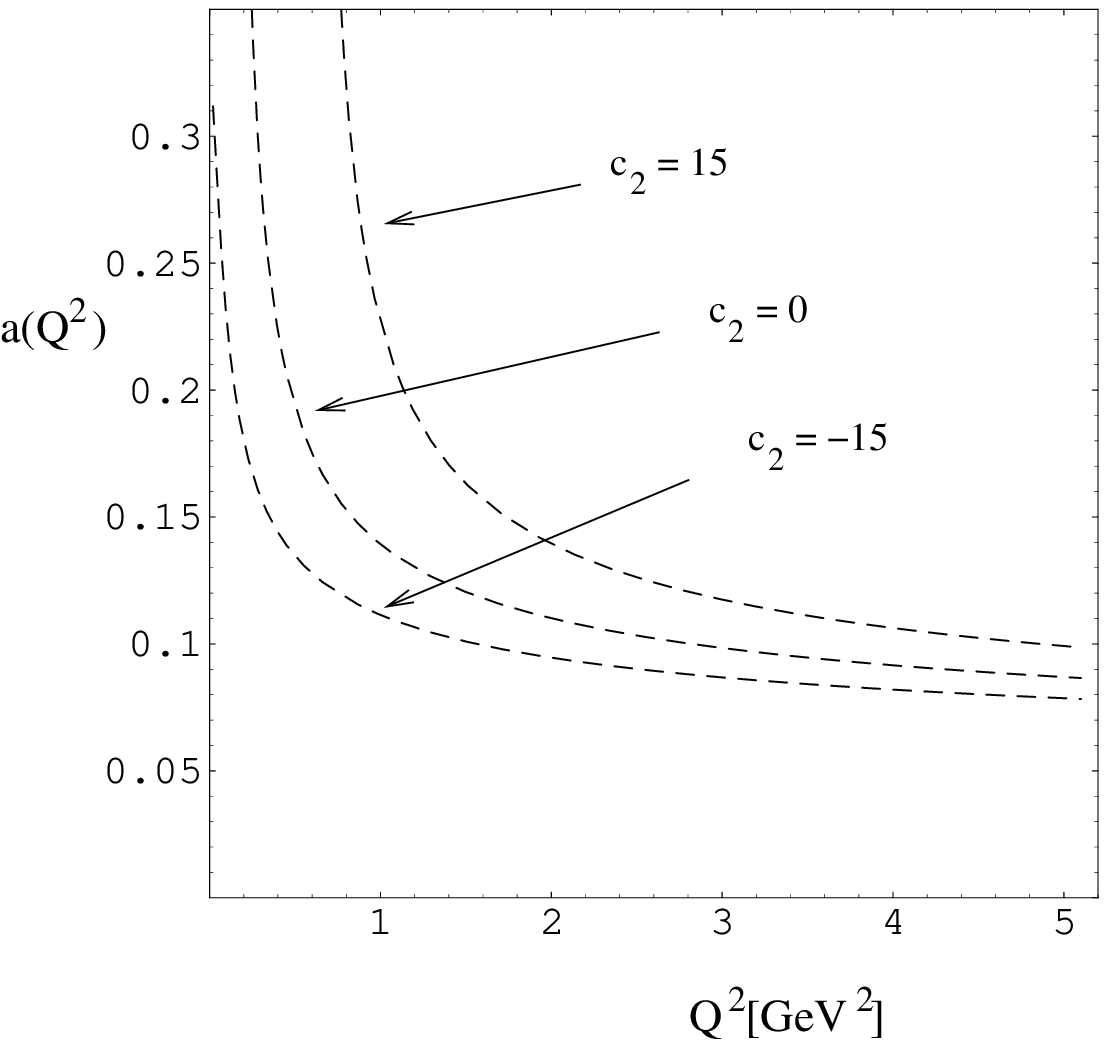}
\end{center}
  \vspace{-.7cm}
\caption{\small The conventional couplant $a(Q^{2})$ in the
  NNL order, as given 
by Equation  (\ref{eq:205murgint}) with appropriate functions $F^{(N)}$,
$r_{1}=r_{1}^{\overline{{{\rm MS}}}}$ and
$\Lambda_{\overline{{{\rm MS}}}}^{(3)}=0.35\,\mbox{GeV}$, 
for three values of $c_{2}$. 
\label{fig:09conrcqd} }
\end{figure}

In order to determine the $Q^{2}$-dependence of the modified couplant
we have to calculate the function $F^{(N)}(a,c_{2},...,c_{N},...)$
defined by Equation  (\ref{eq:207betint}), with $\beta^{(N)}(a)$ replaced
by $\tilde{\beta}^{(N)}(a)$. The relevant integral in the NL order
may be calculated in closed form. The function $F^{(1)}(a,\eta,\kappa)$
in the Equation  (\ref{eq:205murgint}) takes the form: 
\begin{eqnarray}
F^{(1)}(a,\eta,\kappa) & = &
 -E_{2}\ln\left[1+(c_{1}+3\eta)a+\eta^{3}\kappa
 a^{2}\right]+\nonumber \\ 
 &  &
 +E_{3}\arctan\left[\frac{c_{1}+
3\eta}{\sqrt{-c_{1}^{2}-6c_{1}\eta-9\eta^{2}+
4\eta^{3}\kappa}}\right]-\nonumber \\  
 &  & -E_{3}\arctan\left[\frac{c_{1}+3\eta+2\eta^{3}\kappa
     a}{\sqrt{-c_{1}^{2}-6c_{1}\eta-9\eta^{2}+4\eta^{3}\kappa}}\right],
\label{eq:325f1E1}
\end{eqnarray}
 where 
\begin{eqnarray*}
E_{2} & = & \frac{1+c_{1}\kappa}{2\kappa}\\
E_{3} & = &
\frac{-c_{1}-3\eta+c_{1}^{2}\kappa+3c_{1}\eta\kappa+
6\eta^{2}\kappa-2\eta^{3}\kappa^{2}}
{\kappa\sqrt{-c_{1}^{2}-6c_{1}\eta-9\eta^{2}+4\eta^{3}\kappa}}.   
\end{eqnarray*}
 Strictly speaking, this expression is valid for 
\[
-c_{1}^{2}-6c_{1}\eta-9\eta^{2}+4\eta^{3}\kappa>0.
\]
 The expression for $F^{(1)}(a,\eta,\kappa)$ for other values of
the parameters is obtained by analytic continuation. The integrals
in the NNL order are more complicated, but still may be handled by
\textsc{Mathematica}.

Examples of the plots of the modified couplant $\tilde{a}(Q^{2})$
in the NL and NNL order for a particular value of $\eta$ are shown
in Figure  \ref{fig:08imE1rcqd}; these plots should be compared with the
corresponding plots for the conventional couplant, shown in
Figure  \ref{fig:09conrcqd}. 
We clearly see that  the modified couplant is much less scheme and
order dependent than the conventional couplant.

\chapter{Modified perturbation expansion}

\section{Perturbative predictions with a modified couplant}

In the previous chapter we constructed a modified couplant
$\tilde{a}(Q^{2})$ that is free from Landau singularity and is
much less RS dependent than the conventional couplant. The
modified couplant is determined in the implicit way by the Equation 
(\ref{eq:207betint})
\begin{eqnarray}
\frac{b}{2}\ln\frac{Q^{2}}{
\Lambda_{\overline{{{\rm MS}}}}^{2}}&=&r_{1}^{(0)\overline{{{\rm
	MS}}}}-r_{1}+c_{1}\ln\frac{b}{2}+ 
	  \nonumber \\
&& +\, \frac{1}{\tilde{a}}+c_{1}\ln
	  \tilde{a}+F^{(N)}(\tilde{a},c_{2},...,c_{N},\eta, \kappa), 
\label{eq:400intrgeimp}
\end{eqnarray}
which is obtained by integrating (with an appropriate boundary
condition) the  modified 
renormalization group 
equation (\ref{eq:225q2rge})  
\begin{equation}
Q^{2}\frac{d\tilde{a}}{dQ^{2}}=\tilde{\beta}^{(N)}(\tilde{a}), 
\label{eq:400imp-rge}
\end{equation}
with a modified generator $\tilde{\beta}^{(N)}$ given by
Equation~(\ref{eq:317modbetE1}).  

We shall now
use this new couplant to construct modified perturbative expressions
for some physical quantities. We will show that 
the predictions obtained in the modified expansion are much less
RS dependent than the predictions in the conventional expansion. 

The simplest way to
obtain an improved perturbative expression for a 
physical quantity 
$\delta$ is to replace the conventional couplant $a(Q^{2})$
in the usual RG improved expansion for $\delta^{(N)}$ by the
modified couplant:
\begin{eqnarray}
\lefteqn{\tilde{\delta}^{(N)}(Q^{2})=\tilde{a}(Q^{2})\left[1+
r_{1}\,\tilde{a}(Q^{2})+
r_{2}\,\tilde{a}^{2}(Q^{2})+\right.}
  \qquad\qquad\qquad\qquad\qquad
 \nonumber \\
  &&\qquad\qquad \left. ...+ r_{(N)}\,\tilde{a}^{N}(Q^{2})\right].
\label{eq:400dn-imp}
\end{eqnarray}
 The coefficients $r_{i}$ in the improved approximant are
 constrained by the requirement of the consistency with the
 conventional expansion. In fact, they may be taken to be the
 same as in the conventional approximant. This follows from the
 fact that the function $\tilde{\beta}^{(N)}$ satisfies the
 condition (I) from Section~3.2, which in turn implies, that the
 difference between $\tilde{a}_{(N)}(Q^{2})$ and $a_{(N)}(Q^{2})$
 is of the 
 order $a^{N+2}$, so expanding the improved approximant
 $\tilde{\delta}^{(N)}$ given by (\ref{eq:400dn-imp}) up to and including
 the order $a^{N+1}$ we recover the conventional N-th order
 approximant $\delta^{(N)}$. This means, that we may characterize
 the RS dependence of the modified perturbative results using the
 same scheme parameters as in the case of the conventional
approximants. 

In the following we
shall usually omit the tildas, when there is no doubt that we
are dealing with the modified couplant and the modified
expression for the physical quantity.

\section{PMS scheme in the modified expansion}

In our analysis of the RS dependence of conventional
approximants in Chapter~2 the PMS scheme played an important
role. The PMS scheme may be defined also for the modified
approximants, as we describe below. 

To obtain the modified PMS approximant in the NL order we apply
the PMS condition (\ref{eq:240d1pmseq}) to the modified NL order
approximant $\tilde{\delta}^{(1)}$. In this way we obtain an
equation, which has the same general form as the corresponding
equation for the the conventional expansion, except that the
$\beta^{(1)}$ is replaced by $\tilde{\beta}^{(1)}$: 
\begin{equation}
\bar{a}^{2}+(1+2\bar{r}_{1}\bar{a})\frac{2}{b}\tilde{\beta}^{(1)}(\bar{a})=0 
\label{eq:401dd1dr1imp}
\end{equation}
Solving this for $\bar{r}_1$ we find  
\begin{equation}
\bar{r}_{1}(\bar{a})=-\frac{b\bar{a}^2+
2\tilde{\beta}^{(1)}(\bar{a})}{4\bar{a}\tilde{\beta}^{(1)}(\bar{a})}. 
\label{eq:401r1pmsimpgen}
\end{equation}
For $\tilde{\beta}^{(1)}$ defined by
Equation~(\ref{eq:317modbetE1}), with the mapping
(\ref{eq:316confmap}) and the coefficients (\ref{eq:318ctilda1}),
this implies
\begin{equation}
\bar{r}_{1}(\bar{a})=\frac{-c_{1}+
\eta^{2}(3-\eta\kappa)\bar{a}+
\eta^{3}\bar{a}^{2}}{2\left[1+(c_{1}+3\eta)\bar{a}+
\eta^{3}\kappa\bar{a}^{2}\right]}.
\label{eq:401d1e1pmsr1}\end{equation}
 Inserting this into the Equation~(\ref{eq:225intrgim}) for the
 NL order modified couplant (with the function
$F^{(1)}$ given by (\ref{eq:325f1E1})) we obtain certain
 transcendental 
 equation for $\bar{a}$; solving this equation and inserting the
 relevant values of $\bar{a}$ and $\bar{r}_1$ into the 
 expression (\ref{eq:400dn-imp}) for $\delta^{(1)}$ we then obtain
 the NL order PMS prediction 
 for $\delta$ in the modified expansion. 

For larger values of $Q^{2}$ we may obtain an approximate
 expression for $\bar{r}_{1}$ in the form of expansion in powers
 of $\bar{a}$.  It is interesting that the leading order term in
 this expansion is independent of the concrete form of
 $\tilde{\beta}^{(1)}$ and coincides with the value
 (\ref{eq:241nlpmsapp}) obtained in the case of the conventional
 NL order PMS approximant:
\begin{equation}
\bar{r}_{1}=-\frac{c_{1}}{2}+O(\bar{a}), 
\label{eq:403d1e1r1pmsappr}\end{equation}
as may be easily verified by inserting
$\tilde{\beta}^{(1)}(a)=\beta^{(1)}(a)+O(a^4)$ into
(\ref{eq:401r1pmsimpgen}). This approximate solution does not
have much practical value for our considerations, since we are
interested in the region of moderate $Q^2$. However, it is
interesting from the theoretical point of view and may serve as a
cross check for various numerical procedures. 

The PMS conditions (\ref{eq:243dd2dr1}) and (\ref{eq:245dd2dc2})
for the NNL order approximant in the modified expansion generate
equations, which in the general appearance are similar to the equations
obtained for the conventional expansion, although of course they are more
complicated. The partial derivative needed for the PMS equation
(\ref{eq:243dd2dr1}) 
has the form: 
\begin{equation}
\frac{\partial}{\partial r_{1}}\tilde{\delta}^{(2)}= a^2 + (c_1 +
2r_1) a^3 + (1+ 
2r_1 a + 3r_2 a^2)\frac{2}{b}\tilde{\beta}^{(2)}(a),  
\label{eq:404dd2dr1mod} 
\end{equation}
where we used the notation $\tilde{\delta}^{(2)}$ to emphasize 
that we are dealing with the modified expression for $\delta^{(2)}$. 
A more explicit form of this derivative is not very
instructive, except for the leading term in the expansion in
powers of $a$, which is important for further
considerations: 
\begin{equation}
\frac{\partial}{\partial r_{1}}\tilde{\delta}^{(2)}= 
-(3\rho_{2} - 2c_{2}+ 5c_{1}r_{1}+
 3r_{1}^{2})a^{4}+O(a^{5}), 
\label{eq:405modpms2appA}\end{equation}
It is interesting that  this term is independent of the concrete
form of $\tilde{\beta}^{(2)}$ and has the same form as the
corresponding term in the 
case of the conventional expansion (Equations
(\ref{eq:243dd2dr1-c}) and (\ref{eq:247d2pmseqr1})). This may be
easily verified by 
inserting $\tilde{\beta}^{(2)}(a)=\beta^{(2)}(a)+O(a^5)$ into the 
Equation (\ref{eq:404dd2dr1mod}).

The partial derivative needed for the PMS equation
(\ref{eq:245dd2dc2}) has the form 
\begin{equation}
\frac{\partial}{\partial c_{2}}\tilde{\delta}^{(2)}= -a^{3}+
(1+2r_{1}a+3r_{2}a^{2})\tilde{\beta}^{(2)}\int_0^a\,
\frac{1}{(\tilde{\beta}^{(2)})^2}\frac{\partial 
  \tilde{\beta}^{(2)}}{\partial c_2},  
\label{eq:406dd2dc2mod}\end{equation}
which coincides with the Equation (\ref{eq:248dd2dc2calc}),
except that $\beta^{(2)}$ is replaced by $\tilde{\beta}^{(2)}$. 
Again, the explicit form of this derivative is not very
illuminating, except for the leading order term in the expansion in
powers of $a$. This term depends on the leading order
difference between $\tilde{\beta}^{(2)}$ and $\beta^{(2)}$, which
we shall write in the following form:  
\begin{equation}
\tilde{\beta}^{(2)}(a)=\beta^{(2)}-\frac{b}{2}d_3 a^5+ O(a^6).
\label{eq:407d3def} 
\end{equation}
Using this notation we obtain: 
\begin{equation}
\frac{\partial\tilde{\delta}^{(2)}}{\partial c_{2}}=
(2r_{1}+\frac{1}{2}\frac{\partial d_3}{\partial
  c_2})a^{4}+O(a^{5}). 
\label{eq:407modpms2appB}\end{equation}
In the concrete case of $\tilde{\beta}^{(2)}$ given by
(\ref{eq:317modbetE1}), (\ref{eq:316confmap}),
(\ref{eq:318ctilda1}) and (\ref{eq:319ctilda2}) we have  
\begin{equation}
\frac{\partial\tilde{\delta}^{(2)}}{\partial c_{2}}=
(2r_{1}-2\eta)a^{4}+O(a^{5}).
\label{eq:407modpms2appB-b}
\end{equation}

The PMS prediction for $\delta^{(2)}$ in the modified expansion
is obtained in the same way, as in the conventional expansion:
solving the PMS equations (\ref{eq:247d2pmseqr1}) and
(\ref{eq:253d2pmseqc2}) and the implicit equation for the NNL
order couplant (\ref{eq:233a2con}) (with function $F^{(2)}$
appropriate for the modified expansion) for the chosen value of
$Q^2$ we obtain the parameters $\bar{a}$, $\bar{r}_{1}$ and
$\bar{c}_{2}$ singled out by the PMS method; inserting these
values into the expression (\ref{eq:400dn-imp}) for
$\delta^{(2)}$ we obtain the modified PMS prediction for
$\delta^{(2)}$ at this value of $Q^2$.

Similarly as in the case of the conventional PMS approximant, for
 small $\bar{a}$ (i.e.\ 
 large $Q^2$) we may obtain an approximate solutions of the PMS
 equations in the leading order in the expansion in powers of
 $\bar{a}$, using the formulas  (\ref{eq:405modpms2appA}) and
 (\ref{eq:407modpms2appB}).  Taking
 into account (\ref{eq:407modpms2appB-b}) we immediately obtain from
 Equation (\ref{eq:245dd2dc2}) that  
\begin{equation}
\bar{r}_{1}=\eta+O(\bar{a}).
\label{eq:409d2e1pmsapp-a}\end{equation}
Inserting this into the Equation (\ref{eq:243dd2dr1}) and taking
into account (\ref{eq:405modpms2appA}) we then
find: 
\begin{equation}
\bar{c}_{2}=
\frac{3}{2}\rho_{2}+\frac{5}{2}c_{1}\eta+
\frac{3}{2}\eta^{2}+O(\bar{a}).
\label{eq:409d2e1pmsapp-b}
\end{equation}
 In the limit $\eta\rightarrow0$ these approximate
 solutions 
coincide with the corresponding approximate expressions for the
 NNL order PMS 
parameters in the conventional expansion. As we already mentioned
 in the case of the NL order approximant,
 these approximate solutions are of little 
 practical significance at moderate $Q^2$, where $\bar{a}$ is
 rather large, but they serve as a useful cross check for various
 numerical routines. They are  also interesting from the theoretical
 point of view, because they show that  --- at least in the weak
 coupling limit --- the 
 solutions 
of the PMS equations for the modified expansion indeed exist and
 are 
unique.

Let us note that  the PMS scheme for the modified expansion
stands 
theoretically on a better footing than in the case of the
conventional 
expansion, because in the modified expansion the perturbative
approximants 
remain {\em finite} for \emph{all} scheme parameters and \emph{all}
positive 
values of $Q^{2}$.

\section{Choosing the value of $\eta$}

By introducing the modified couplant we obtained an improved
perturbation expansion that does not suffer from the Landau
singularity and (as we shall see in the following sections) is
much less sensitive to the choice of the renormalization scheme.
However, in this new expansion the finite order perturbative
predictions acquire a dependence on the parameter\footnote{They
depend also on the parameter $\kappa$, but to simplify the
analysis we have initially assumed $\kappa=2/b$.} $\eta$. From
the way we introduced the modified couplant it is clear that the
sum of the perturbation series should be independent of $\eta$,
but this statement requires some comments. First of all, the
perturbation expansion in QCD is divergent, so in order to
achieve the $\eta$-independence we would have to use an
appropriate summation method. As mentioned in Section~2.2, there
is a possibility, that in the case of QCD such a summation method
may be provided by the PMS approach, which motivates us to give
some preference to the modified PMS approximants when considering
phenomenological applications of the the modified
expansion. Secondly, in fact we {\em do} expect some
$\eta$-dependence even if the perturbation series is resummed to
all orders. This is because for $\eta=0$, i.e.\ in the
conventional expansion, there are whole classes of schemes, for
which resummation of the perturbation series for the physical
quantity cannot give a satisfactory result, because it is
unlikely that it would remove the Landau singularity in the
$Q^2$-dependence of the couplant; one example of such a scheme is
the so called 't~Hooft scheme, in which all the $\beta$-function
coefficients beyond the NL order are chosen to be identically
equal to zero: $c_i=0$ for $i\geq 2$. On the other hand, if
$\eta\neq 0$, then perturbative predictions are {\em finite} for
{\em all} positive values of $Q^2$, and one may reasonably expect that the
resummed series would give an expression that does not suffer
from the Landau singularity problem.

Even if we use some resummation procedure, the low order
approximants do depend on $\eta$. It should be stressed that such
dependence may be in 
fact seen as an advantage of the modified expansion, because it
may be used to improve the accuracy of the predictions by
incorporating some information outside of perturbation
theory. This may be achieved for example by adjusting $\eta$ to
match some phenomenological or nonperturbative results for some
important physical quantity. Ideally one would want to use some
results obtained from first principles, for example in the
lattice approach, but this is a complicated subject, so as a
first try we use in this report a much simpler condition that
relies on a phenomenological formula for $\delta_{{{\rm V}}}$
\cite{0461bgt80,0463bt81} that has had some success in
correlating the experimental data for heavy quarkonia. This
formula has the form of an implicit equation for $\delta_{{\rm
V}}$ (denoted as $\delta_{{{\rm BGT}}}$):
\begin{eqnarray}
\frac{b}{2}\ln\frac{Q^{2}}{(\Lambda_{\overline{{{\rm
	  MS}}}}^{eff})^{2}} 
& = & r_{1}^{\overline{{{\rm MS}}},{{\rm V}}}+
\frac{b}{2}\ln\left[\exp\left(\frac{2}{b\delta_{{\rm
	  BGT}}}\right)-1\right]+
\nonumber \\
 &  & +\,c_{1}\left[\ln\frac{2b}{\lambda}-\gamma_{{\rm E}}-
\mbox{E}_{1}(\frac{\lambda\delta_{{\rm BGT}}}{4})\right],
\label{eq:411bgt}\end{eqnarray}
 where $\gamma_{{\rm E}}$ is the Euler constant, $\mbox{E}_{1}(x)$ denotes the
exponential integral function, $b$ and $c_{1}$ are the usual
renormalization group coefficients for $n_{f}=3$ and
$r_{1}^{\overline{{{\rm MS}}},{{\rm V}}}$ is the first expansion
coefficient for $\delta_{{{\rm V}}}$ in the
$\overline{\mbox{MS}}$ scheme. The recommended values of 
the parameters in this expression are $\lambda=24$ and
$\Lambda_{\overline{{{\rm MS}}}}^{eff}=500\,\mbox{MeV}$. The
expression (\ref{eq:411bgt}) is very convenient from our point of
view, because it refers directly to the $Q^{2}$ space and by
construction it is consistent with the NL order perturbative
asymptotic behaviour of $\delta_{{{\rm V}}}(Q^{2})$ at large
$Q^{2}$. We propose to choose $\eta$ in such a way that at
moderate $Q^{2}$ the modified predictions for $\delta_{{{\rm
V}}}$ would match the phenomenological expression
(\ref{eq:411bgt}) as closely as possible.\footnote{It should be noted
that a generalization of the expression (\ref{eq:411bgt}) was
recently discussed in \cite{0467kisel01}. However, authors of
\cite{0467kisel01} incorporate in their formula a large NNL order
correction and use as a constraint in the fit a rather large
value of $\alpha_{s}(M_{z}^{2})$, so we decided to use the older
expression (\ref{eq:411bgt}). In any case, we have verified that
the fitted value of $\eta$ is not affected in a significant way.}

The concrete procedure that we use to fix $\eta$ is the
following: we assume that the modified predictions for
$\delta_{{{\rm V}}}^{(N)}$ coincide with the value given by the
phenomenological expression (\ref{eq:411bgt}) at
$Q^{2}=9\,\mbox{GeV}^{2}$, i.e.\ at the upper boundary of the
$n_{f}=3$ region (which is achieved by adjusting the parameter
$\Lambda_{\overline{{{\rm MS}}}}$ in the expression for
$\delta_{{{\rm V}}}$) and we adjust $\eta$ so as to obtain the
best possible agreement between these two expressions at lower
$Q^{2}$, down to $Q^{2}=1\,\mbox{GeV}^{2}$. To perform this
matching procedure we use the predictions for $\delta_{{\rm V}}$
in the PMS scheme. (We justified our preference for this scheme
above.) Putting aside the problems of resummation, we could 
of course fix $\eta$ using predictions in other renormalization
schemes, and the 
resulting value would of course depend to some extent on the
scheme; however, this dependence would be very small, because the
scheme dependence of the modified predictions is small (as we
shall see), and the difference to large extent would cancel
away, if we would calculate the predictions for other physical
quantities in the same renormalization scheme.

\begin{figure}[tb]
\begin{center}  
\includegraphics[width=0.55\textwidth]{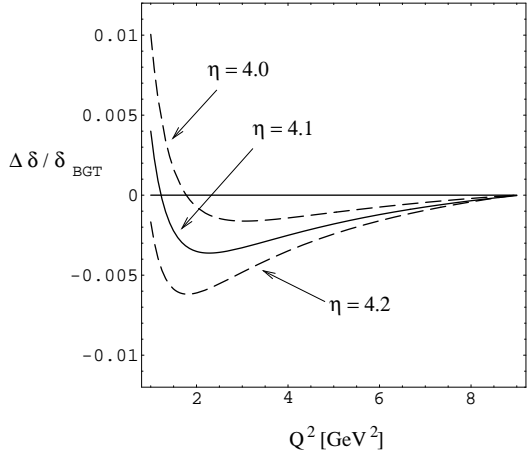}\end{center}
\vspace{-.7cm}
\caption{\small The relative difference between the modified NNL
  order PMS prediction 
for $\delta_{{{\rm V}}}$ and the value obtained from the
  phenomenological expression (\ref{eq:411bgt}), 
as a function of $Q^{2}$, for three values of $\eta$. In this
  calculation  the parameter $\Lambda_{\overline{{{\rm MS}}}}$ in
the expression for $\delta_{{{\rm V}}}$ is adjusted in such a
way that  it coincides with $\delta_{{{\rm BGT}}}$ coincide at
  $Q^{2}=9\,\mbox{GeV}^{2}$. 
\label{fig:10deveta2e1}}
\end{figure}

The curves illustrating the relative deviation of the modified
PMS predictions for $\delta_{{{\rm V}}}$ in the NNL order from
the phenomenological expression (\ref{eq:411bgt}) are shown in
Figure \ref{fig:10deveta2e1} for three values of $\eta$. It
appears that the best fit is obtained for $\eta\approx 4.1$ ---
for this value of $\eta$ the relative deviation is less than 1\%
down to $Q^{2}=1\,\mbox{GeV}^{2}$. This is the value that we
shall use in further analysis.

Fixing the value of $\eta$ we should take into account the fact
that the phenomenological expression (\ref{eq:411bgt}) has
limited precision.  In order to estimate, how this might affect
the preferred value of $\eta$ we repeated the fitting procedure,
using $\Lambda_{\overline{{{\rm MS}}}}^{eff}=400\,\mbox{MeV}$ in
the phenomenological expression (\ref{eq:411bgt}). (This value
has been quoted in \cite{0461bgt80} as a lower limit on this
effective parameter.) We found that this change has only slight
effect on the fitted value of $\eta$ --- the general picture is
unchanged and a good match is obtained for $\eta=4.0$. This
indicates that the uncertainty in the fitted value of $\eta$
arising from the uncertainty in the phenomenological expression
is of the order of $\pm 0.1$.

\section{$\delta_{{{\rm V}}}$ and $\delta_{{{\rm GLS}}}$ in the modified
expansion}

\begin{figure}[p] 
\begin{center}  
\includegraphics[width=0.55\textwidth]{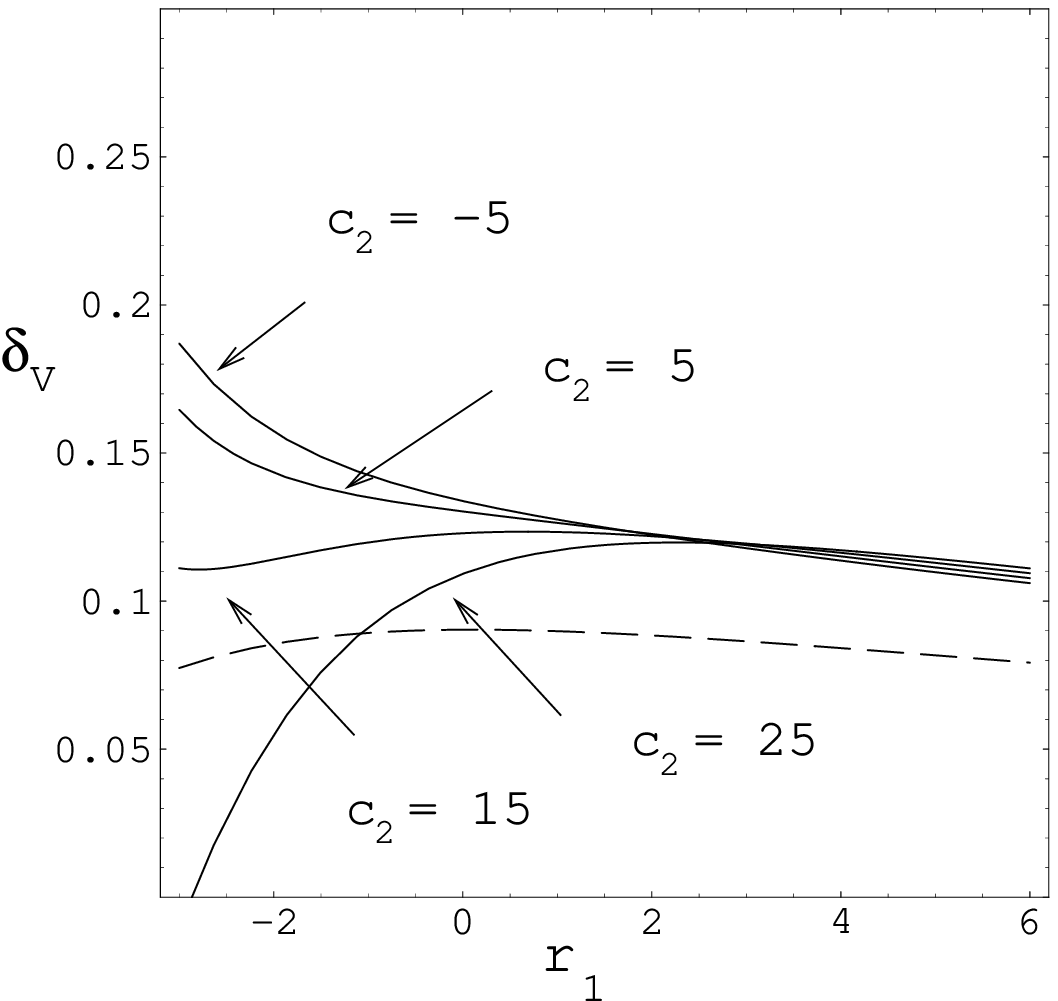}
\end{center}
\vspace{-.7cm}
  \caption{\small $\delta_{{{\rm V}}}$ at
    $Q^{2}=3\,\mbox{GeV}^{2}$ (for $n_{f}=3$) 
as a function of $r_{1}$, for several values of $c_{2}$, obtained
in the NNL order in the \emph{modified} perturbation expansion with
the $\beta$-function (\ref{eq:317modbetE1}) and $\eta=4.1$. Dashed
line indicates the NL order prediction. \label{fig:11dve1r1}}
\end{figure}

\begin{figure}[p] 
\begin{center}  
\includegraphics[width=0.55\textwidth]{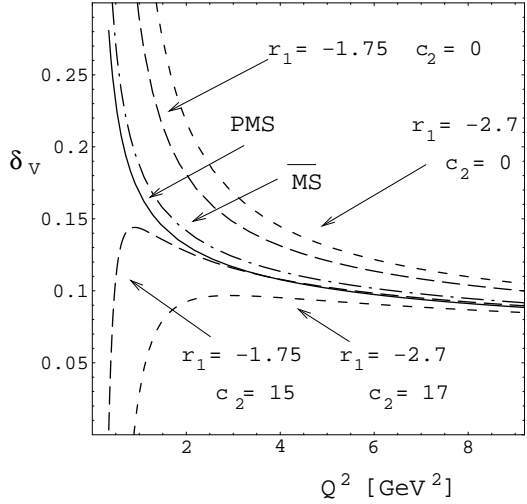}
\end{center}
\vspace{-.7cm}
\caption{\small $\delta_{{{\rm V}}}$ as a function of $Q^{2}$ (for
    $n_{f}=3$), obtained in the NNL order in the \emph{modified}
    perturbation expansion 
with the $\beta$-function (\ref{eq:317modbetE1}) and $\eta=4.1$
in several renormalization schemes, including the $\overline{\mbox{MS}}$
scheme (dash-dotted line) and the PMS scheme (solid line).  
\label{fig:12dve1qd}}
\end{figure}

\begin{figure}[p] 
\vspace{-.2cm}
\begin{center}  
\includegraphics[width=0.55\textwidth]{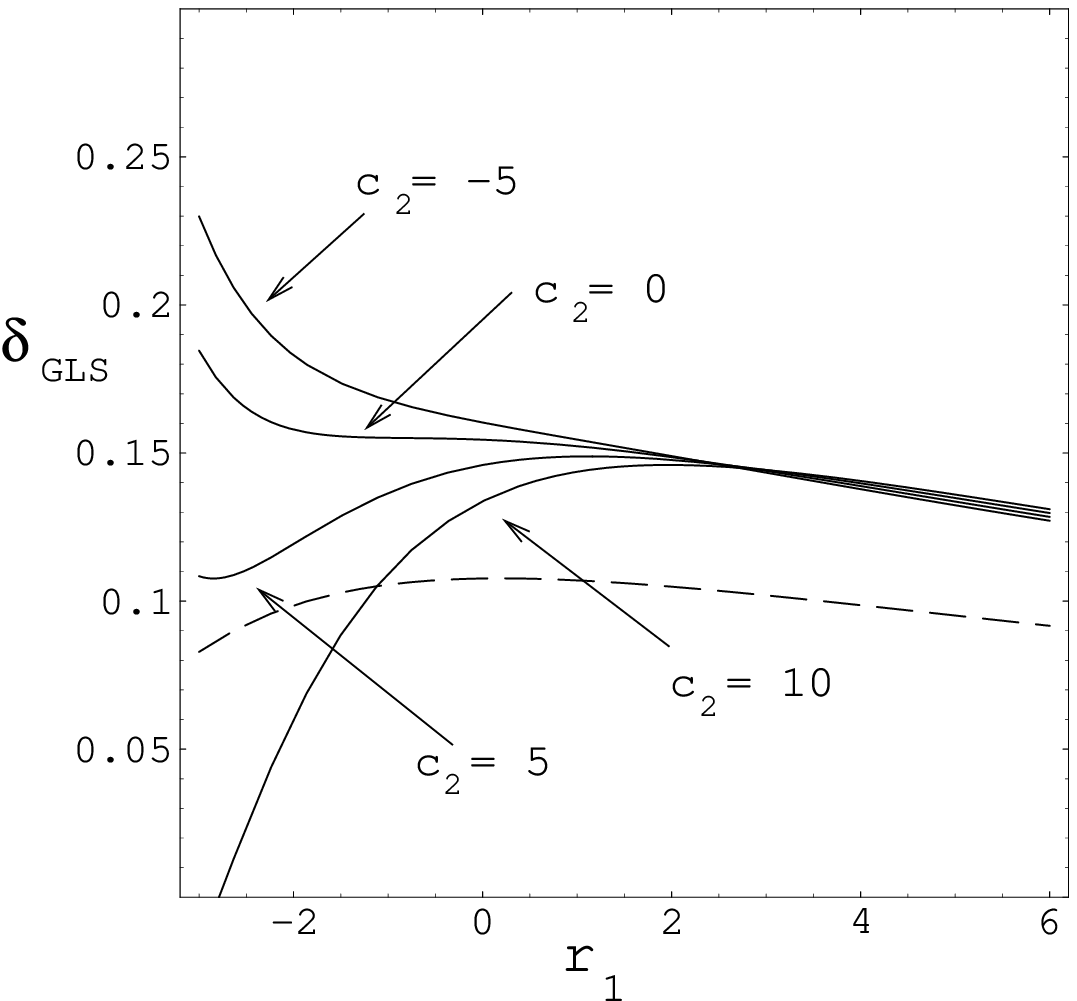}
\end{center}
\vspace{-.9cm}
  \caption{\small $\delta_{{{\rm GLS}}}$ at
    $Q^{2}=3\,\mbox{GeV}^{2}$ (for $n_{f}=3$), 
as a function of $r_{1}$, for several values of $c_{2}$, obtained
in the NNL order in the \emph{modified} perturbation expansion with
$\beta$-function (\ref{eq:317modbetE1}) and $\eta=4.1$. Dashed
line indicates the NL order prediction. \label{fig:14dge1r1}}
\end{figure}

\begin{figure}[p]
\vspace{-.2cm}
\begin{center}  
\includegraphics[width=0.55\textwidth]{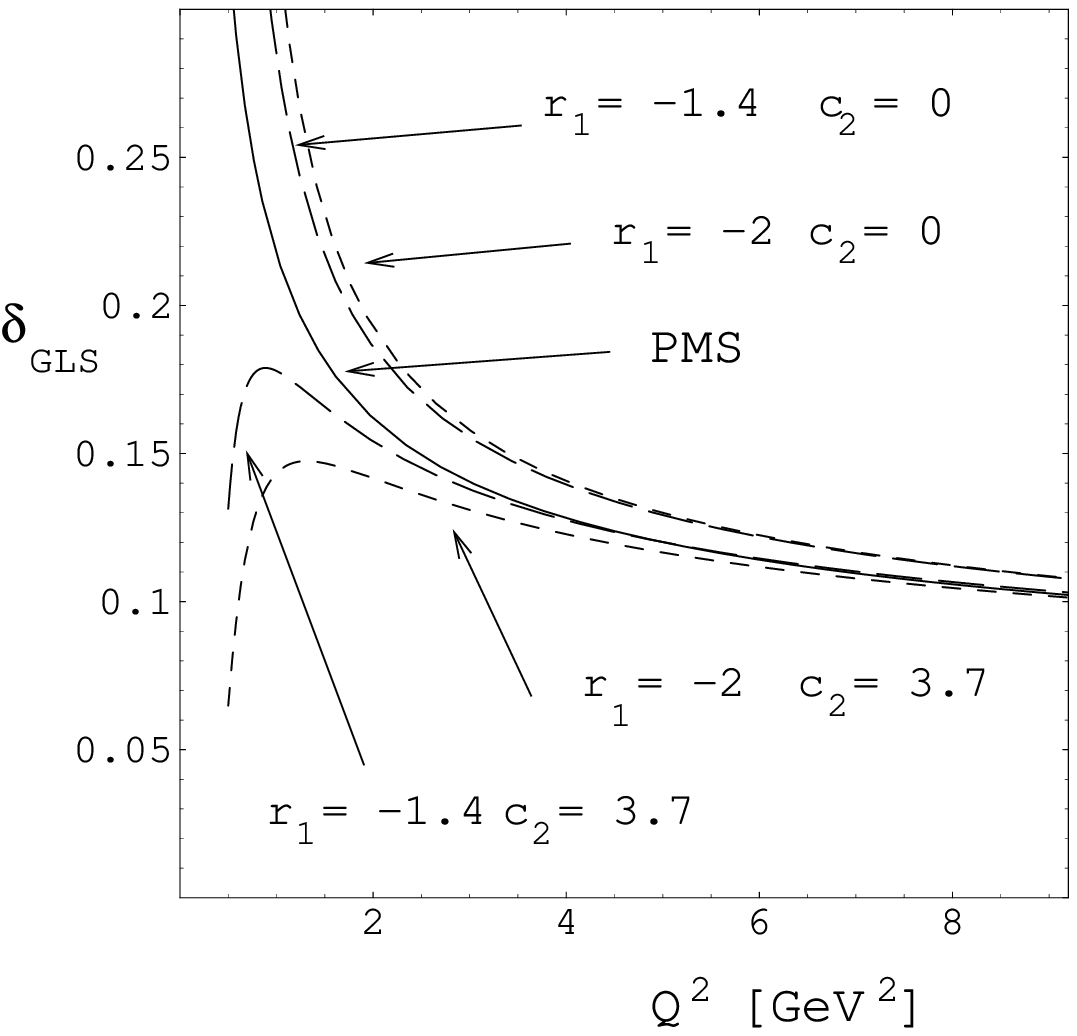}
\end{center}
  \vspace{-.9cm}\caption{\small $\delta_{{{\rm GLS}}}$ as a function of $Q^{2}$
    (for $n_{f}=3$), 
obtained in the NNL order in the \emph{modified} perturbation expansion
with $\beta$-function (\ref{eq:317modbetE1}) and $\eta=4.1$, in
several renormalization schemes, including the PMS scheme (solid line).
The curve corresponding to the $\overline{\mbox{MS}}$ scheme is
indistinguishable 
from the PMS curve in the scale of this figure. 
\label{fig:15dge1qd}}
\end{figure}

Having fixed $\eta$ we may now verify, whether the use of the modified
couplant indeed results in a better stability of the predictions.
In Figure ~\ref{fig:11dve1r1} we show the modified predictions for
$\delta_{{{\rm V}}}$ as a function of $r_{1}$, for $Q^{2}=3\,\mbox{GeV}^{2}$
and $\eta=4.1$. In Figure ~\ref{fig:12dve1qd} we show the modified
predictions for the same quantity, in the NNL approximation, as a
function of $Q^{2}$, for several values of $r_{1}$ and $c_{2}$.
In Figure ~\ref{fig:14dge1r1} and Figure ~\ref{fig:15dge1qd} we show
the corresponding plots for $\delta_{{{\rm GLS}}}$. Comparing
these figures with Figures~\ref{fig:01condvr1}, \ref{fig:02condvqd},
\ref{fig:05condgr1} and \ref{fig:06condgqd}, respectively, we may
clearly see that  the scheme dependence of the modified predictions
is substantially smaller (at least for this value of $\eta$) than
the scheme dependence of the predictions obtained from the conventional
expansion.\footnote{Strictly speaking, in our comparison of the
  RS dependence of the modified and conventional perturbative
  predictions the modified predictions should be plotted with a
  slightly different value of $\Lambda_{\overline{{{\rm
	  MS}}}}$. Indeed, if we take the world average 
of   $\alpha_{s}(M_{Z}^{2})=0.1182\pm0.0027$ found in
  \cite{0565beth04} and calculate the value of  
$\Lambda^{(3)}_{\overline{{{\rm MS}}}}$, for which the same
value is obtained for the {\em modified} strong coupling
constant in the $\overline{\mbox{MS}}$ scheme at the $M_Z$ scale
(calculated with $\eta = 4.1$, using  
the matching procedure described in detail in Chapter~6), we
get $\Lambda^{(3)}_{\overline{{{\rm
	MS}}}}=0.37\,\mbox{GeV}\pm^{0.05}_{0.04}$. However,
changing from $\Lambda^{(3)}_{\overline{{{\rm 
	MS}}}}=0.35\,\mbox{GeV}$ to $0.37\,\mbox{GeV}$ does not
affect the plots shown in this section in any significant way,
so for the sake of simplicity we use $\Lambda^{(3)}_{\overline{{{\rm 
	MS}}}}=0.35\,\mbox{GeV}$ in all the plots throughout this
report.}
This means that our initial 
hypothesis concerning the origin of the strong RS dependence of
conventional approximants (that it is due to strong RS dependence
of the conventional couplant itself) was correct.

\section{PMS predictions in the modified expansion}

\begin{figure}[p]
\begin{center}  
\includegraphics[width=0.55\textwidth]{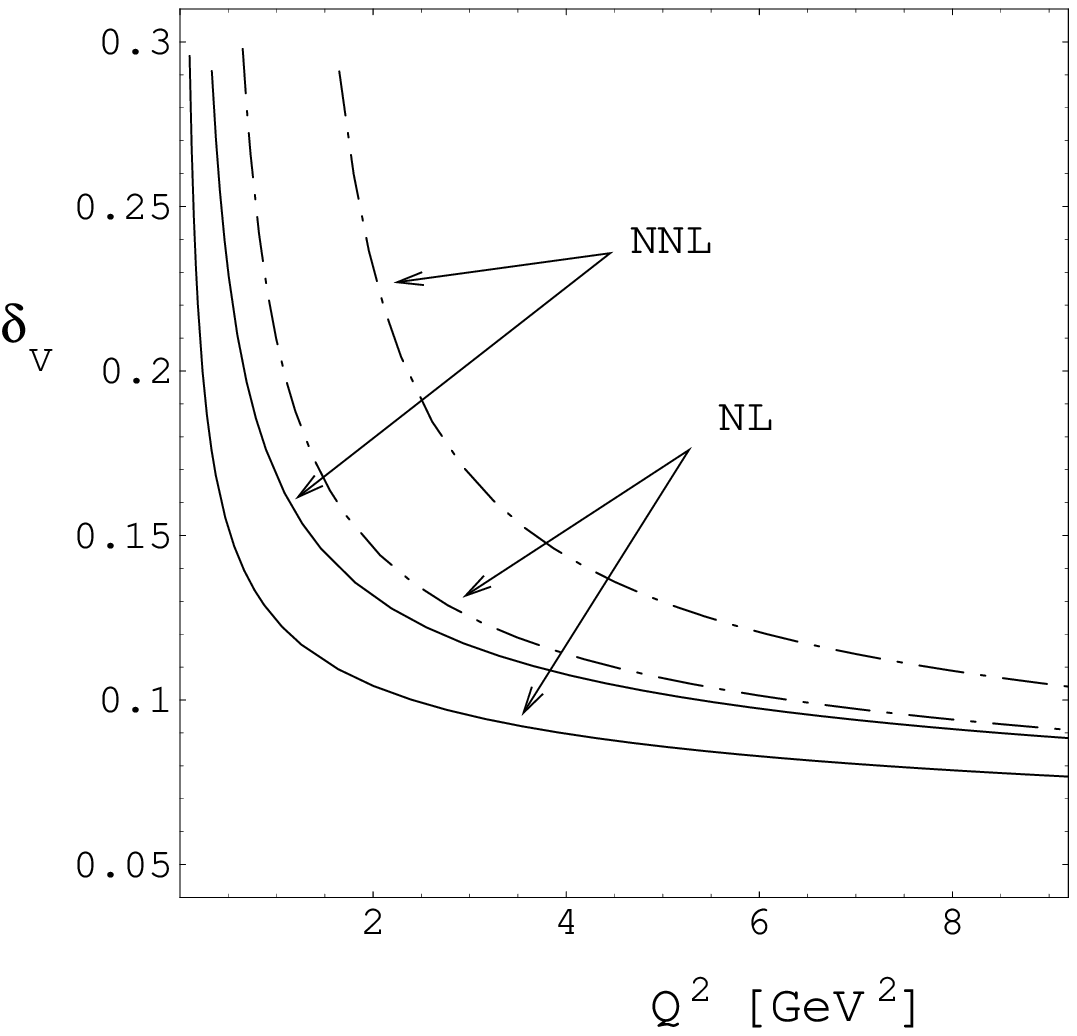}
\end{center}
  \vspace{-.7cm}
\caption{\small The NL and NNL order PMS predictions for $\delta_{{{\rm V}}}$,
as a function of $Q^{2}$, obtained in the modified expansion with
$\eta=4.1$ (solid lines) and in the conventional RG improved
expansion (dash-dotted 
lines). 
\label{fig:13dve1pmsqd}}
\end{figure}

\begin{figure}[p]
\begin{center}  
\includegraphics[width=0.55\textwidth]{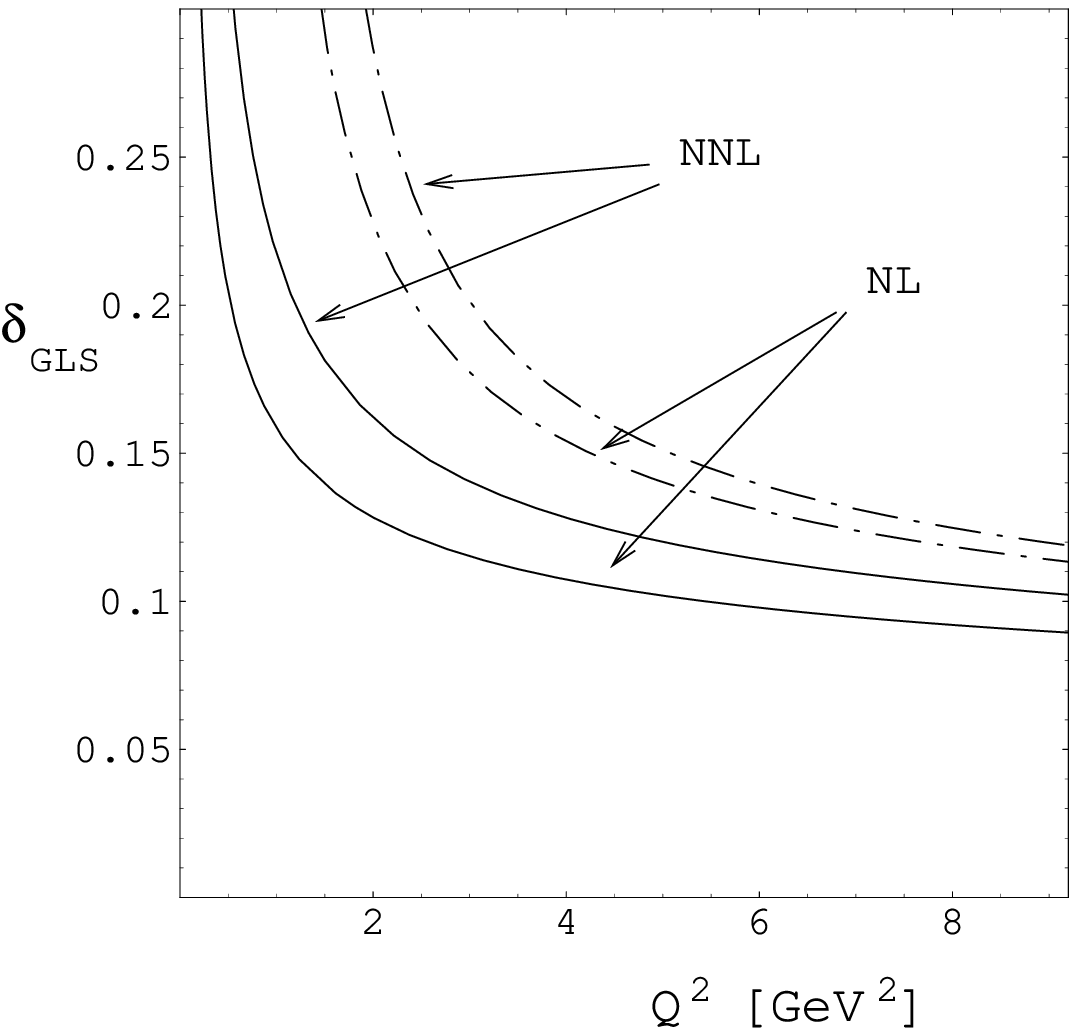}
\end{center}
  \vspace{-.7cm}
\caption{\small The NL and NNL order PMS predictions for $\delta_{{\rm GLS}}$,
as a function of $Q^{2}$, obtained in the modified expansion with
$\eta=4.1$ (solid lines) and the conventional RG improved expansion
(dash-dotted lines). \label{fig:16dge1pmsqd}}
\end{figure}

In Sections~2.2 and~4.3 we mentioned some arguments in favor of
the PMS scheme, so it is of some interest to discuss the
PMS predictions for $\delta_{{{\rm V}}}$ and $\delta_{{{\rm
GLS}}}$ in more detail.  Looking at the Figures~4.2
and~4.3, which show the reduced RS dependence of
the predictions in the modified expansion, we make an interesting
observation concerning the 
predictions in the PMS scheme. In the case of the conventional
expansion at moderate $Q^{2}$ the variation of predictions for
{\em finite} (i.e.\ not infinitesimal) deviations of the scheme
parameters from the PMS values is quite strong, as may be seen in
Figures~2.1 and~2.5, which could raise doubts about validity of
this prescription to choose the scheme. In the case of the
modified expansion the scheme dependence in the vicinity of PMS
parameters much weaker, simply because the modified predictions
are generally much less sensitive to the choice of the scheme
parameters. This further supports our
conclusion at the end of Section~4.2 that the PMS scheme in the 
modified expansion stands on a better footing than in the
conventional expansion.

In Figure \ref{fig:13dve1pmsqd} we show
the curves representing the modified NL and NNL order predictions
for $\delta_{{{\rm V}}}$ in the PMS scheme, compared with the
conventional PMS predictions for this quantity. Strictly
speaking, the NNL order curve for $\delta_{{{\rm V}}}$ 
is not a pure prediction, because we used this quantity to fix the
value of the parameter $\eta$ (although this fitting had a rather
indirect character --- we fitted only the energy dependence, not
the numerical values of the predictions themselves). However,
having done that, we are 
in position to make some genuine improved predictions for any other
physical quantity, for example $\delta_{{\rm GLS}}$. In Figure
\ref{fig:16dge1pmsqd} we show the NL and NNL order modified 
predictions for $\delta_{{{\rm GLS}}}$ in the PMS scheme, compared
with the PMS predictions for this quantity obtained in the conventional
expansion.

One characteristic feature shown by these figures is that the PMS
predictions in the modified expansion lie below the PMS
predictions in the conventional expansion, both in the NL and NNL
order. This effect is perhaps not very surprising, given the way
we constructed the modified couplant, but a moment of thought
shows, that it is also nontrivial, since we are dealing with
''floating'' optimal scheme parameters that in the modified
expansion have different value than in conventional expansion,
as shown for example by  the
Equations~(\ref{eq:409d2e1pmsapp-a})
and~(\ref{eq:409d2e1pmsapp-b}). We discuss this issue in greater 
detail in Section~6.1.

\chapter{Modified expansion with a different couplant }

\section{Alternative definitions of the modified couplant }

Before we apply the modified expansion to any phenomenological
problems, 
we have to address an important question, to what extent the predictions
obtained in the modified expansion depend on the exact form of the
chosen modified $\beta$-function, since the set of functions satisfying
the conditions (I)-(V) from Section~3.2 is still rather large. In
order to investigate 
this problem in a quantitative way let us consider an alternative class
of the modified $\beta$-functions, resulting from a different application
of the mapping method, as described in Section~3.3. We start with
the expression 
\begin{equation} 
a^{2-k}(1+c_{1}a+...+c_{N}a^{N}),
\label{eq:501}\end{equation}
 where $k\leq1$; we then substitute $a(u)$, extract the factor
$u^{2-k}$ in front of the expression (because for general $k$ it
would give an essential singularity) and expand the rest in the
powers of $u$. This gives
\begin{equation}
u^{2-k}(1+\tilde{c}_{1}u+...+\tilde{c}_{N}u^{N}+...)
\label{eq:503}\end{equation}
 where $\tilde{c}_{n}$ are appropriately modified expansion coefficients.
As a modified $\beta$-function we now take the following expression:
\begin{equation}
\tilde{\beta}^{(N)}(a)=-\frac{b}{2}a^{k}\left(u(a)\right)^{2-k}
\left[1+\tilde{c}_{1}u(a)+...+\tilde{c}_{N}(u(a))^{N}\right] 
\label{eq:505altmodbetgen}\end{equation}

To make further progress we have to specify explicitly the form of
$u(a)$. If we use the same mapping as previously
(Equation~(\ref{eq:316confmap})), we find 
\begin{equation}
\tilde{c}_{1}=c_{1}+(2-k)\eta,\qquad\tilde{c}_{2}=
c_{2}+\frac{1}{2}\eta(3-k)\left[(2-k)\eta+2c_{1}\right]. 
\label{eq:507altmodgenck}\end{equation}
 In this way for any chosen $k\leq1$ we obtain a very simple modification
of $\beta^{(N)}(a)$ that  in contrast to (\ref{eq:317modbetE1})
contains only \emph{one} free parameter $\eta$. It is obvious from
the above construction that  $\beta$-functions constructed in
this way indeed satisfy the conditions
(I)--(III) and (V), formulated in Section 3.2.

Re-expanding $\tilde{\beta}(a)$ in powers of $a$ we find in the NL
order 
\begin{equation}
\tilde{\beta}^{(1)}(a)-\beta^{(1)}(a)=
\frac{b}{4}\left[2c_{1}(3-k)\eta+(3-k)(2-k)\eta^{2}\right]a^{4}+O(a^{5}),
\label{eq:509modbetacorr1}
\end{equation}
and in the NNL order 
\begin{eqnarray}
\lefteqn{\tilde{\beta}^{(2)}(a)-\beta^{(2)}(a)=
  \qquad\qquad\qquad\qquad\qquad\qquad} 
\nonumber \\
&&\frac{b}{12}(4-k)\eta\left[6c_{2}+3c_{1}(3-k)\eta+
(3-k)(2-k)\eta^{2}\right]a^{5}+\, O(a^{6}) \qquad 
\label{eq:511modbetacorr2}
\end{eqnarray}
 As we see, the condition (IV) is also satisfied, at least up to
 the NNL
order.

There is a price we must pay for the simplicity of these models
for $\tilde{\beta}^{(N)}(a)$, namely that the coefficient in the
large-$a$ behaviour of $\tilde{\beta}^{(N)}$ cannot be freely
adjusted and is uniquely fixed by $\eta$ and the coefficients of
the small-$a$ expansion. This is however perfectly consistent
with our general approach, since our goal is only to improve the
predictions at \textit{moderate} $Q^{2}$, and from this point of
view the exact behaviour of the couplant at \textit{very low}
$Q^{2}$ is not very important.

\begin{figure}[tb]
\begin{center}  
\includegraphics[width=0.55\textwidth]{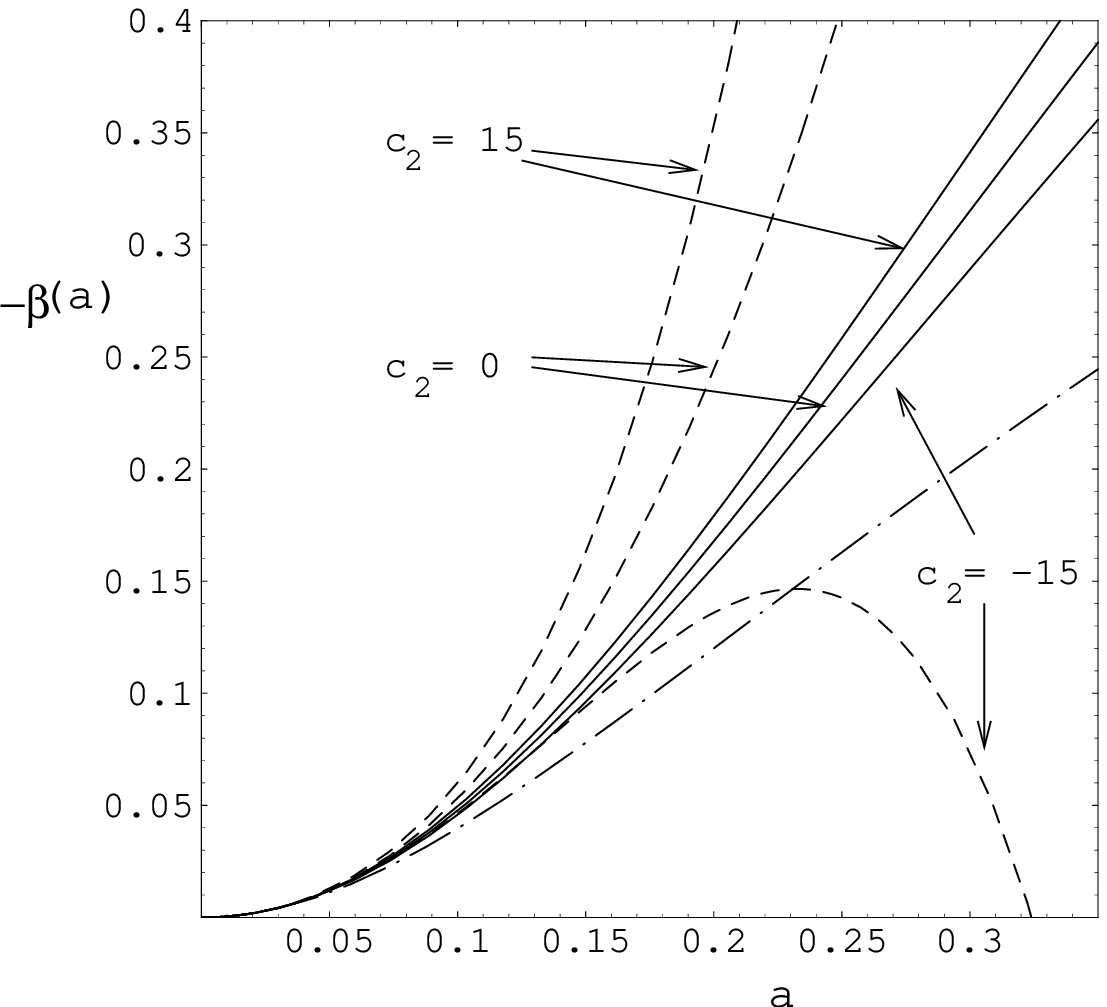}\end{center}
  \vspace{-.7cm}
\caption{\small Alternative model of a modified $\beta$-function, as
    given by Equation  
(\ref{eq:515bet2b8}) with $\eta=3.8$, in the NL order (dash-dotted
line) and the NNL order for three values of $c_{2}$ (solid lines),
compared with the conventional $\beta$-function (dashed
    lines). \label{fig:17betb8}} 
\end{figure}

For concrete numerical calculations we choose $\tilde{\beta}^{(N)}(a)$
of the form (\ref{eq:505altmodbetgen}) with\footnote{The $\beta$-function
of this form has been first discussed by the present author in
\cite{0530par00}.} $k=0$: 
\begin{eqnarray}
\tilde{\beta}^{(2)}(a)&=&-\frac{b}{2}\frac{a^{2}}{(1+\eta
  a)^{2}}\left[1+(c_{1}+2\eta)\frac{a}{1+\eta
    a}+\right. \nonumber \\
&& \left. +\,(c_{2}+3c_{1}\eta+3\eta^{2})\left(\frac{a}{1+\eta
    a}\right)^{2}\right].
\label{eq:515bet2b8}
\end{eqnarray}
 The plot of this $\beta$-function in the NL and NNL order for a
particular value of $\eta$ is shown in Figure
\ref{fig:17betb8} (in this plot we assumed $\eta=3.8$, which is
justified in Section~5.2). It is interesting that although this
$\beta$-function has different large-$a$ asymptotics than the
function given by Equation (\ref{eq:317modbetE1}), the qualitative
behaviour of these functions in the range
of $a$ which is important at moderate $Q^{2}$ is quite similar.

The corresponding function $F^{(N)}$ in the Equation  (\ref{eq:225intrgim}),
determining the $Q^{2}$-dependence of the couplant, has in this case
in the NL order a very simple form: 
\begin{equation}
F^{(1)}(a,c_{1},\eta)=-\frac{(c_{1}+2\eta)^{3}}{(c_{1}+3\eta)^{2}}
\ln\left[1+(c_{1}+3\eta)a\right]-\frac{\eta^{3}}{c_{1}+3\eta}\,
a.
\label{eq:517b8f1}\end{equation}
 The expression for $F^{(2)}$ in the NNL order for this $\beta$-function
is lengthy, but more manageable than the corresponding formula for
the function (\ref{eq:317modbetE1}):
\begin{eqnarray}
\lefteqn{F^{(2)}(a,c_{2},\eta) = -\frac{a\,\eta^{4}}{c_{2}+4\,
 c_{1}\,\eta+6\,\eta^{2}}-} \nonumber \\
 &  & -\,
 E_{2}\,\ln\left[1+a\,\left(c_{1}+4\,\eta\right)+a^{2}\,\left(c_{2}+
4\, c_{1}\,\eta+6\,\eta^{2}\right)\right]+\nonumber \\
 &  & +\,
 E_{3}\,\arctan\left(\frac{c_{1}+2\,\eta}{\sqrt{-c_{1}^{2}+
4\, c_{2}+8\, c_{1}\,\eta+8\,\eta^{2}}}\right)-\nonumber \\
 &  & -\, E_{3}\arctan\left(\frac{c_{1}+2\,\eta+a\,\left(2\,
 c_{2}+7\,
 c_{1}\,\eta+8\,\eta^{2}\right)}{\left(1+a\,\eta\right)\,\sqrt{-c_{1}^{2}+4\, 
 c_{2}+8\, c_{1}\,\eta+8\,\eta^{2}}}\right)
\label{eq:519b8f2}
\end{eqnarray}
where 
\begin{eqnarray}
E_{2} & = & \frac{c_{1}\, c_{2}^{2}+8\, c_{1}^{2}\, c_{2}\,\eta+
\left(16\, c_{1}^{3}+12\, c_{1}\,
c_{2}\right)\,\eta^{2}}{2\,\left(c_{2}+
4\, c_{1}\,\eta+6\,\eta^{2}\right)^{2}}+\nonumber \\
 &  & +\, \frac{\left(48\, c_{1}^{2}+4\,
 c_{2}\right)\,\eta^{3}+51\,
 c_{1}\,\eta^{4}+20\,\eta^{5}}{2\,\left(c_{2}+4\,
 c_{1}\,\eta+6\,\eta^{2}\right)^{2}},
\end{eqnarray}
 and 
\begin{eqnarray}
\lefteqn{E_{3} = } \nonumber \\
&&\frac{\left(c_{1}^{2}-2\, c_{2}\right)\, c_{2}^{2}+
4\, c_{1}\,\left(2\, c_{1}^{2}-5\, c_{2}\right)\, c_{2}\,\eta+
4\,\left(4\, c_{1}^{4}-13\, c_{1}^{2}\, c_{2}-6\,
c_{2}^{2}\right)\,\eta^{2}}
{\left(c_{2}+4\,
  c_{1}\,\eta+6\,\eta^{2}\right)^{2}\,\sqrt{-c_{1}^{2}+
4\, c_{2}+8\, c_{1}\,\eta+8\,\eta^{2}}}-\nonumber \\
&&\nonumber\\
 && -\,\frac{4\,\left(4\, c_{1}^{3}+37\, c_{1}\,
    c_{2}\right)\,\eta^{3}+9\,\left(19\, c_{1}^{2}+10\,
    c_{2}\right)\,\eta^{4}+232\,
    c_{1}\,\eta^{5}+92\,\eta^{6}}{\left(c_{2}+4\,
    c_{1}\,\eta+6\,\eta^{2}\right)^{2}\,\sqrt{-c_{1}^{2}+4\,
      c_{2}+8\, c_{1}\,\eta+8\,\eta^{2}}}
\end{eqnarray}
Strictly speaking, this form is valid for $a$, $c_{2}$
and $\eta$ satisfying the condition
\[
-c_{1}^{2}+4\, c_{2}+8\, c_{1}\,\eta+8\,\eta^{2}>0\,,
\]
 which also guarantees that  
\[
1+(c_{1}+4\eta)a+(c_{2}+4c_{1}\eta+6\eta^{2})a^{2}>0.
\]
 For other values of the parameters $c_{2}$ and $\eta$ the function
$F^{(2)}$ is obtained via analytic continuation.

Similarly as in the previous case we define the modified
expansion for a physical quantity $\delta$ by replacing  the
conventional couplant in the series expansion for $\delta$ with the
modified couplant determined by (\ref{eq:515bet2b8}). The PMS
approximants are also defined in a similar way as in Section~4.2. The NL order
approximant is defined by the equation (\ref{eq:401dd1dr1imp}),
where $\tilde{\beta}^{(1)}$ is now given by
(\ref{eq:515bet2b8}). Solving this equation with respect to 
$\bar{r}_{1}$ we obtain: 
\begin{equation}
\bar{r}_{1}(\bar{a})=\frac{-c_{1}+
3\eta^{2}\bar{a}+\eta^{3}\bar{a}^{2}}{2(1+(c_{1}+3\eta)\bar{a})} 
\label{eq:521b8pms1r1}\end{equation}
The PMS prediction for $\delta^{(1)}$ in this modified expansion
is then obtained in the same way as in Section~4.2. 

The NNL order PMS approximant for this model of the modified couplant is
defined by the Equations 
(\ref{eq:243dd2dr1}) and (\ref{eq:245dd2dc2}), with the relevant
partial derivatives given by (\ref{eq:404dd2dr1mod}) and
(\ref{eq:406dd2dc2mod}), where $\tilde{\beta}^{(2)}$ is now given by
(\ref{eq:515bet2b8}). Similarly as in the previous case, these
equations are too cumbersome to be reproduced here in the
full form, but it is easy to obtain an explicit approximate
solution for them in the leading
order of expansion in $\bar{a}$. Using (\ref{eq:407modpms2appB})
and taking into 
account the form of the coefficient 
$d_3$, which follows from the expansion
(\ref{eq:511modbetacorr2}), we find: 
\begin{equation}
\frac{\partial\tilde{\delta}^{(2)}}{\partial c_{2}}=
(2r_{1}-2\eta)a^{4}+O(a^{5}).
\label{eq:523b8pms2appB}
\end{equation}
Somewhat surprisingly, this coincides with the expression
(\ref{eq:407modpms2appB-b}), obtained for the previous model of
the modified couplant. This means that in the leading order in
$\bar{a}$ the approximate values of the PMS parameters for the
coupling defined by (\ref{eq:515bet2b8}) are the same as the
approximate PMS parameters (\ref{eq:409d2e1pmsapp-a}) and
(\ref{eq:409d2e1pmsapp-b}), obtained for the couplant defined by
(\ref{eq:317modbetE1}). However, looking at the
Equation~(\ref{eq:407modpms2appB}) for the approximate value of
the derivative of $\tilde{\delta}^{(2)}$ over $c_2$ and taking
into account the form of the relevant coefficient in 
 the Equation (\ref{eq:511modbetacorr2}) we immediately see that
 this is just a 
coincidence, i.e.\ for other values of $k$ in the modified
$\beta$-function (\ref{eq:505altmodbetgen}) the approximate values
of PMS
parameters would be different. As mentioned previously, this
approximate solution does not have much practical importance for
our discussion, but it shows that also for this type of the
modified couplant the solutions 
of the PMS equations in the NNL order exist and are unique, at
least in the weak coupling 
limit.

\section{Choosing $\eta$ for the alternative modified couplant}

\begin{figure}[tb]
\begin{center}  
\includegraphics[width=0.55\textwidth]{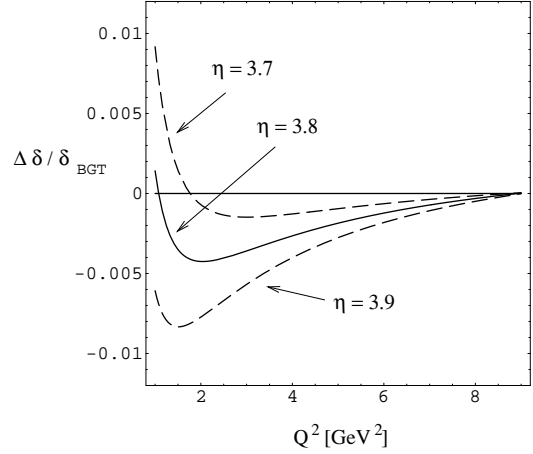}\end{center}
  \vspace{-.7cm}
\caption{\small The relative difference between the modified NNL order
    PMS prediction 
for $\delta_{{{\rm V}}}$, obtained with the couplant
    corresponding to (\ref{eq:515bet2b8}),
and the prediction $\delta_{{\rm BGT}}$ from the phenomenological
expression (\ref{eq:411bgt}), as a function of $Q^{2}$, for several
values of $\eta$. For this comparison the parameter
    $\Lambda_{\overline{{{\rm MS}}}}$ 
in the expression for $\delta_{{{\rm V}}}$ is adjusted in such
a way that  these quantities coincide at $Q^{2}=9\,\mbox{GeV}^{2}$.
\label{fig:18deveta2b8}}
\end{figure}

Similarly as in the previous case, the modified predictions in
low orders of the expansion depend on the value of the parameter
$\eta$, and in order to obtain meaningful predictions we have to
fix somehow this parameter. We use exactly the same procedure as
described in Section~4.3.  In Figure  \ref{fig:18deveta2b8}
we show the relative difference between the modified NNL order PMS
predictions for $\delta_{{{\rm V}}}$ evaluated with the $\beta$-function
(\ref{eq:515bet2b8}) and the phenomenological expression (\ref{eq:411bgt}).
As we see, the value giving the best fit appears to be
approximately 
$\eta=3.8$;
for this value the relative deviation is  
within $1\%$ down to $Q^{2}=1\,\mbox{GeV}^{2}$.

\section{$\delta_{{{\rm V}}}$ and $\delta_{{{\rm GLS}}}$ for 
    alternative modified couplant}

\begin{figure}[p]
\begin{center}  
\includegraphics[width=0.55\textwidth]{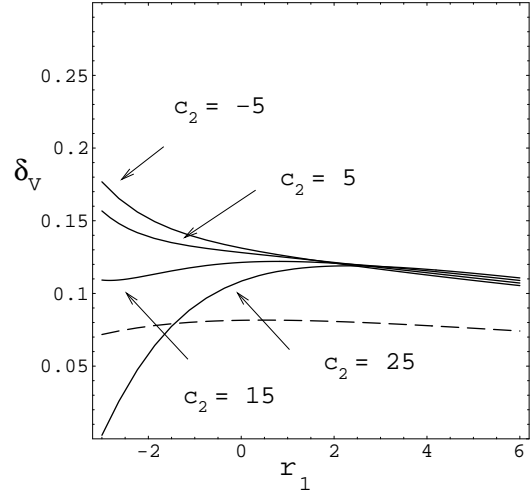}\end{center}
  \vspace{-.7cm}
\caption{\small $\delta_{{{\rm V}}}$ at $Q^{2}=3\,\mbox{GeV}^{2}$ 
(for $n_{f}=3$),
as a function of $r_{1}$, for several values of $c_{2}$, obtained
using the alternative modified couplant defined by Equation
(\ref{eq:515bet2b8}) 
with $\eta=3.8$. Dashed line indicates the NL order
prediction. \label{fig:19dvb8r1}} 
\end{figure}

In Figure \ref{fig:19dvb8r1} we show $\delta_{{{\rm V}}}$ at
$Q^{2}=3\,\mbox{GeV}^{2}$, as a function of $r_{1}$, for several
values of $c_{2}$, obtained with an alternative couplant defined
by the $\beta$-function (\ref{eq:515bet2b8}). As we
see, this figure is very similar to the corresponding Figure
\ref{fig:11dve1r1}, obtained with the previous model of the 
$\beta$-function. The same is true with respect to
$r_1$ dependence of the NNL order predictions for 
$\delta_{{{\rm
      GLS}}}$ at fixed $Q^2$ and the 
$Q^2$ dependence of the NNL order predictions for both quantities for
various (fixed) scheme 
parameters. 

\begin{figure}[p]
\begin{center}  
\includegraphics[width=0.55\textwidth]{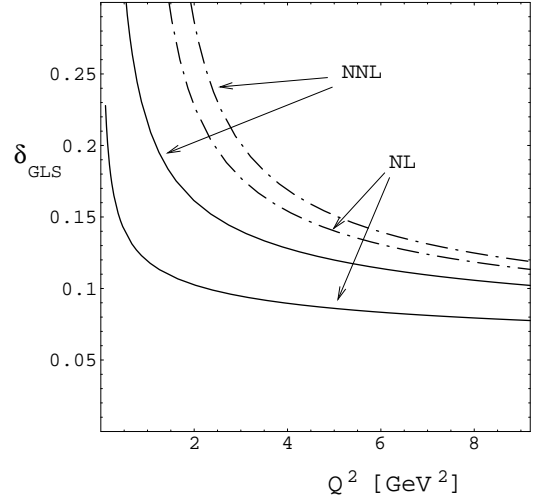}
\end{center}
  \vspace{-.7cm}\caption{\small The NL and NNL order PMS predictions for
    $\delta_{{{\rm GLS}}}$, 
as a function of $Q^{2}$, obtained with the alternative modified
couplant defined by Equation  (\ref{eq:515bet2b8}) with $\eta=3.8$ (solid
lines), compared with the conventional PMS predictions (dash-dotted
lines). \label{fig:20dgb8pmsqd}}
\end{figure}

In Figure \ref{fig:20dgb8pmsqd} we show the PMS predictions for
$\delta_{{{\rm GLS}}}$, obtained in the modified expansion with
the couplant defined by Equation (\ref{eq:515bet2b8}). This
figure should be compared with Figure \ref{fig:16dge1pmsqd},
representing PMS predictions for the same quantity, obtained with
the modified couplant defined by the Equation  
(\ref{eq:317modbetE1}).  We find that the NNL order predictions
obtained in both cases are practically identical. However, the NL
order modified predictions do not agree so well, which is not
surprising; they are nevertheless consistent, provided we take
into account the fact, that one should assign to those
predictions the theoretical error of the order of the difference
between NL and NNL order predictions in each case. Let us note
that the couplant defined by Equation (\ref{eq:515bet2b8}) leads
to larger NL/NNL difference, which means that the modified
expansion in this case is less precise.

\section{Modified expansion with other couplants}

All the calculations reported above have been performed for yet
another modified couplant, defined by the $\tilde{\beta}$ of the
form (\ref{eq:505altmodbetgen}) with $k=1$. This couplant is
interesting from the technical point of view, because the NL and
NNL order function $F^{(N)}$ in the Equation
(\ref{eq:205murgint}) for this couplant has a very simple form
(we give detailed formulas in the Appendix B), so it may be
useful also for other considerations in the QCD phenomenology. In
order to match the modified PMS predictions for $\delta_{{{\rm
V}}}$ to the phenomenological expression (\ref{eq:411bgt}) a
somewhat larger value of $\eta$ had to be used, but the modified NNL
order predictions for $\delta_{{{\rm GLS}}}$ were again in very
good agreement with the results obtained with  $\tilde{\beta}$
of the form (\ref{eq:317modbetE1}).

The consistency of the predictions obtained with different
modified couplants is rather encouraging, because it shows, that
despite some freedom in choosing the modified $\beta$-function
our approach leads to consistent physical predictions even in low orders
of perturbation expansion.  We see
two factors that might be responsible  for this ''universality''
of the predictions.  
One is that we have an adjustable parameter
$\eta$, which to some extent absorbs possible differences. Second
factor is the use of the PMS approximant for physical predictions:
using PMS we no longer choose the RS parameters by hand --- the values
of these parameters ``adjust themselves'' in an ``automatic''
way; presumably the exact behavior of the running couplant at low energies
is not so important, as long as it stays finite for all positive
$Q^{2}$. This gives additional motivation for using PMS in the modified
approach.

We might expect that this behaviour would be a general feature of the
modified perturbation expansion, i.e.\ concretely if we use the NNL order
expression for some physical quantity to fix the adjustable
parameters in $\tilde{\beta}$, then the NNL order predictions for
all other physical quantities should not be significantly altered
by changing the form of the modified $\beta$-function, at least
when the predictions are evaluated in the PMS scheme. However,  
the differences between predictions in successive orders for the
same physical quantity may vary, depending on $\tilde{\beta}$.

\chapter{Consequences for phenomenology}

\section{Reduced $Q^{2}$-dependence of the modified predictions }

Our analysis of the perturbative predictions for $\delta_{\rm V}$
and $\delta_{\rm GLS}$ has shown an interesting effect: the PMS
predictions in the modified expansion lie below the PMS
predictions in the conventional expansion. This effect is of
course not unexpected, since the initial motivation for our
construction of the modified expansion was exactly to reduce the
rate of growth of the couplant with decreasing $Q^2$ in some
schemes and to eliminate the 
singularity at positive $Q^{2}$. On the other hand, this effect
is not entirely trivial, because there is no immediate relation
between the low-$Q^{2}$ behaviour of the couplant in any
particular scheme and the low-$Q^{2}$ behaviour of the PMS
approximants: firstly, in some renormalization schemes the
modified predictions lie \textit{above} the conventional
predictions, as may be seen for example in Figure
\ref{fig:11dve1r1}, and secondly, the PMS scheme parameters for
the modified approximants are different from the PMS parameters
of the conventional approximants. If such a pattern would be
observed for all physical quantities, it might have important
consequences for phenomenology. It is therefore of some interest
to investigate, whether we may support our observation by some
general arguments.

An ideal solution would be to give some rigorous bounds on the
PMS predictions in the modified expansion, but this seems to be
a rather difficult problem. We shall therefore adopt a less
ambitious approach: in order to compare the modified PMS
predictions with the conventional PMS predictions in a given
order we shall expand both expressions in terms of a new couplant
$a_{0}$, defined by the implicit equation:
\begin{equation}
\frac{b}{2}\ln\frac{Q^{2}}{\Lambda_{\overline{{{\rm MS}}}}^{2}}=
r_{1}^{(0)\overline{{{\rm MS}}}}+
c_{1}\ln\frac{b}{2}+\frac{1}{a_{0}}+c_{1}\ln
a_{0}-c_{1}^{2}a_{0},  
\label{eq:601}\end{equation}
where $r_{1}^{(0)\overline{\rm MS}}$ denotes as usual the first
expansion coefficient for the considered physical quantity in the
$\overline{\mbox{MS}}$ scheme. The usefulness of the couplant
$a_0$ lies in the fact that it is nonsingular for all real
positive $Q^{2}$ and at the same time it has correct weak
coupling behaviour up to NL order. It satisfies the
renormalization group equation of the form 
\begin{equation}
Q^2\frac{da_0}{dQ^2}= 
-\frac{b}{2}\frac{a_0^2}{1-c_1 \,a_0 + c_1^2 \,a_0^2}
\label{eq:602}
\end{equation}
For the $N$-th order modified PMS approximants the first $N$
correction terms in the expansion in terms of $a_0$ must coincide
with the corresponding terms in the expansion of the conventional
PMS approximants (this follows from the general properties of our
modified expansion), so in order to make an estimate of the
difference between these two types of approximants we have to 
compare the first correction terms beyond the formal order of the
approximants.

Let us first consider the modified NL order PMS approximant,
defined by the Equation (\ref{eq:401d1e1pmsr1}). We want to calculate
the $O(a_{0}^{3})$ term in the expansion of this approximant in
powers of $a_0$. To this end we insert (\ref{eq:401d1e1pmsr1}) into
(\ref{eq:229a1con}) and we assume that the modified couplant
$\bar{a}(Q^2)$ in the PMS scheme is related to $a_0(Q^2)$ via the
relation
\begin{equation}
\bar{a}=a_0\left[1 + A_1 \,a_0 + A_2 \,a_0^2 +
  A_3 \,a_0^3 + 
  ... \right].
\label{eq:605}
\end{equation}
Making use of the defining equation (\ref{eq:601}) for
$a_0(Q^2)$ we find: 
\begin{equation}
A_1 = \frac{1}{2}\,c_1,\quad\quad A_2=\frac{1}{4}\,c_1^2 -
\frac{9}{2}\,c_1 \,\eta -\frac{9}{2}\, \eta^2 + 
\frac{3}{2}\,\kappa \eta^3. 
\label{eq:607}
\end{equation}
Taking this into account and expanding the expression for
$\bar{\delta}^{(1)}$ 
in terms of $a_0$ we obtain:
\begin{equation}
\bar{\delta}^{(1)}=a_0 + \left(\frac{1}{4}\,c_1^2 -3 c_1 \,\eta -3
\eta^2 + \kappa \eta^3\right) a_0^3 + O(a_0^4).
\label{eq:611}
\end{equation}
The relevant expansion for the conventional NL order PMS
approximant is obtained from this result by setting $\eta=0$. We
see therefore that the difference between the NL order
PMS approximants in the modified expansion and the conventional
expansion is given by  
\begin{equation}
\bar{\delta}^{(1)}_{\rm mod} - \bar{\delta}^{(1)}_{\rm con}=
\left(-3c_1 \,\eta -3\eta^2 + \kappa \eta^3 \right)a_0^3+O(a_0^4).
\label{eq:612}
\end{equation}
In our calculations we assumed $\kappa=2/b=4/9$; for this
value of $\kappa$ the coefficient of the leading term in this
difference is
negative 
for 
$\eta <8.21$. This shows,  that the modified
NL order PMS predictions  would indeed lie below the conventional PMS
predictions for all physical quantities that may be written in the form
(\ref{eq:201deltamu}), at least for small $a_0$.

In a similar way we obtain the expansion of the NNL order
modified PMS predictions up to the $O(a_0^4)$ term. We first look
for the solutions of the system of two equations
\begin{equation}
\frac{\partial}{\partial
  r_{1}}\delta^{(2)}(a(Q^{2},r_{1},c_{2}),r_{1},c_{2})\mid_{r_{1}= 
\bar{r}_{1},\,c_{2}=\bar{c}_{2}}=0, 
\end{equation}
 and
\begin{equation}
\frac{\partial}{\partial
  c_{2}}\delta^{(2)}(a(Q^{2},r_{1},c_{2}),r_{1},c_{2})\mid_{r_{1}=
\bar{r}_{1},\,   
  c_{2}=\bar{c}_{2}}=0, 
\end{equation}
in the form of the power series in $a_0$. We find:
\begin{equation}
\bar{r}_1=\eta + B_1 \,a_0 + B_2 \,a_0^2 + ...,
\label{eq:613}
\end{equation}
where
\begin{eqnarray}
B_1 & =& \frac{1}{2} \,\rho_2 + \frac{3}{2} \,c_1 \,\eta
+ \frac{5}{6} \,\eta^2 \nonumber \\
B_2 &=& -\rho_2 \,\eta -\frac{27}{8} \,c_1 \,\eta^2 - \frac{29}{12}\,\eta^3
+ \frac{3}{8}\,\kappa \eta^4, 
\label{eq:615}
\end{eqnarray}
and 
\begin{equation}
\bar{c}_2 = \frac{3}{2}\,\rho_2 + \frac{5}{2}\,c_1 \,\eta +
\frac{3}{2}\,\eta^2 + D_1 \,a_0 + ...,
\label{eq:617}
\end{equation}
where 
\begin{equation}
D_1 = \frac{1}{2}\,c_1 \,\rho_2 +\frac{3}{2}\, c_1^2 \,\eta + \frac{1}{3}
\,c_1 \,\eta^2 -\eta^3 +\frac{1}{2}\,\kappa \eta^4.
\label{eq:619}
\end{equation}
We then insert these values of $\bar{r}_1$ and $\bar{c}_2$ into
the defining equation (\ref{eq:400intrgeimp}) for the NNL order modified
PMS couplant 
$\bar{a}$, and we look for the solution for $\bar{a}$ in the form of
the expansion (\ref{eq:605}). Taking into account the defining
equation for $a_0$ we now find:
\begin{eqnarray}
A_1&=& -\eta \nonumber \\
A_2&=& \rho_2 + \frac{5}{3} \,\eta^2 \nonumber \\
A_3&=& \frac{1}{2}\,c_1^3-\frac{11}{2}\,\rho_2 \,\eta - 
\frac{57}{8}\,c_1\,
\eta^2 -
\frac{89}{12} \,\eta^3 + \frac{5}{8} \,\kappa \eta^4 
\label{eq:621}
\end{eqnarray}
Inserting the expressions for $\bar{r}_1$, $\bar{c}_2$ and
$\bar{a}$ into the expression for
$\bar{\delta}^{(2)}$ we find:
\begin{equation}
\bar{\delta}^{(2)}=a_0 + \rho_2 \,a_0^3+ \left(\frac{1}{2}\,c_1^3
-3\rho_2 \,\eta -\frac{11}{2}\,c_1 \,\eta^2 -3\eta^3 +
\frac{1}{2}\,\kappa \eta^4\right) a_0^4 + O(a_0^5).
\label{eq:623}
\end{equation}
Again, the relevant expansion for the conventional NNL order PMS
approximant is obtained from this expression by setting
$\eta=0$. We see therefore that the difference between the NNL
order PMS predictions in the modified expansion and the
conventional expansion is equal to
\begin{equation}
\bar{\delta}^{(2)}_{\rm mod} - \bar{\delta}^{(2)}_{\rm con}=
\left(-3\rho_2 \eta -\frac{11}{2}\,c_1 \,\eta^2 -3\eta^3 +
\frac{1}{2}\,\kappa \eta^4\right)a_0^4+O(a_0^5).
\label{eq:625}
\end{equation}
For $\kappa=2/b$ the coefficient in the leading term is negative
for $\eta <16.40$. This shows that (at least for small $a_0$)
also in the NNL order  the modified PMS 
predictions  would lie below the conventional
PMS predictions for all physical quantities that may be expressed
in the form (\ref{eq:201deltamu}).

\section{Some fits for the GLS sum rule }

The fact that  the predictions for physical quantities obtained in
the modified expansion lie below the predictions obtained in conventional
expansion and evolve less rapidly with $Q^{2}$ may have interesting
consequences for phenomenology. To illustrate this, let us again consider
the GLS sum rule and let us see, how the use of the modified expansion
affects the results of the fits to experimental data. We shall rely
on the result reported in \cite{0305kim98}, where it was found that 
measurements of deep inelastic neutrino-nucleon scattering for 
$1<Q^{2}<15\,\mbox{GeV}^{2}$
imply 
\begin{equation}
\alpha_{s}^{\overline{{{\rm MS}}}}(3\,\mbox{GeV}^{2})=
0.23\pm0.035(\mbox{stat})\pm0.05(\mbox{sys}). 
\end{equation}
Using the $n_{f}=3$ NNL order formula\footnote{To be precise, one
  should take into account at each value of $Q^2$ the
  contributions from different values of $n_f$, as discussed in  
\cite{0317chyl92}, but this makes the whole discussion much more
  complicated. In order to see clearly possible  effects
  associated with the use of the modified couplant instead of the
  conventional couplant we limit our discussion to the
  case of $n_f=3$.}
 for $\delta_{{{\rm GLS}}}$
we may interpret this result as 
$\delta_{{{\rm GLS}}}^{{{\rm exp}}}(3\,\mbox{GeV}^{2})=0.131\pm0.40$,
where we have added the statistical and systematic errors in quadrature.
This would be the starting point of our phenomenological analysis.
We fit $\Lambda_{\overline{{{\rm MS}}}}^{(3)}$ to this number
using various theoretical expressions for $\delta_{{{\rm GLS}}}$,
and then we convert the result into the corresponding value of 
$\alpha_{s}^{\overline{{{\rm MS}}}}(M_{Z}^{2})$.
To relate the values of the couplant $a=\alpha_{s}/\pi$ for
  different numbers of 
quark flavors we use the matching formula 
\cite{0549wetz82,0551bern82,0553bern83,0555bern83b,0557larin95,0559chet97}:
\begin{eqnarray}
\lefteqn{a(\mu^{2},n_{f}) = a(\mu^{2},n_{f}-1)\left[1+
\frac{1}{3}\ln\frac{\mu}{m_{q}}\,a(\mu^{2},n_{f}-1)\,+\right. 
\qquad\qquad}\nonumber \\
 &  & \left.+\frac{1}{9}\,\left(\ln^{2}\frac{\mu}{m_{q}}+
\frac{57}{4}\ln\frac{\mu}{m_{q}}-
\frac{11}{8}\,\right)\,a^{2}(\mu^{2},n_{f}-1)\right]
\label{eq:603}\end{eqnarray}
 where $m_{q}$ is the running mass of the decoupled quark, evaluated
at the scale $\mu^{2}=m_{q}^{2}$. (For evolution of the couplant
in the NL order we use the above formula restricted to the first two
terms.) We use $m_{c}=1.3\,\mbox{GeV}$ and $m_{b}=4.3\,\mbox{GeV}$
and choose as the matching points $\mu^{2}=4m_{q}^{2}$. We use the
same matching formula both in the conventional and the modified expansion
(which is justified, because the difference between the modified
perturbative expression and the
conventional perturbative expression at given order is formally
of higher order). 

As a consistency check of our approach we first perform the fit in
the $\overline{\mbox{MS}}$ scheme. Using the conventional NNL order
expression we find
$\alpha_{s}(M_{Z}^{2})=0.1126\pm_{0.0116}^{0.0072}$. (We keep an
extra digit in the quoted numbers and present 
asymmetric errors in order to minimize the roundoff errors and 
illustrate the sensitivity of the fitted values in a better way.) 
This number is slightly smaller than the result quoted in the original
paper \cite{0305kim98}; the difference may be presumably attributed
to the different method of matching the couplant across the thresholds
and/or different way of calculating the $Q^{2}$-evolution of the
couplant. 
If we use the conventional NL order 
approximant, we find $\alpha_{s}(M_{Z}^{2})=0.1162\pm_{0.0127}^{0.0081}$,
which in turn is slightly higher than the value quoted in \cite{0305kim98}.
Note that both numbers are below the world average 
$\alpha_{s}(M_{Z}^{2})=0.1176\pm0.0020$
obtained in \cite{0563pdg04} and the world average 
$\alpha_{s}(M_{Z}^{2})=0.1182\pm0.0027$
calculated in \cite{0565beth04}, and in the case of the NNL order
approximant the difference is significant.

Now if we perform the same fit in the $\overline{\mbox{MS}}$ scheme,
using the modified perturbative expression with the  coupling
defined by the Equation~(\ref{eq:317modbetE1}) 
with $\eta=4.1$, we find in the NNL order 
$\tilde{\alpha}_{s}(M_{Z}^{2})=0.1147\pm_{0.0126}^{0.0084}$. 
(This is the \emph{modified} coupling at the scale $\mu^{2}=M_{Z}^{2}$,
and we used the \emph{modified} renormalization group equation to
evolve from $\mu^{2}=3\,\mbox{GeV}^{2}$ to $\mu^{2}=M_{Z}^{2}$;
it is meaningful to compare this value with $\alpha_{s}(M_{Z}^{2})$
from other sources, because it is exactly this value that we would
insert for example into the expression for the QCD corrections to
$\Gamma(Z\rightarrow\mbox{hadrons})$, if we would want to make a direct
comparison with experimental data at $\mu^{2}=M_{Z}^{2}$.) When the
NL order approximant is used, the result is 
$\tilde{\alpha}_{s}(M_{Z}^{2})=0.1224\pm_{0.0154}^{0.0108}$.
As we see, if the modified expansion is used in the fit, then 
somewhat higher values of $\alpha_{s}(M_{Z}^{2})$ are obtained.

If we perform the same fits using the conventional PMS
approximants, we obtain $\alpha_{s}(M_{Z}^{2})=0.1097\pm_{0.0102}^{0.0058}$
in the NNL order and $\alpha_{s}(M_{Z}^{2})=0.1099$ $\pm_{0.0104}^{0.0061}$
in the NL order. Note that  these values are smaller than values
obtained from the fit in the $\overline{\mbox{MS}}$ scheme and
much smaller than the 
world average quoted above. However, if we use the PMS approximants
in the modified expansion with the $\beta$-function (\ref{eq:317modbetE1})
and $\eta=4.1$, we obtain 
$\tilde{\alpha}_{s}(M_{Z}^{2})=0.1150\pm_{0.0127}^{0.0084}$
in the NNL order and 
$\tilde{\alpha}_{s}(M_{Z}^{2})=0.1156\pm_{0.0132}^{0.0089}$
in the NL order. Similarly as in the case of the $\overline{\mbox{MS}}$
scheme, the use of the modified expansion gives higher values of 
$\alpha_{s}(M_{Z}^{2})$, that is close to the present world
average. In the case 
of the PMS approximants this effect is much more pronounced than in
the case of the approximants in the $\overline{\mbox{MS}}$ scheme.
Interestingly, the difference between the NL and NNL order results,
which may be taken as an estimate of the theoretical error, is in
this case much
smaller in the PMS scheme than in the $\overline{\mbox{MS}}$ scheme,
both in the conventional and the modified expansion.

Both in the $\overline{\mbox{MS}}$ scheme and the PMS scheme we
find that the value of $\alpha_{s}(M_{Z}^{2})$ 
arising from the fit to the experimental data is higher when the
modified expansion is used.  This
is a welcome effect, because it makes the  low and high energy
determinations of 
the strong coupling parameter more consistent. (The world
  averages for  $\alpha_{s}(M_{Z}^{2})$ quoted above are  dominated by
  measurements at higher energies, corresponding to $n_f=4,5$.) 
The shift in the central value is not
very big in comparison with the experimental error for the GLS
sum rule; however, for other physical quantities measured at
moderate energies the experimental errors may be much smaller,
while the shift in central value would remain the same.

We see therefore that  within the modified perturbative approach it
could be easier to accommodate in a consistent way some measurements
giving relatively small value of the strong coupling parameter at
moderate energies, as compared with the measurements of the
strong coupling parameter 
at high energies. It would be interesting for example to analyze in
detail within the modified approach the QCD corrections to the decay
rates of the heavy quarkonia, which tend to give rather low values of
the strong coupling constant, as has been recently discussed for example
in \cite{0567field02}.

\chapter{Summary and outlook}

In this report we considered the problem of reliability of
perturbative QCD predictions at moderate values of $Q^2$ (i.e.\
of the order of few GeV$^2$). Using as an example the
next-to-leading and next-to-next-to-leading expressions for the
effective charge $\delta_{{\rm V}}$, related to the static
interquark potential, and the QCD correction $\delta_{{\rm GLS}}$
to the Gross-Llewellyn-Smith sum rule in the deep inelastic
neutrino-nucleon scattering, we have shown that QCD predictions
at moderate energies obtained with conventional renormalization
group improved perturbation expansion depend strongly on the
choice of the renormalization scheme (Sections~2.3--2.5), even
for a conservative choice of the renormalization scheme
parameters (condition (\ref{eq:263el2})). This casts doubt on the
reliability of the commonly used perturbative expressions in this
energy range. Taking closer look at the conventional expansion we
observed that one of the possible sources of the strong
renormalization scheme dependence of perturbative predictions may
be the strong scheme dependence of the conventionally used
running QCD coupling parameter (the couplant) itself; an extreme
manifestation of the scheme dependence of the conventional
couplant is the presence of a scheme dependent singularity
(Landau singularity) at nonzero value of $Q^2$ (Section~3.1).

As a step towards improving the reliability of perturbative
predictions we proposed to replace the conventional couplant in
the perturbative approximant of a given order by the modified
couplant, obtained by integrating the renormalization group
equation with a modified, nonpolynomial $\beta$-function that in
each order is a generalization of the polynomial expression for
the $\beta$-function used in the conventional expansion. We 
formulated some general criteria that the sequence of modified
$\beta$-functions has to satisfy in order to generate a useful
modification of the conventional perturbation expansion
(Section~3.2).  We constructed a concrete model
sequence of the modified $\beta$-functions that satisfies our
criteria, using the so called mapping method
(Equation~(\ref{eq:317modbetE1})).  One of the properties of our
model is that in each order of perturbation
expansion the modified couplant is free from the Landau
singularity, despite the fact that the $\beta$-function does not
contain any  essentially nonperturbative
terms. We then performed a detailed
analysis of the modified NL and NNL order predictions for
$\delta_{{\rm V}}$ and $\delta_{{\rm GLS}}$ in various
renormalization schemes. In particular, we
generalized to the case of the modified perturbation expansion
the equations defining the renormalization scheme distinguished
by the Principle of Minimal Sensitivity (Section~4.2).  The
sequence of modified $\beta$-functions in our approach contains 
two parameters: parameter $\kappa$, related to the asymptotic
behaviour of the modified couplant in the limit $Q^2 \rightarrow
0$, and the parameter $\eta$, which characterizes the value of
the modified couplant, for which the nonpolynomial character of
the modified $\beta$-function becomes important. The presence of
these parameters gives us the opportunity to improve the accuracy
of low order perturbative approximants by using use some information from
outside of the perturbation theory. We proposed to fix the
parameter $\kappa$ in such a way that the modified couplant in
every order has a $1/Q^2$ behaviour in the limit $Q^2 \rightarrow
0$, and to adjust the parameter $\eta$ so that the modified NNL
order PMS predictions for $\delta_{{\rm V}}$ would match certain
phenomenological expression for this effective charge
(Section~4.3). We have 
then shown that for the chosen values of the these parameters the
modified perturbative predictions for $\delta_{{\rm V}}$ and
$\delta_{{\rm GLS}}$ are much less sensitive to the choice of the
renormalization scheme than the predictions obtained in the
conventional expansion (Section~4.4). The observed pattern of the
scheme
dependence of the modified expansion indicates in particular that the
PMS predictions in this expansion stand on a better footing
than in the conventional expansion.

The general conditions on the sequence of the modified
$\beta$-functions in our approach leave some freedom in choosing
the concrete form of these functions. In order to see, how this
freedom might affect the modified predictions in low orders of
the expansion we considered an
alternative model sequence of $\beta$-functions
(Equation~(\ref{eq:515bet2b8})) and we repeated the whole
analysis of $\delta_{{\rm V}}$ and $\delta_{{\rm GLS}}$ for this
new model (Section~5.3). We found that the results agree quite
well with the predictions obtained in the previous model. We
argued that in the PMS scheme this is likely to be true for any
other $\beta$-function 
satisfying our criteria.

The modified predictions for $\delta_{{\rm V}}$ and $\delta_{{\rm
GLS}}$ considered in this report have an interesting property
that their evolution with $Q^2$ is less rapid than that of the
conventional predictions. In the case of the NL and NNL order PMS
predictions we have given a general argument, that this would be
true for all physical quantities in QCD of the form considered in
this report, at least for relatively large values of $Q^2$ (i.e.\
small values of the modified couplant).  This has been achieved
by expanding the PMS approximants in terms of certain regular
couplant to one order beyond the formal order of the
approximant (Section~6.1). To see, what might be the
phenomenological 
significance of this effect we made a simple fit to the
experimental data for $\delta_{{\rm GLS}}$ (Section~6.2). After
extrapolation 
to the energy scale of $M_Z$ we found that the obtained values of
$\alpha_{s}(M_{Z}^{2})$ are shifted upwards (by about $5\%$) and
are closer the world average (dominated by measurements at higher
energies) than the values
obtained with the conventional expansion. This suggests that
within the modified expansion it might be easier to accommodate in
a consistent way the relatively low values of the couplant obtained from
some low-$Q^2$ experiments and  relatively large values of the
couplant from some experiments at high $Q^2$.

In our discussion we have given explicit formulas describing the 
$Q^2$ dependence of the modified couplant for various model
$\beta$-functions that are perturbatively consistent with the
conventional expansion, but are free from Landau
singularity (Equations~(\ref{eq:325f1E1}), (\ref{eq:519b8f2}),
(\ref{eq:903altbet2}), 
(\ref{eq:601})). These very simple expression are interesting 
in their own right and might be useful in other theoretical
considerations, unrelated to the concepts presented in this
report.

Our considerations in this report had exploratory character ---
they should be regarded more as a ''feasibility study'' of
certain idea for 
the modification of the QCD perturbation expansion, than a very
precise quantitative analysis of concrete predictions. It seems
however that the results are sufficiently encouraging to justify
a comprehensive study of various perturbative QCD predictions at
moderate energies within the modified perturbation expansion
approach. Such study could include the perturbative QCD effects
in the physics 
of heavy quarkonia, in hadronic decays of the $\tau$ leptons and
in the deep inelastic lepton-nucleon scattering at the lower end
of the energy 
scale. It would be then interesting to compare the results of
such comprehensive analysis with the high energy measurements of
the strong coupling parameter $\alpha_{s}$. As a first step
towards such an analysis one should perform a more detailed study
of possible methods to constrain the parameters $\eta$ and
$\kappa$ in the proposed sequence of the modified
$\beta$-functions, generalizing the very simple approach adopted
in this report.  Another interesting line of research would be to
study the resummation of the QCD perturbation expansion in the
modified approach --- the stability of the modified couplant with
respect to change of the renormalization scheme should have
favorable effect on the stability and the rate of convergence of
the results obtained with various summation methods. In this
context it would be interesting to investigate, whether the PMS
prescription applied to the modified expansion does indeed act as
an automatic resummation procedure for divergent perturbation
series.

\appendix

\chapter{Solution of the inequality $\sigma_{2}(r_{1},c_{2})\leq l|\rho_{2}|$}

In this Appendix we give a detailed description
of the sets of the NNL order renormalization scheme parameters
$r_{1}$ and $c_{2})$ that satisfy 
the condition 
\begin{equation}
\sigma_{2}(r_{1},c_{2})\leq l|\rho_{2}|,
\label{eq:801sig-el}\end{equation}
 where $l\geq1$. Let us recall, that $\rho_2$
 is given by the Equation (\ref{eq:219rho2})
\[
\rho_{2}=c_{2}+r_{2}-c_{1} r_{1}- r_{1}^{2},
\]
and $\sigma_2(r_{1},c_{2})$ is given by the Equation (\ref{eq:261sigma2})
\[
\sigma_{2}(r_{1},c_{2})=|c_{2}|+|r_{2}(r_{1},c_{2})|+c_{1}|r_{1}|+r_{1}^{2}.
\]
We assume of course $\rho_2\neq 0$. We have to consider several
cases, depending on the value of $\rho_2$ and $l$. 

\vspace{11pt}

\noindent 1. For $\rho_{2}\geq c_{1}^{2}/4$ we
 first define the 
following parameters:
\begin{eqnarray}
r_{1}^{{{\rm min}}}&=&-\sqrt{\rho_{2}(l-1)/2}, \\
r_{1}^{{{\rm max}}}&=&\frac{1}{2}\left(-c_{1}+\sqrt{c_{1}^{2}+
2(l-1)\rho_{2}}\right),\\
c_{2}^{{{\rm min}}}&=&-\rho_{2}(l-1)/2,\\
c_{2}^{{{\rm max}}}&=&\rho_{2}(l+1)/2,\\
c_{2}^{{{\rm int}}}&=&c_{1}r_{1}^{{{\rm min}}}+c_{2}^{{{\rm max}}}.
\end{eqnarray}
 For $c_{2}>0$ the set of scheme parameters in the $(r_{1},c_{2})$
plane satisfying the condition (\ref{eq:801sig-el}) is bounded by
 the segments joining points $(r_{1}^{{{\rm min}}},0)$, 
$(r_{1}^{{{\rm min}}},c_{2}^{{{\rm int}}})$, $(0,c_{2}^{{{\rm max}}})$,
$(r_{1}^{{{\rm max}}},c_{2}^{{{\rm max}}})$, $(r_{1}^{{{\rm max}}},0)$.
For $c_{2}<0$ this set is bounded by the curves 
\begin{equation}
c_{2}(r_{1})=r_{1}^{2}+c_{2}^{{{\rm min}}}\quad\mbox{for}
\quad r_{1}^{{{\rm min}}}\leq r_{1}\leq0,
\end{equation}
\begin{equation}
c_{2}(r_{1})=r_{1}^{2}+c_{1}r_{1}+c_{2}^{{{\rm
      min}}}\quad\mbox{for}
\quad0\leq r_{1}\leq r_{1}^{{{\rm max}}}.
\end{equation}

\vspace{11pt}

\noindent 2. In the case $0<\rho_{2}<c_{1}^{2}/4$ and 
$l\notin<\frac{4-d-4\sqrt{1-d}}{d},\frac{4-d+4\sqrt{1-d}}{d}>$,
where $d=4\rho_{2}/c_{1}^{2}$, the description of the relevant set
of parameters is the same as in the case $\rho_{2}\geq c_{1}^{2}/4$. For
$l\in<\frac{4-d-4\sqrt{1-d}}{d},\frac{4-d+4\sqrt{1-d}}{d}>$ we have
to redefine the parameters $r_{1}^{{{\rm min}}}$ and $c_{2}^{{{\rm int}}}$:
\begin{eqnarray}
r_{1}^{{{\rm min}}}&=&-\rho_{2}(l+1)/2c_{1}\\
c_{2}^{{{\rm int}}}&=&(r_{1}^{{{\rm min}}})^{2}+c_{2}^{{{\rm
      min}}}.
\end{eqnarray}
 For $c_{2}>0$ the set of parameters satisfying the condition 
(\ref{eq:801sig-el})
is bounded by the segments joining  points $(r_{1}^{{{\rm min}}},0)$,
$(0,c_{2}^{{{\rm max}}})$, $(r_{1}^{{{\rm max}}},c_{2}^{{{\rm max}}})$,
$(r_{1}^{{{\rm max}}},0)$. For $c_{2}<0$ this set is bounded by
the curves 
\begin{equation}
c_{2}(r_{1})=r_{1}^{2}+c_{2}^{{\rm min}}\quad\mbox{for}
\quad r_{1}^{{{\rm min}}}\leq r_{1}\leq0,
\end{equation}
 \begin{equation}
c_{2}(r_{1})=r_{1}^{2}+c_{1}r_{1}+c_{2}^{{{\rm
      min}}}\quad\mbox{for}
\quad0\leq r_{1}\leq r_{1}^{{{\rm max}}},
\end{equation}
 supplemented by a segment  joining points $(r_{1}^{{{\rm min}}},0)$
and $(r_{1}^{{{\rm min}}},c_{2}^{{{\rm int}}})$.

\vspace{11pt}

\noindent 3. For $\rho_{2}<0$ and
$|\rho_{2}|\geq 2c_{1}^{2}(l+1)/(l-1)^{2}$ we define 
the parameters 
\begin{eqnarray}
r_{1}^{{{\rm min}}}&=&-\sqrt{|\rho_{2}|(l+1)/2},\\
r_{1}^{{{\rm max}}}&=&\frac{1}{2}\left(-c_{1}+
\sqrt{c_{1}^{2}+2(l+1)|\rho_{2}|}\right),\\
c_{2}^{{{\rm min}}}&=&-|\rho_{2}|(l+1)/2,\\
c_{2}^{{{\rm max}}}&=&|\rho_{2}|(l-1)/2,\\
c_{2}^{{{\rm int}}}&=&c_{1}r_{1}^{{{\rm min}}}+c_{2}^{{{\rm
      max}}}.
\end{eqnarray}
 For $c_{2}>0$ the set of scheme parameters in the $(r_{1},c_{2})$
plane satisfying the condition (\ref{eq:801sig-el}) is bounded by
the segments joining  points $(r_{1}^{{{\rm min}}},0)$, 
$(r_{1}^{{{\rm min}}},c_{2}^{{{\rm int}}})$, $(0,c_{2}^{{{\rm max}}})$,
$(r_{1}^{{{\rm max}}},c_{2}^{{{\rm max}}})$, $(r_{1}^{{{\rm max}}},0)$.
For $c_{2}<0$ this set is bounded by the curves  
\begin{equation}
c_{2}(r_{1})=r_{1}^{2}+c_{2}^{{{\rm min}}}\quad\mbox{for}
\quad r_{1}^{{{\rm min}}}\leq r_{1}\leq0,
\end{equation}
\begin{equation}
c_{2}(r_{1})=r_{1}^{2}+c_{1}r_{1}+c_{2}^{{{\rm
      min}}}\quad\mbox{for}
\quad0\leq r_{1}\leq r_{1}^{{{\rm max}}}.
\end{equation}
 
\vspace{11pt}

\noindent 4. For $\rho_{2}<0$ and $|\rho_{2}|<2c_{1}^{2}(l+1)/(l-1)^{2}$ we
have to redefine again some parameters: 
\begin{eqnarray}
r_{1}^{{{\rm min}}}&=&-|\rho_{2}|(l-1)/2c_{1},\\
c_{2}^{{{\rm int}}}&=&(r_{1}^{{{\rm min}}})^{2}+c_{2}^{{{\rm
      min}}}.
\end{eqnarray}
 For $c_{2}>0$ the set of scheme parameters in the $(r_{1},c_{2})$
plane satisfying the condition(\ref{eq:801sig-el}) is bounded by
the segments joining points $(r_{1}^{{{\rm min}}},0)$, 
$(0,c_{2}^{{{\rm max}}})$, $(r_{1}^{{{\rm max}}},c_{2}^{{{\rm max}}})$,
$(r_{1}^{{{\rm max}}},0)$. For $c_{2}<0$ this set is bounded
by the segment joining points $(r_{1}^{{{\rm min}}},0)$
and $(r_{1}^{{{\rm min}}},c_{2}^{{{\rm int}}})$ and the curves
\begin{equation}
c_{2}(r_{1})=r_{1}^{2}+c_{2}^{{{\rm min}}}\quad\mbox{for}
\quad r_{1}^{{{\rm min}}}\leq r_{1}\leq0,
\end{equation}
\begin{equation}
c_{2}(r_{1})=r_{1}^{2}+c_{1}r_{1}+c_{2}^{{{\rm
      min}}}\quad\mbox{for}\quad0\leq r_{1}\leq r_{1}^{{{\rm
              max}}}.
\end{equation}

\vspace{11pt}

One way to obtain these solutions is to consider 
 a {\em three}
dimensional space $(r_1, c_2, r_2)$ and carefully analyze all possible
intersections of a (closed) surface defined by the equation  
\begin{equation}
|c_{2}|+|r_{2}|+c_{1}|r_{1}|+r_{1}^{2}=2|\rho_2|
\end{equation}
with the surface
\begin{equation}
r_2=\rho_2 - c_2 +c_1 r_1 +r_1^2. 
\end{equation}
Calculations are straightforward, but tedious.

\chapter{Yet another model of a modified couplant}

In this Appendix we give an explicit solution for a particular model
of the modified couplant that  has not been used in the main text,
but is so simple that  it may be interesting in its own right. For
$\tilde{\beta}^{(1)}(a)$ of the form: 
\begin{equation}
\tilde{\beta}^{(1)}(a)=-\frac{b}{2}\frac{a^{2}}{1+\eta
  a}\left[1+(c_{1}+\eta)\frac{a}{1+\eta a}\right],
\label{eq:901altbet}
\end{equation}
 the function $F^{(1)}(a,\eta)$ in the Equation  (\ref{eq:225intrgim})
has a very simple form: 
\begin{equation}
F^{(1)}(a,\eta)=-\frac{(c_{1}+\eta)^{2}}{(c_{1}+2\eta)}
\ln\left[1+(c_{1}+2\eta)a\right].
\end{equation}
 For $\tilde{\beta}^{(2)}(a)$ of the form: 
\begin{eqnarray}
\tilde{\beta}^{(2)}(a)&=&-\frac{b}{2}\frac{a^{2}}{1+\eta
  a}\left[1+(c_{1}+\eta)\frac{a}{1+\eta
    a}+ \right. \nonumber \\
&& \left. +\, (c_{2}+2c_{1}\eta+\eta^{2})\left(\frac{a}{1+\eta
    a}\right)^{2}\right],
\end{eqnarray}
 the function $F^{(2)}(a,c_{2},\eta)$ in the Equation
 (\ref{eq:225intrgim}) 
is slightly more complicated: 
\begin{eqnarray}
F^{(2)}(a,c_{2},\eta) & = &
 -E_{2}\ln\left[1+(c_{1}+3\eta)a+
(c_{2}+3c_{1}\eta+3\eta^{2})a^{2}\right]+\nonumber
 \\
 &  & +\, E_{3}\arctan\left[\frac{c_{1}+\eta+\frac{2a(c_{2}+2\eta
       c_{1}+\eta^{2})}{1+\eta
       a}}{\sqrt{-c_{1}^{2}+4c_{2}+6c_{1}\eta+3\eta^{2}}}\right]-
 \nonumber \\
 &  &
 -\, E_{3}\arctan\left[\frac{c_{1}+\eta}{\sqrt{-c_{1}^{2}+4c_{2}+
6c_{1}\eta+3\eta^{2}}}\right]
\label{eq:903altbet2}
\end{eqnarray}
 where
\begin{equation}
E_{2}=\frac{(c_{1}c_{2}+3c_{1}^{2}\eta+3c_{1}\eta^{2}+
\eta^{3})}{2(c_{2}+3c_{1}\eta+3\eta^{2})}
\end{equation}
and 
\begin{equation}
E_{3}=\frac{c_{2}(-c_{1}^{2}+2c_{2})-3c_{1}\eta(c_{1}^{2}-3c_{2})+
6(c_{1}^{2}+c_{2})\eta^{2}+10c_{1}\eta^{3}+3\eta^{4}}{(c_{2}+
3c_{1}\eta+3\eta^{2})\sqrt{-c_{1}^{2}+4c_{2}+6c_{1}\eta+3\eta^{2}}}\,.
\end{equation}

\end{document}